\documentclass[prd,onecolumn,nofootinbib,aps,tightenlines,preprintnumbers,notitlepage,longbibliography,superscriptaddress,12pt]{revtex4-1}

\usepackage[dvipsnames]{xcolor}
\usepackage{color}
\usepackage{tikz}
\usepackage[caption=false]{subfig}
\usepackage{hyperref}
\usepackage{multirow}
\usepackage{siunitx}
\usepackage[export]{adjustbox}
\usepackage{graphicx}
\usepackage{dcolumn}
\usepackage{bm}
\usepackage[normalem]{ulem}
\usepackage{epstopdf}
\usepackage{tabularx}
\usepackage{suffix}
\usepackage{mathtools}

\graphicspath{ {figures/} }
\usetikzlibrary{arrows.meta,
                chains,
                positioning,
                quotes,
                shapes.geometric}

\tikzset{FlowChart/.style={
startstop/.style = {rectangle, rounded corners, draw, fill=blue!10,
                    minimum width=3cm, minimum height=1cm, align=center,
                    on chain, join=by arrow},
  process/.style = {rectangle, draw, fill=orange!10,
                    minimum width=3cm, minimum height=1cm, align=center,
                    on chain, join=by arrow},
 decision/.style = {diamond, aspect=1.5, draw, fill=green!30,
                    minimum width=3cm, minimum height=1cm, align=center,
                    on chain, join=by arrow},
       io/.style = {trapezium, trapezium stretches body,   
                    trapezium left angle=70, trapezium right angle=110,
                    draw, fill=blue!30,
                    minimum width=3cm, minimum height=1cm,
                    text width =\pgfkeysvalueof{/pgf/minimum width}-2*\pgfkeysvalueof{/pgf/inner xsep},
                    align=center,
                    on chain, join=by arrow},
    arrow/.style = {thick,-Triangle}
                        }
        }

\newcommand{\apjl}{Astrophys. J. Lett.}
\newcommand{\aap}{Astron. Astrophys.}
\newcommand{\apjs}{Astrophys. J. Suppl. Ser.}

\newcommand{\lensed}{\mathrm{len}}

\newcommand{\delensed}{\dd }
\newcommand{\unlensed}{\mathrm{u}}
\newcommand{\observed}{\mathrm{obs}}
\def\gl{{\rm gl}}
\def\Neff{N_{\rm eff}}
\newcommand{\dd}{{\rm d}}

\newcolumntype{C}{>{\centering\arraybackslash}X}
\newcolumntype{R}{>{\raggedleft\arraybackslash}X}

\newcommand{\bn}{\boldsymbol{n}}
\newcommand{\be}{\begin{eqnarray}}

\newcommand{\ee}{\end{eqnarray}}

\definecolor{colorA}{HTML}{9467bd}
\definecolor{colorB}{HTML}{bcbd22}
\definecolor{colorC}{HTML}{2ca02c}

\newcommand{\ucsd}{Department of Physics,
University of California San Diego, UC San Diego 9500 Gilman Dr. La Jolla, CA 92093, USA}

\newcommand{\smu}{Department of Physics,
Southern Methodist University, 3215 Daniel Ave, Dallas, TX 75275, USA}

\newcommand{\asu}{School of Earth and Space Exploration, Arizona State University, Tempe, AZ 85287, USA}

\newcommand{\jhu}{Department of Physics \& Astronomy, Johns Hopkins University, Baltimore, MD 21218, USA}

\begin{document}

\title{The Benefits of CMB Delensing}

\author{Selim~C.~Hotinli}
\affiliation{\jhu}

\author{Joel~Meyers}
\affiliation{\smu}

\author{Cynthia~Trendafilova}
\affiliation{\smu}

\author{Daniel~Green}
\affiliation{\ucsd}

\author{Alexander~van~Engelen}
\affiliation{\asu}

\date{\today}

\begin{abstract}

The effects of gravitational lensing of the cosmic microwave background (CMB) have been measured at high significance with existing data and will be measured even more precisely in future surveys.
Reversing the effects of lensing on the observed CMB temperature and polarization maps provides a variety of benefits.
Delensed CMB spectra have sharper acoustic peaks and more prominent damping tails, allowing for improved inferences of cosmological parameters that impact those features.
Delensing reduces $B$-mode power, aiding the search for primordial gravitational waves and allowing for lower variance reconstruction of lensing and other sources of secondary CMB anisotropies.
Lensing-induced power spectrum covariances are reduced by delensing, simplifying analyses and improving constraints on primordial non-Gaussianities.
Biases that result from incorrectly modeling nonlinear and baryonic feedback effects on the lensing power spectrum are mitigated by delensing.
All of these benefits are possible without any changes to experimental or survey design.
We develop a self-consistent, iterative, all-orders treatment of CMB delensing on the curved sky and demonstrate the impact that delensing will have with future surveys.

\end{abstract}

\maketitle

\section{Introduction}
\label{sec:introduction}

Gravitational lensing of the cosmic microwave background (CMB) is both a help and hindrance to our understanding of the history and contents of the universe.
The deflection of CMB photons by the large scale structure intervening between the last scattering surface and our telescopes allows us to utilize measurements of the CMB to learn about the distribution of matter in the late universe, billions of years after recombination.
Yet, the lensing manifests itself as a distortion of the primary CMB anisotropies and also functions as an obstacle to analyses which rely on a pristine view of the last scattering surface.

The theoretical aspects of CMB lensing are thoroughly understood (see \cite{Lewis:2006fu} for a comprehensive review).
Lensing smooths the acoustic peaks of the CMB power spectra, transfers power from large angular scales to small scales, and converts $E$-mode polarization to $B$-mode polarization.  
It also generates non-stationary statistics of CMB fluctuations, leading to the coupling of modes with different wavenumber.

Fortunately, the off-diagonal mode couplings induced by lensing allow us to reconstruct maps of the lensing potential~\cite{Hu:2001kj,Okamoto:2003zw}.  In this sense, the late-time information can be isolated from the primary CMB independently of the cosmological parameters.  The effects of CMB lensing have been measured at high significance, including a $40\sigma$ measurement with data from the Planck satellite~\cite{Planck:2018lbu}.  Upcoming surveys with experiments like Simons Observatory~\cite{SimonsObservatory:2018koc}, CMB-S4~\cite{CMB-S4:2016ple}, PICO~\cite{NASAPICO:2019thw}, and CMB-HD~\cite{Sehgal:2019ewc} will map the CMB sky with unprecedented precision.  
The role of CMB lensing as both a nuisance and a tool will be enhanced at the high fidelity of these upcoming surveys.

The reconstruction of the CMB lensing potential is a valuable cosmological probe in its own right, especially for measuring the growth of structure in the late universe.  
Cosmological measurements of the neutrino mass~\cite{Dolgov:2002wy,Kaplinghat:2003bh,Lesgourgues:2006nd,Dvorkin:2019jgs,Green:2021xzn} and constraints on the dark energy equation of state~\cite{Frieman:2008sn} provide two prominent examples where this information is crucial to our understanding of fundamental physics.
In addition, cross-correlating maps of CMB lensing with galaxy surveys can offer advantages by breaking degeneracies and canceling cosmic variance, thereby improving parameter constraints~\cite{Seljak:2008xr,Schaan:2016ois,Schmittfull:2017ffw,Yu:2018tem,Yu:2021vce}.

Unfortunately, reconstruction of the lensing map alone does not mitigate the obstructive influence it has on the primary CMB.  Lensing presents a serious obstacle to the search for primordial gravitational waves; while scalar fluctuations do not produce $B$-mode polarization at linear order~\cite{Kamionkowski:1996zd,Zaldarriaga:1996xe,Seljak:1996gy,Kamionkowski:1996ks}, gravitational lensing converts $E$-mode polarization to $B$-mode polarization, thereby acting as a source of confusion for primordial gravitational wave searches.
It is therefore essential that we also {\it delens} the primary CMB to remove the influence of lensing in order to significantly improve constraints on the primordial gravitational wave amplitude~\cite{Knox:2002pe, Kesden:2002ku, Seljak:2003pn, Smith:2010gu}.  The most widely used technique, iterative $EB$ lensing reconstruction~\cite{Seljak:2003pn,Smith:2010gu}, will play an important part in allowing experiments like CMB-S4 to achieve their targets for primordial gravitational waves~\cite{Abazajian:2016yjj,Abazajian:2019eic,CMB-S4:2020lpa}.

\begin{figure}[t!]
    \vspace{-0.25cm}
    \includegraphics[width=\columnwidth]{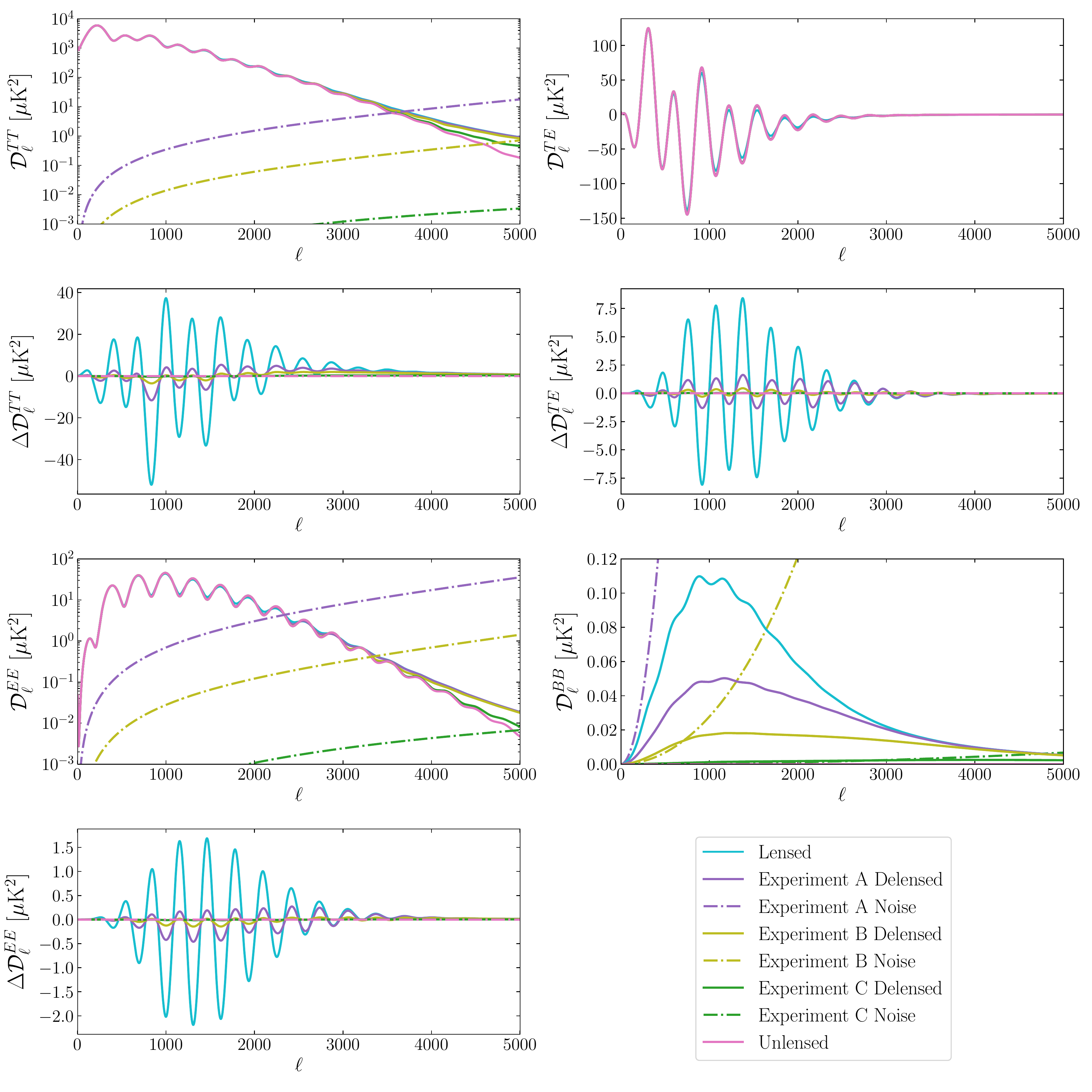}
    \vspace{-0.75cm}
    \caption{
    Delensed CMB power spectra for each of three experimental configurations defined in Table~\ref{table:experiments} compared to lensed and unlensed spectra, plotted in terms of $\mathcal{D}_\ell \equiv \frac{\ell(\ell+1)}{2\pi} C_\ell$, with $\Delta \mathcal{D}_\ell$ computed as the lensed or delensed spectra minus the unlensed spectra.
    Lensing causes smoothing of acoustic peaks, transfer of power from large scales to small scales, and conversion of $E$ modes to $B$ modes. 
    Delensing reduces each of these effects by an amount that depends upon the experimental configuration.
    }
    \label{fig:spectra}
\end{figure}

However, the role of lensing in obscuring cosmological information is not limited to the $B$-modes.
Delensing of the $T$ and $E$ modes can sharpen acoustic peaks, tighten parameter constraints, and reduce lensing-induced off-diagonal power spectrum covariances~\cite{Green:2016cjr}.
In this paper, we expand on all of these benefits and further explore the value of delensing to all primary CMB science.  Some of these benefits are illustrated in Fig.~\ref{fig:spectra}, where we show the effect of delensing on the $TT$, $TE$, $EE$, and $BB$ spectra for each of three experimental configurations defined in Table~\ref{table:experiments}.
Delensing of the small-scale $T$ and $E$ spectra has been demonstrated with real data, including with Planck~\cite{Larsen:2016wpa,Carron:2017vfg,Planck:2018lbu}, the Atacama Cosmology Telescope~\cite{ACT:2020goa}, and the South Pole Telescope~\cite{Millea:2020iuw}, in all cases yielding a more pristine view of the CMB at the last scattering surface, including sharper acoustic peaks.  Ref.~\cite{ACT:2020goa} additionally obtained the first cosmological constraints from delensed power spectra, finding them to be consistent with the fully lensed case.

Here, we provide a quantitative discussion of the improvements offered by delensing with an eye towards the lower-noise experiments that are expected in the near future.
We demonstrate how delensing helps in measuring peak positions and peak heights and recovering the damping scale.
We show that iterative delensing of all spectra leads to lower lensing reconstruction noise than other techniques, including iterative $EB$ reconstruction.
We provide a more complete treatment of non-Gaussian lensing-induced power spectrum covariances and their reduction with delensing.
We demonstrate with self-consistent forecasts that delensing improves constraints on cosmological parameters, and we also show that it reduces biases that arise from incorrect modeling of the lensing power spectrum.

Additionally, the implementation of delensing in this paper has several improvements compared to that in Ref.~\cite{Green:2016cjr}.
We formulate lensing and delensing on the curved sky rather than using the flat-sky approximation.
We employ iterative delensing of all spectra.
The non-Gaussian covariances for lensed and delensed spectra are treated more precisely, including contributions that were previously neglected.
The numerical implementation is stable, efficient, and publicly available as a modification of the \texttt{CLASS} Boltzman code~\cite{Blas:2011rf} that we call \texttt{CLASS\_delens}\footnote{\url{https://github.com/selimhotinli/class_delens}}. 
This code can be used as a forecasting tool to compute the best possible constraints on cosmological parameters from a given CMB dataset, consistently including the beneftis of delensing.


The paper is organized as follows.
In Sec.~\ref{sec:delensing}, we outline our iterative delensing and lensing reconstruction procedure, leaving the details to two appendices: Appendix~\ref{sec:correlation_functions} focuses on the calculation of delensed spectra, and Appendix~\ref{sec:lensing_reconstruction} gives the details of lensing reconstruction.
We describe the phenomenology of delensing in Sec.~\ref{sec:benefits}, discussing broadly the various benefits that delensing provides.
In Sec.~\ref{sec:forecasts}, we provide explicit forecasts to demonstrate the quantitative benefits of delensing for cosmological parameter inferences.
We conclude in Sec.~\ref{sec:conclusion}.

\section{Delensing}
\label{sec:delensing}

Gravitational lensing deflects CMB photons such that the lensed CMB temperature and polarization in line-of-sight direction $\bn$ are given by the unlensed CMB in a direction that differs from the line-of-sight direction by the lensing deflection $\mathbf{d}(\bn)$.
At lowest order, the deflection angle is a pure gradient $\mathbf{d}(\bn)=\boldsymbol{\nabla}\phi(\bn)$ where $\phi$ is the lensing potential. 
For example, the lensed temperature field $T^\lensed$ is given in terms of the unlensed temperature field $T^\unlensed$ by
\be
    T^\lensed(\bn)=T^\unlensed(\bn+\mathbf{d}(\bn))\simeq T^\unlensed(\bn) + \mathbf{d}(\bn)\cdot\boldsymbol{\nabla}T^\unlensed(\bn) + \ldots \, .
    \label{eq:T_lens}
\ee
The aim of delensing is to manipulate observed CMB maps (such as $T^\observed$) and estimates of the lensing deflection $\mathbf{d}^\observed$ to reverse this remapping in order to recover an estimate of the unlensed CMB.

\subsection{All-Orders Delensing on the Full-Sky}
\label{sub:allordersdelensing}

There are many inequivalent implementations of delensing that could be employed in principle.
Here we will follow the strategy described in Ref.~\cite{Green:2016cjr}.
Specifically we require the following:
\begin{itemize}
    \item Delensing should be accurate in the limit where the noise vanishes, $T^\delensed(\bn)\simeq T^\unlensed(\bn)$.
    \item The delensing procedure must conserve total power.
    \item Maps should be filtered to minimize the impact of noisy modes on the observables. 
\end{itemize}
With this procedure, the delensed temperature $T^\delensed$ can be expressed as
\begin{align}
    T^\delensed(\bn) = \bar{h} \star T^\observed(\bn) + h \star T^\observed \left( \bn - g \star \mathbf{d}^\observed(\bn) \right) \, ,
    \label{eq:delensedT}
\end{align}
where the star denotes a convolution on the 2-sphere of the sky, and $g$, $h$, and $\bar{h}$ are filters most straightforwardly defined in harmonic space.
We take the same set of filters as was used in Ref.~\cite{Green:2016cjr},
\begin{equation}
    g_\ell = \frac{C_\ell^{\phi\phi}}{C_\ell^{\phi\phi,\observed}} \, , \qquad
    h_\ell = \frac{C_\ell^{TT}}{C_\ell^{TT} + N_\ell^{TT}} \ ,
\end{equation}
with $\bar{h}_\ell$ fixed by the condition that delensing should conserve total power.
There is a separate set of filters $h_\ell^P$ and $\bar{h}_\ell^P$ for the polarization maps.  
If the lensing field is obtained from an external tracer,  then the $g$ filter can be obtained from the cross-correlation between the tracer and the lensing potential; see e.g.~Refs.~\cite{Smith:2010gu,Sherwin:2015baa}.  
The denominator of the $h$ filter in principle includes contributions from astrophysical foregrounds, which here we include in the noise power spectra $N_\ell^{TT}$ for simplicity.   
The \texttt{CAMB} software~\cite{Lewis:1999bs} includes an option to calculate partially lensed CMB spectra, which in our notation corresponds to delensing with $h=1$ and $\bar{h}=0$.
It was discussed in Ref.~\cite{Green:2016cjr} how choosing not to filter noisy modes leads to a sub-optimal delensing procedure.

The effects of lensing at all orders on the CMB power spectra can be computed using  curved sky correlation functions~\cite{Challinor:2005jy}.
As in Ref.~\cite{Green:2016cjr}, we utilize a correlation function approach to calculate the delensed spectra, thereby including the residual lensing to all orders.  
In this work we additionally take into account effects of the curved sky.
For example, the delensed temperature auto spectrum is given by
\begin{equation}
    C^{TT, \delensed}_\ell  = 2\pi\int_{-1}^{1}\xi^{TT,\delensed}(\beta)d^\ell_{00}(\beta) \, \dd\cos{\beta} \, ,
    \label{eq:delensed_ClTT}
\end{equation}
where $\xi^{TT,\delensed}(\beta)$ is the delensed temperature correlation function at angular separation $\beta$ and $d^\ell_{m m'}(\beta)$ denotes a Wigner $d$-matrix.
Expressions for the lensed and delensed correlation functions are shown in Appendix~\ref{sec:correlation_functions}.


\begin{table}[t!]
    \begin{center}
     \begin{tabular}{l @{\hskip 12pt} c@{\hskip 12pt}c@{\hskip 12pt}c@{\hskip 12pt}c} 
     \toprule
       Label & $\Delta_T$ ($\mu$K-arcmin)    &   $\theta_{\rm FWHM}$ (arcmin) & \multicolumn{2}{c}{Color} \\ [0.5ex] 
     \hline
    Experiment A & 5 & 1.4  & \textcolor{colorA}{\rule{1cm}{0.75mm}} & Purple \\ 
    Experiment B & 1 & 1.4  & \textcolor{colorB}{\rule{1cm}{0.75mm}} & Yellow \\
    Experiment C & 0.1 & 0.1  & \textcolor{colorC}{\rule{1cm}{0.75mm}} & Green \\
      \hline
    \end{tabular}
    \caption{
    Parameters defining the three experimental configurations and the colors used for them in plots.
    }
    \label{table:experiments}
    \end{center}
\end{table}


\subsection{Lensing Reconstruction and Iterative Delensing}
\label{sub:iterativedelensing}

Lensing leads to non-stationary CMB statistics and off-diagonal mode coupling which can be used to reconstruct the lensing deflection field, most commonly through the use of quadratic estimators~\cite{Hu:2001kj}.
Each of the six pairings of the $T$, $E$, and $B$ fields can be used to construct a quadratic estimator.
The $TT$ quadratic estimator gives the lowest variance lensing reconstruction with current CMB data~\cite{Planck:2018jri}.
At the low noise and high resolution expected from future observations, the $EB$ estimator will provide a better reconstruction due to the low observed $B$-mode power.
Delensing can further reduce $B$-mode power by reversing the effects of lensing that convert $E$-mode polarization to $B$-mode polarization.
This reduction in $B$-mode power allows for an improved lensing reconstruction, which can in turn be used for improved delensing.
This iterative $EB$ delensing procedure~\cite{Seljak:2003pn,Smith:2010gu} will improve the lensing reconstruction in upcoming experiments.

\begin{figure}
    \begin{center}
    \begin{tikzpicture}[FlowChart,      
                        node distance = 5mm and 7mm,
                        start chain = A going below   
                        ]
        \node[startstop] {$T^\observed(\bn)$, $E^\observed(\bn)$, $B^\observed(\bn)$};
        \node[process] {$\mathbf{d}^\observed(\bn)$};
        \node[startstop] {$T^\delensed(\bn)$, $E^\delensed(\bn)$, $B^\delensed(\bn)$};
        \draw [arrow] (A-3.east) to ["iteration"] ++ (2,0) |- (A-2);
    \end{tikzpicture}
    \hspace{2cm}
    \begin{tikzpicture}[FlowChart,      
                        node distance = 5mm and 7mm,
                        start chain = A going below   
                        ]
        \node[startstop] {$C_\ell^{TT,\observed}$, $C_\ell^{TE,\observed}$, $C_\ell^{EE,\observed}$, $C_\ell^{BB,\observed}$};
        \node[process] {$N_\ell^{\mathrm{mv},\phi\phi}$};
        \node[startstop] {$C_\ell^{TT,\delensed}$, $C_\ell^{TE,\delensed}$, $C_\ell^{EE,\delensed}$, $C_\ell^{BB,\delensed}$};
        \draw [arrow] (A-3.east) to ["iteration"] ++ (2,0) |- (A-2);
    \end{tikzpicture}
    \caption{Schematic diagram of the iterative delensing procedure at map level (\textit{left}) and spectrum level (\textit{right}). The observed CMB maps of temperature and polarization are used to estimate the lensing deflection, which can be used to delens the temperature and polarization. Delensed maps can be used to improve the estimate of the lensing deflection, which can in turn be utilized to further delens the CMB maps. The procedure can be iterated to convergence.  While in practice, delensing necessarily occurs at the level of the maps, we estimate the effects of delensing at the level of the spectra.
    } 
    \label{fig:schematic_iterative}
    \end{center}
\end{figure}
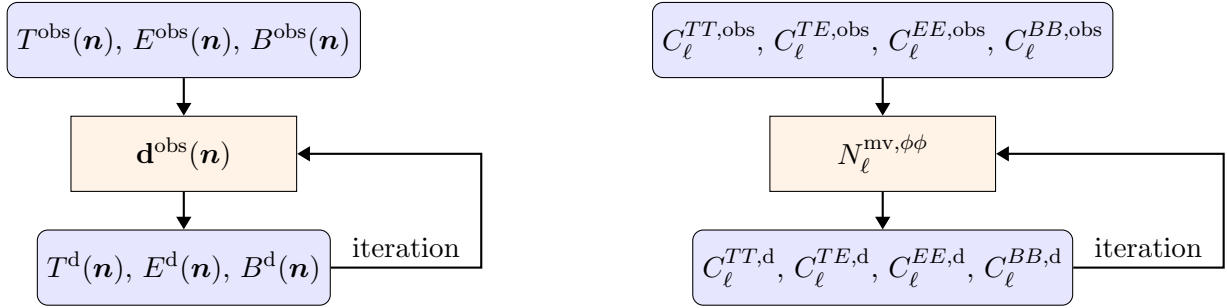

Lensing-induced temperature and $E$-mode power on small angular scales can also hinder lensing reconstruction in low noise experiments.
As we will demonstrate, iteratively delensing all of $T$, $E$, and $B$ (rather than just removing $B$-mode power) can thereby improve the fidelity of lensing reconstruction.
This iterative delensing procedure is shown schematically in Fig.~\ref{fig:schematic_iterative}.
We implement a procedure to estimate the effects of iterative delensing on the CMB power spectra and the noise power of the reconstructed lensing field.  
Our forecasts are based on the expectations of what would be achieved from map-level delensing, but our calculations are performed at the level of the power spectrum. 
Our results may serve as a useful point of comparison for map-level implementations of lensing reconstruction that go beyond the quadratic estimator, such as those of Refs.~\cite{Hirata:2002jy,Seljak:2003pn,Carron:2017vfg,Millea:2017fyd,Horowitz:2017iql,Caldeira:2018ojb,Hadzhiyska:2019cle,Millea:2020cpw}.
The details of the lensing reconstruction procedure are provided in Appendix~\ref{sec:lensing_reconstruction}.

\subsection{Software Implementation}
\label{sec:code_description}

We implement our delensing procedure as a modification of the lensing routine in the \texttt{CLASS} Boltzmann code~\cite{Blas:2011rf}. 
Our goal is to provide a tool that allows accurate, stable, and efficient computation of the delensed CMB spectra and lensing reconstruction noise.
In addition to benefiting from the specialised numerical routines available in the \texttt{CLASS} code, we use efficient real-space expressions introduced in Refs.~\citep{Dvorkin:2009ah,Smith:2010gu} for the delensed spectra and lensing reconstruction quadratic estimators.
These expressions can be evaluated at the cost of $\mathcal{O}(\ell_{\rm max}^2)$ rather than $\mathcal{O}(\ell_{\rm max}^3)$, where $\ell_{\rm max}$ is the maximum multipole used in the quadratic estimator calculation.
This provides a significant increase in speed which is valuable for repeated calculations of iterative delensing as well as rapid exploration of the parameter space for a Markov chain Monte Carlo analysis, for example.
This simplification is due to the identities shown in Eqs.~\eqref{eq:fdef} and \eqref{eq:fdef2}, and involves computing products of Wigner $3j$-symbols in terms of Wigner $d$-matrices.

In addition to its efficiency, \texttt{CLASS\_delens} is flexible, allowing users to either generate lensing reconstruction noise internally for a given experimental noise or use externally calculated lensing reconstruction noise curves for the purpose of delensing. 
Users can also choose the lensing reconstruction scheme from options such as employing no delensing, iterative $EB$ reconstruction, or iterative delensing on all spectra, and the code can also be used to calculate the covariance between various lensing quadratic estimators.

\section{Phenomenology of Delensing}
\label{sec:benefits}

\subsection{Acoustic Peaks}
\label{sub:peaklocations}

The patterns of acoustic peaks in the CMB power spectra are a striking signature of sound waves that propagated through the primordial plasma~\cite{Peebles:1970ag,Hu:1994jd}. 
The angular scale and amplitude of these peaks carry a wealth of information about the history and contents of the universe~\cite{Pan:2016zla}.
Data from the Planck satellite has been used to measure 7 peaks in the $TT$ spectrum, 6 peaks in the $TE$ spectrum, and 5 peaks in the $EE$ spectrum~\cite{Planck:2018nkj}.
Measurements of acoustic peak properties translate to tight constraints on cosmological parameters~\cite{Planck:2018vyg}.

Gravitational lensing smooths CMB acoustic peaks.
This can be understood qualitatively by considering that features of a fixed angular size can be either magnified or de-magnified by the deflection of photons, leading sharp features in the power spectrum to be blurred over a range of scales.
The angular scale of a sharp feature in the power spectrum is easier to measure than a broad hump, and gravitational lensing therefore weakens our ability to precisely measure acoustic peak positions in the CMB power spectra.
Delensing reverses this peak smoothing, providing sharper peaks whose angular scale can be more precisely measured.
Similar comments apply to measurements of peak heights.

\begin{figure}[t!]
    \vspace{-0.25cm}
    \includegraphics[width=\columnwidth]{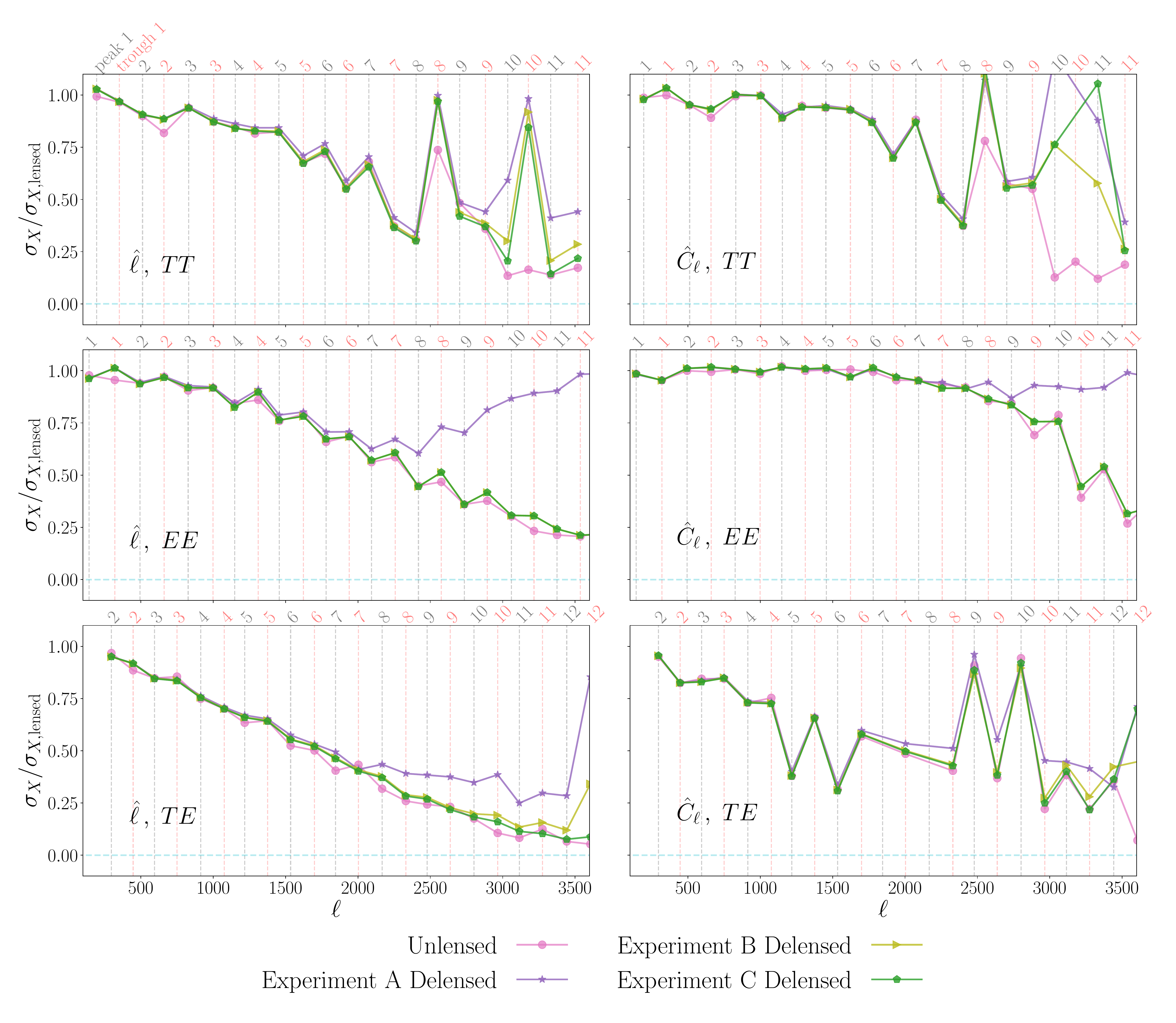}
    \vspace{-0.75cm}
    \caption{Fractional improvement from delensing on the $TT$, $EE$, and $TE$ spectrum peak and trough positions $\hat{\ell}$ (\textit{left}) and amplitudes $\hat{C_\ell}$ (\textit{right}). The uncertainty obtained with unlensed and delensed spectra for each of the experimental configurations defined in Table~\ref{table:experiments} are compared to those from the lensed spectra.  Peaks are indicated with vertical gray dashed lines and troughs with red dashed lines, with the peak or trough number shown at the top.
    }
    \label{fig:peak_fitting_improvement}
\end{figure}

We show in Fig.~\ref{fig:peak_fitting_improvement} the fractional improvement on the $TT$, $TE$, and $EE$ spectrum peak and trough positions and amplitudes that is provided by delensing.
We calculate the uncertainty on the estimated peak locations $\hat{\ell}$ and amplitudes $\hat{C_\ell}$ by making measurements on $\num{40000}$ realizations of simulated unlensed, lensed, and delensed CMB spectra.  
For each realization, the value of $C_\ell$ at every $\ell$ is drawn from a Gaussian distribution, taking into account cosmic variance and the experimental noise for each of the three experimental configurations described in Table~\ref{table:experiments}.  
In order to make the peaks abide closer to a Gaussian profile, we multiply the CMB spectra with an exponential function of the form $\exp[A(\gamma\ell)^{b}]$ with parameters $A = 0.68$ and $b = 1.3$.  For the $TT$ and $EE$ spectra we use $\gamma = 1.6\times10^{-3}$ while for $TE$ we use $\gamma = 1.8\times10^{-3}$.
We then fit Gaussian profiles to each peak and trough and calculate the variances from the distributions of best fit values.  
For all spectra and up to the smallest scales we consider, the improvement from delensing is evident from comparing the uncertainty of the peak and trough positions and amplitudes obtained with the unlensed and delensed spectra with the uncertainty from the lensed spectra. 
The relative improvement from delensing increases at lower noise, as can be seen by comparing the results for the three experimental configurations.
The trends seen for the peak heights are complicated by the fact that sharpening peaks increases the power at the peak and thus increases the cosmic variance when treating the lensed (and delensed) CMB as Gaussian random fields.
Despite this subtlety, it is clear from Fig.~\ref{fig:peak_fitting_improvement} that delensing improves our ability to measure peak positions and peak heights, and our estimates of the improvement are likely somewhat conservative since we neglected lensing-induced non-Gaussian covariances here.

One cosmological parameter that is particularly closely related to peak positions is the angular size of the sound horizon at last scattering $\theta_s$, which can be measured from the spacing between acoustic peaks.
Measurement of $\theta_s$ plays an important role in the inference of the present Hubble rate $H_0$ from CMB measurements~\cite{Knox:2019rjx}.
The comoving angular diameter distance to the surface of last scattering $D_A$ can be measured from the ratio $D_A = r_s / \theta_s$, where $r_s$ is the comoving size of the sound horizon at last scattering.
The value of $r_s$ can be computed if the baryon density (which affects the sound speed in the photon-baryon plasma) and the matter density (which affects the expansion rate prior to last scattering) are known.
The baryon density is tightly constrained by measuring the relative heights of odd and even peaks in the power spectra, and thus measurement of baryon density also benefits from delensing.
In flat $\Lambda$CDM cosmology with known matter density, one can then infer $H_0$ from the value of $D_A$.
We show explicitly in Sec.~\ref{sec:forecasts} that measurements of both $\theta_s$ and $\Omega_b h^2$ are improved by delensing in a 6-parameter $\Lambda$CDM model.

More generally, measurements of the peak locations are a reflection of our ability to test physics beyond $\Lambda$CDM with the primary CMB.  
For example, new models that are introduced to resolve the Hubble tension are expected to impact the peak locations, perhaps in a way that is not entirely degenerate with $\theta_s$. 
We can anticipate that sharpening the acoustic peaks with delensing will lead to more precise tests of those ideas, even for models that have yet to be defined.
A canonical illustration of this, also discussed in detail in Ref.~\cite{Green:2016cjr}, is the improvement in the measurements of $N_\mathrm{eff}$ made possible by delensing. 
Free-streaming relativistic particles that contribute to $\Neff$ cause a phase shift imprinted on acoustic peaks~\citep{Bashinsky:2003tk,Follin:2015hya,Baumann:2015rya} that distinguishes them from other forms of matter.
The improvement from delensing on $N_\mathrm{eff}$ constraints is especially pronounced in models where both $N_\mathrm{eff}$ and the primordial helium abundance $Y_\mathrm{p}$ are free to vary.
This is due to the fact that both $\Neff{}$ and $Y_\mathrm{p}$ affect the damping scale, and are thus somewhat degenerate~\cite{Hou:2011ec}; however, only $\Neff{}$ causes a shift to the acoustic peaks on small scales.
The sharper peaks and improved determination of peak locations and peak heights enabled by delensing thereby helps to break the degeneracy between $N_\mathrm{eff}$ and $Y_\mathrm{p}$, significantly improving constraints on both~\cite{Green:2016cjr}.

\subsection{Damping Tail}
\label{sub:damptingtail}

The primary CMB anisotropies are exponentially suppressed on small angular scales due to diffusion damping~\cite{1967Natur.215.1155S,Weinberg:1971mx,1983MNRAS.202.1169K,Hu:1996vq,Zaldarriaga:1995gi,Hou:2011ec}.
The damping scale is determined by the expansion rate and free electron density prior to recombination, and is thus sensitive to cosmological parameters such as $\Neff{}$ and $Y_\mathrm{p}$~\cite{Hou:2011ec}.
Lensing transfers power from large angular scales to small angular scales, and lensing-induced power makes the dominant contribution to the $TT$ and $EE$ spectra on the smallest angular scales $\ell \gtrsim 4000$.

\begin{figure}[t!]
    \vspace{-0.25cm}
    \includegraphics[width=\columnwidth]{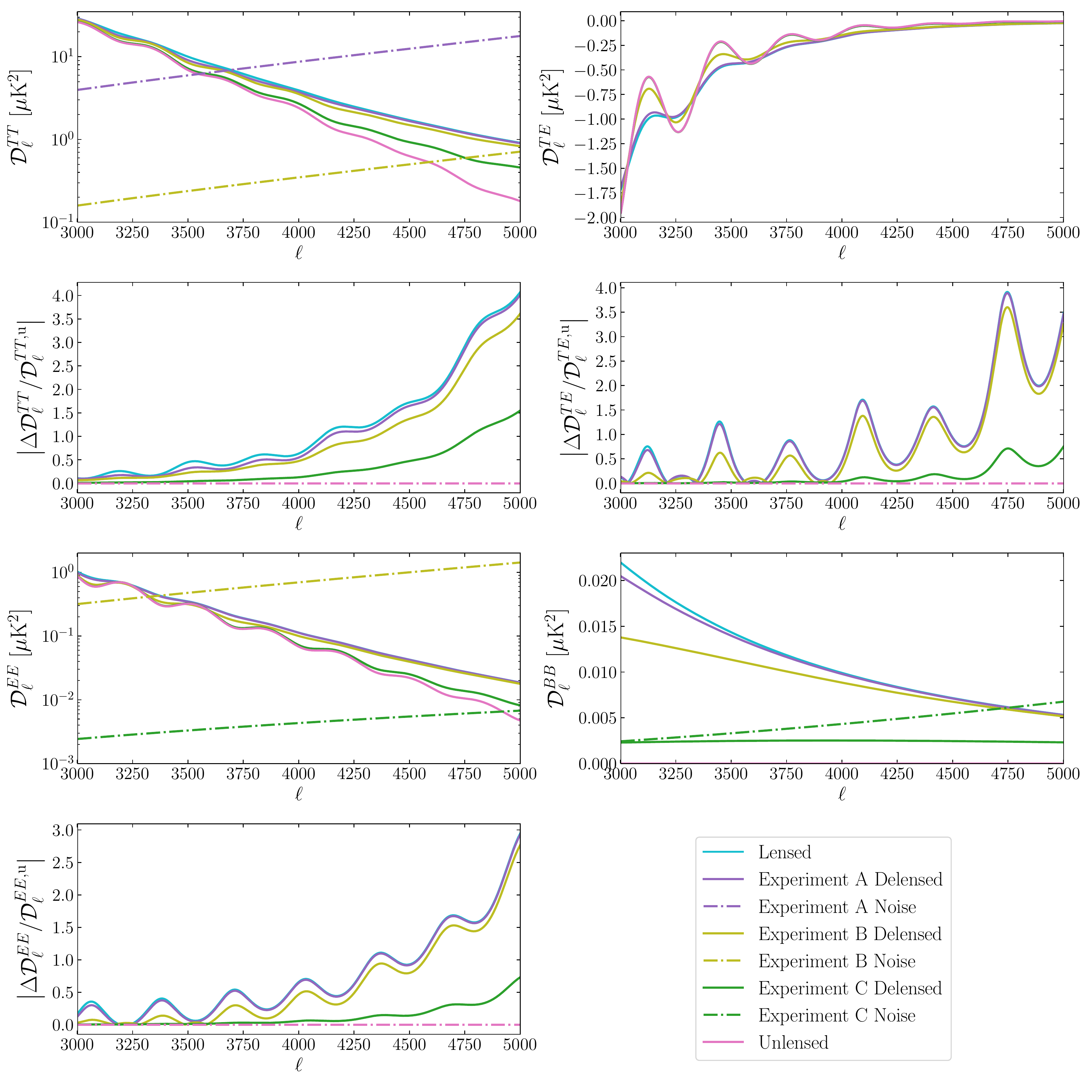}
    \vspace{-0.75cm}
    \caption{
    Effects of lensing and delensing on the small scale power, plotted with the same conventions as in Fig.~\ref{fig:spectra}.
    Delensing, especially with low noise experiments, is capable of recovering some of the damping tail that is obscured in the lensed spectra.
    }
    \label{fig:dampingtail}
\end{figure}

Delensing can reverse the transfer of power to small scales, thereby allowing more precise measurements of the damping scale, and in turn the cosmological parameters that determine it.
However, delensing the damping tail using internal CMB lensing reconstruction is challenging, since the small scale lensing-induced power is due primarily to small scale lensing modes where the lensing reconstruction tends to have lower signal to noise.
This should be contrasted with peak smoothing, which is caused primarily by the large scale lensing modes, which will be reconstructed at high signal to noise in upcoming experiments.
Delensing using external tracers of the lensing potential~\cite{Smith:2010gu,Sherwin:2015baa} could potentially provide better delensing performance in the damping tail than is possible with internal lensing reconstruction.
In Fig.~\ref{fig:dampingtail} we show the effects of delensing on small angular scales.


\subsection{CMB \texorpdfstring{$B$}{B}-modes}
\label{sub:Bmodes}

The most widely studied benefit of delensing is the reduction of lensing-induced $B$-mode power, particularly as it relates to facilitating the search for primordial gravitational gravitational waves~\cite{Knox:2002pe, Kesden:2002ku, Seljak:2003pn, Smith:2010gu}.
Lensing converts $E$ modes to $B$ modes, which act as a source of confusion in searches for primordial gravitational waves.
Delensing reverses this process, reduces the lensing $B$-mode power, and allows for tighter constraints on the amplitude of primordial gravitational waves.
Delensing can also improve constraints on other sources of $B$-mode power, such as anisotropic cosmic polarization rotation~\cite{Williams:2020hqk}.

\subsection{Lensing Reconstruction}
\label{sub:reconstruction}

We have so far focused on how lensing acts as a nuisance to cosmological analyses by obscuring our view of the primary CMB anisotropies.
However, the lensing potential contains a wealth of information since it is sourced by cosmological structure that intervenes between us and the surface of last scattering.
Measurements of the lensing power spectrum are therefore particularly useful for inferring cosmological parameters that affect the growth of structure at late times, including neutrino mass~\cite{Dolgov:2002wy,Kaplinghat:2003bh,Lesgourgues:2006nd,Dvorkin:2019jgs,Green:2021xzn} and dark energy~\cite{Frieman:2008sn}.
The value of CMB lensing maps are enhanced by the opportunities presented by cross-correlations with large scale galaxy surveys~\cite{Seljak:2008xr,Schaan:2016ois,Schmittfull:2017ffw,Yu:2018tem,Yu:2021vce}.

Delensing can improve our ability to reconstruct maps of the lensing potential.
This has been previously studied in detail for iterative $EB$ lensing reconstruction~\cite{Seljak:2003pn,Smith:2010gu}, where the benefit comes from reducing the lensing-induced $B$-mode power that contributes to the variance of the estimator.
One can see from Eq.~\eqref{eq:normalization} that a similar improvement should be expected for each estimator if delensing reduces the observed power for any spectrum.
As discussed above, delensing reduces $TT$ and $EE$ on small angular scales, and it reduces $BB$ on all scales.
We therefore expect that iteratively delensing all spectra should provide a reduction in the lensing reconstruction noise, especially in low-noise high-resolution experiments where the damping tail can be delensed.

\begin{figure}[t!]
    \vspace{-0.25cm}
    \includegraphics[width=1\columnwidth]{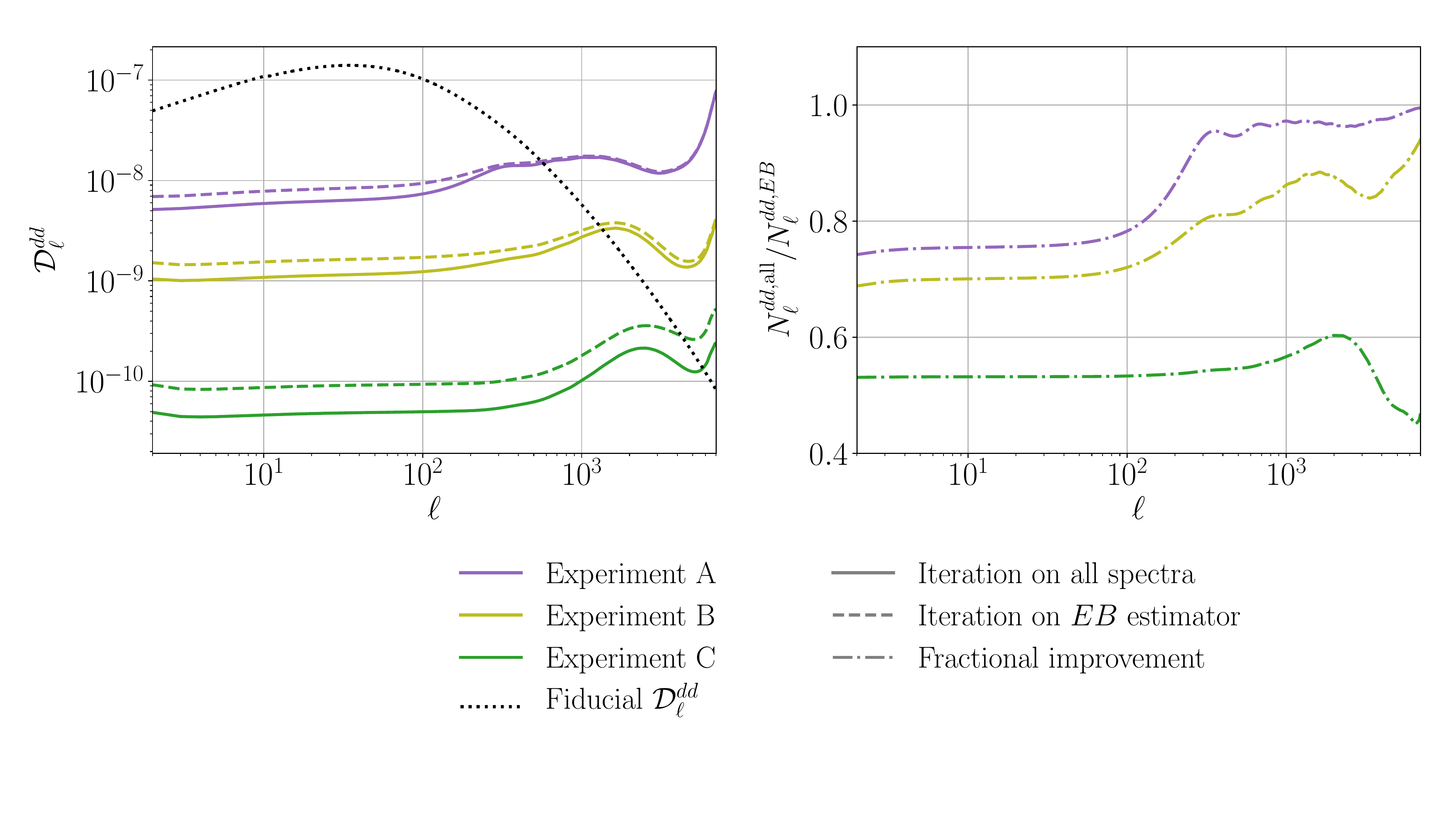}
    \vspace{-1.8cm}
    \caption{Improvement on the lensing reconstruction noise provided by iterative delensing for each of the three experimental configurations described in Table~\ref{table:experiments}. 
    The dashed lines correspond to the minimum variance combination of the $TT$, $TE$, $TB$, and $EE$ estimators using lensed spectra with the iterated $EB$ estimator, while colored solid lines correspond to iteratively delensing all spectra (the $BB$ estimator does not contribute in either case since $C_\ell^{BB,\unlensed}=0$). Gray solid lines correspond to lensing deflection spectra $\mathcal{D}_\ell^{dd}$. The lower right plot shows the fractional reduction in the lensing reconstruction noise for iterative delensing of all spectra compared to the minimum-variance reconstruction with only $EB$ iteration.
    For the purposes of this figure, we neglect the off-diagonal elements of the estimator covariance in constructing the minimum-variance combination.
    }
    \label{fig:delensing_recon_improvement}
\end{figure}

In Fig.~\ref{fig:delensing_recon_improvement} we demonstrate the improvement in the lensing reconstruction noise that comes from iterative delensing for each of the three CMB experiments defined in Table~\ref{table:experiments}. 
In particular, we show that there is a benefit of performing iterative delensing on all spectra compared to iterating only the $EB$ estimator.
In both cases, we compute the reconstruction noise using the minimum variance combination of all estimators (see Eq.~\eqref{eq:Nlmv}).
We find that there is a significant reduction in the lensing reconstruction noise by iteratively delensing all spectra, as compared to combining the $EB$ iterative reconstruction with the other estimators using lensed spectra.
We present the reconstruction noise in terms of the deflection spectra, related to the lensing potential spectrum by $C_\ell^{dd} = \ell(\ell+1)C_\ell^{\phi\phi}$.

\subsection{Reconstruction of Other Fields}
\label{sec:other_reconstruction}

In addition to gravitational lensing, there are other effects that may induce off-diagonal mode coupling and non-stationary CMB statistics.
These effects include patchy reionization~\cite{Santos:2003jb,Zahn:2005fn,McQuinn:2005ce,Dore:2007bz}, anisotropic cosmic polarization rotation~\cite{Carroll:1989vb,Harari:1992ea,Carroll:1998zi,Lue:1998mq,Kosowsky:1996yc}, the polarized Sunyaev-Zel'dovich effect~\cite{Kamionkowski:1997na,Sazonov:1999zp}, and the moving lens effect~\cite{1983Natur.302..315B}, for example.
Estimators have been developed that are aimed at reconstructing these effects; see e.g.~Refs.~\cite{Dvorkin:2008tf,Dvorkin:2009ah,Smith:2016lnt,Kamionkowski:2008fp,Yadav:2009eb,Gluscevic:2009mm,Alizadeh:2012vy,Deutsch:2017cja,Deutsch:2017ybc,Meyers:2017rtf,Hotinli:2018yyc,Hotinli:2020ntd,Hotinli:2021hih,Hotinli:2020csk,Cayuso:2021ljq}.
Just as is the case for lensing reconstruction, reducing the lensing-induced power, especially on small scales and in $B$-mode polarization, will reduce the variance of these estimators; see Refs.~\cite{Guzman:2021nfk,Guzman:2021ygf} for a quantitative discussion of the benefits of delensing for patchy reionization and cosmic polarization rotation.
Delensing also helps to mitigate biases that arise in reconstructing fields whose effects on the CMB are not orthogonal to lensing, though some form of bias-hardening is still required to eliminate the bias due to residual lensing~\cite{Su:2011ff}.

Small-scale off-diagonal mode couplings carry valuable cosmological information about the largest cosmological scales, creating novel opportunities for unique insights. 
Reconstruction of the bulk radial velocity fields through the measurement of the kinetic Sunyaev-Zel'dovich effect, for example, will provide competitive constraints on local-type primordial non-Gaussianity~\cite{Deutsch:2017ybc,Smith:2018bpn,Munchmeyer:2018eey}, deviations from general relativity~\cite{Zhang:2015uta}, specific forms of primordial isocurvature~\cite{Hotinli:2019wdp}, sources of CMB anomalies~\cite{Cayuso:2019hen}, and the reionization history~\cite{Alvarez:2020gvl,Hotinli:2020csk}. 
Reconstruction of the remote temperature quadrupole field through measurements of polarized Sunyaev-Zel'dovich effect provides an opportunity to measure primordial gravitational waves~\cite{Deutsch:2018umo} and to improve the constraints on the mean optical depth to reionization~\cite{Meyers:2017rtf}. 
Reconstruction of the bulk transverse velocity fields from measurements of the moving lens effect can provide precision measurements of the product of the linear-theory growth rate $f$ and amplitude of matter fluctuations $\sigma_8$ and allows the use of kinetic Sunyaev-Zel'dovich effect to learn about astrophysics by breaking the degeneracy between $f$ and various reconstruction biases~\cite{Hotinli:2020csk}. 
Furthermore, by combining these reconstructed large-scale fields with tracers of the density fluctuations, one can constrain parameters such as the scale-dependent galaxy bias without the limitations imposed by cosmic variance that arise when using the galaxy power spectrum~\cite{Seljak:2008xr}. 
Upcoming CMB surveys will allow for the high-precision observations of several of these interesting effects for the first time, and delensing will generally improve our ability to measure them by reducing lensing-induced CMB variance.

\subsection{Lensing-Induced Non-Gaussian Power Spectrum Covariance}\label{sec:ngcov}

While the primary CMB anisotropies are very well approximated by Gaussian random fields, gravitational lensing leads to non-zero connected four-point functions of CMB temperature and polarization anisotropies~\cite{Zaldarriaga:2000ud,Hu:2001fa}.
This is one manifestation of the change to CMB statistics that enables successful reconstruction of the lensing deflection using the observed CMB~\cite{Hu:2001tn,Hu:2001kj,Okamoto:2003zw}.

The lensing-induced CMB trispectrum leads to a non-trivial change to the lensed CMB power spectrum covariance, including covariance between spectra at different values of $\ell$~\cite{Smith:2004up,Smith:2005ue,Smith:2006nk,Li:2006pu,Benoit-Levy:2012dqi,Schmittfull:2013uea,Green:2016cjr,Peloton:2016kbw}.
This non-Gaussian off-diagonal power spectrum covariance effectively reduces the number of independent modes compared to the expectation from purely Gaussian statistics.
Neglecting the non-Gaussian power spectrum covariance therefore leads to overly optimistic parameter forecasts when using lensed CMB spectra along with the reconstructed lensing power spectrum due to a double-counting of information~\cite{Hu:2001fb}.
Accurate parameter forecasts thus require that the lensing-induced non-Gaussian power spectrum covariance is properly taken into account.

The power spectrum covariance matrix including the effects of lensing can be analytically modeled as~\cite{Benoit-Levy:2012dqi}
\be
    {\rm Cov}_{\ell_1\ell_2}^{XY,WZ}=&&  f_{\rm sky}^{-1} \Bigg\{ \frac{\delta_{\ell_1\ell_2}}{2\ell_1+1} \left[ \left( C_{\ell_1}^{XY} + N_{\ell_1}^{XY} \right) \left( C_{\ell_1}^{WZ} + N_{\ell_1}^{WZ} \right) + 
    \left( C_{\ell_1}^{XW} + N_{\ell_1}^{XW} \right) \left( C_{\ell_1}^{YZ} + N_{\ell_1}^{YZ} \right)
    \right] \nonumber \\
    &&+\sum\limits_\ell \left(\frac{\partial C_{\ell_1}^{XY}}{\partial {C}_{\ell}^{XY,\unlensed}} {\rm Cov}_{\ell\ell}^{X^{\unlensed} Y^{\unlensed} , W^{\unlensed} Z^{\unlensed}} \frac{\partial C_{\ell_2}^{WZ}} {\partial C_{\ell}^{WZ, \unlensed}} \right) (1-\delta_{\ell_1 \ell_2})
    \nonumber\\
    &&+\sum\limits_\ell \left(\frac{\partial C_{\ell_1}^{XY}}{\partial {C}_{\ell}^{\phi\phi}}{\rm Cov}_{\ell\ell}^{\phi\phi,\phi\phi}\frac{\partial C_{\ell_2}^{WZ}}{\partial {C}_{\ell}^{\phi\phi}}\right) \Bigg\} \, ,
    \label{eq:covarianceNG}
\ee
where the second and third lines correspond to the non-Gaussian contributions to the covariance matrix. 
These terms can be calculated by first differentiating the correlation functions given in Eqs.~\eqref{eq:Lensed_T_corr}\textendash{}\eqref{eq:Lensed_plus_corr} for the lensed spectra and Eqs.~\eqref{eq:Delensed_T_corr}\textendash{}\eqref{eq:Delensed_plus_corr} with the filters $g_\ell$, $h_\ell$, and $\bar{h}_\ell$ held fixed for the delensed spectra. 
The derivatives of the power spectra can then be computed from the derivatives of the correlation functions by use of Eq.~\eqref{eq:delensed_powspec}.
\texttt{CLASS\_delens} includes functions to calculate the derivatives of the lensed and delensed power spectra with this procedure.
The delta-function factor in the second line avoids double-counting the on-diagonal Gaussian covariance that is present even in the absence of lensing.

\begin{figure}[t!]
    \centering
    \vspace{-0.25cm}
    \subfloat{\includegraphics[width=0.45\columnwidth]{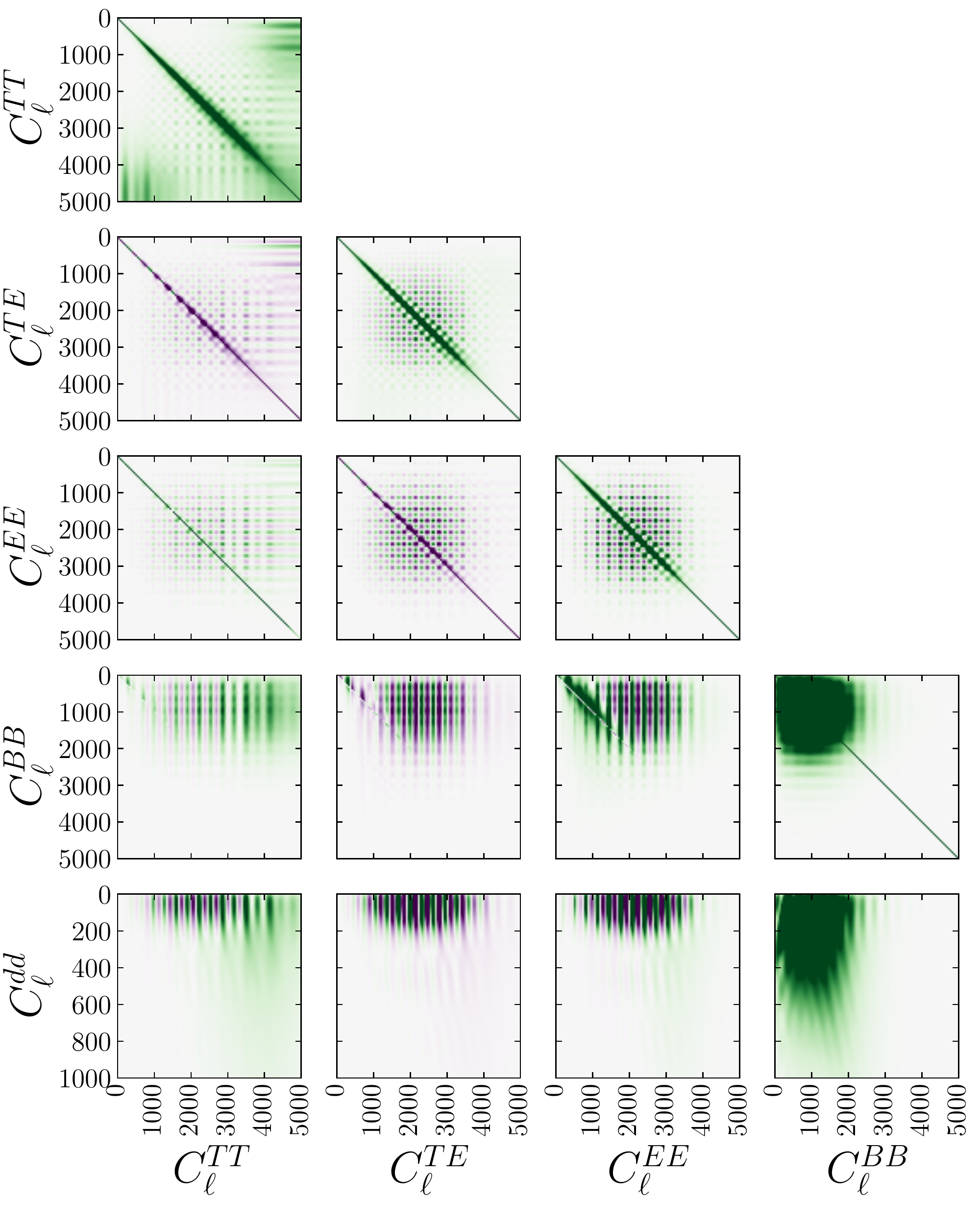}}
    \hfill
    \subfloat{\includegraphics[width=0.45\columnwidth]{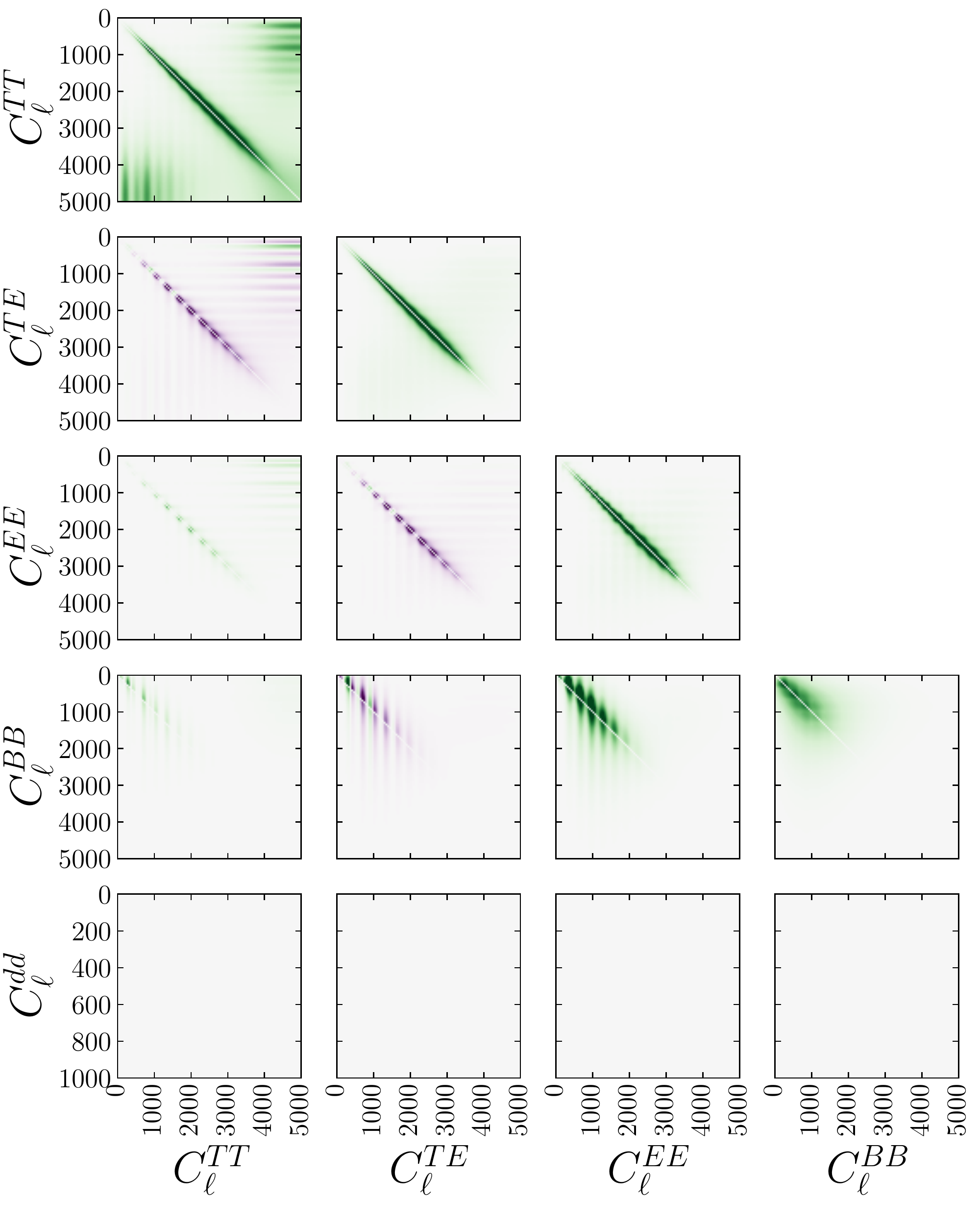}}
    \\
    \subfloat{\includegraphics[width=0.45\columnwidth]{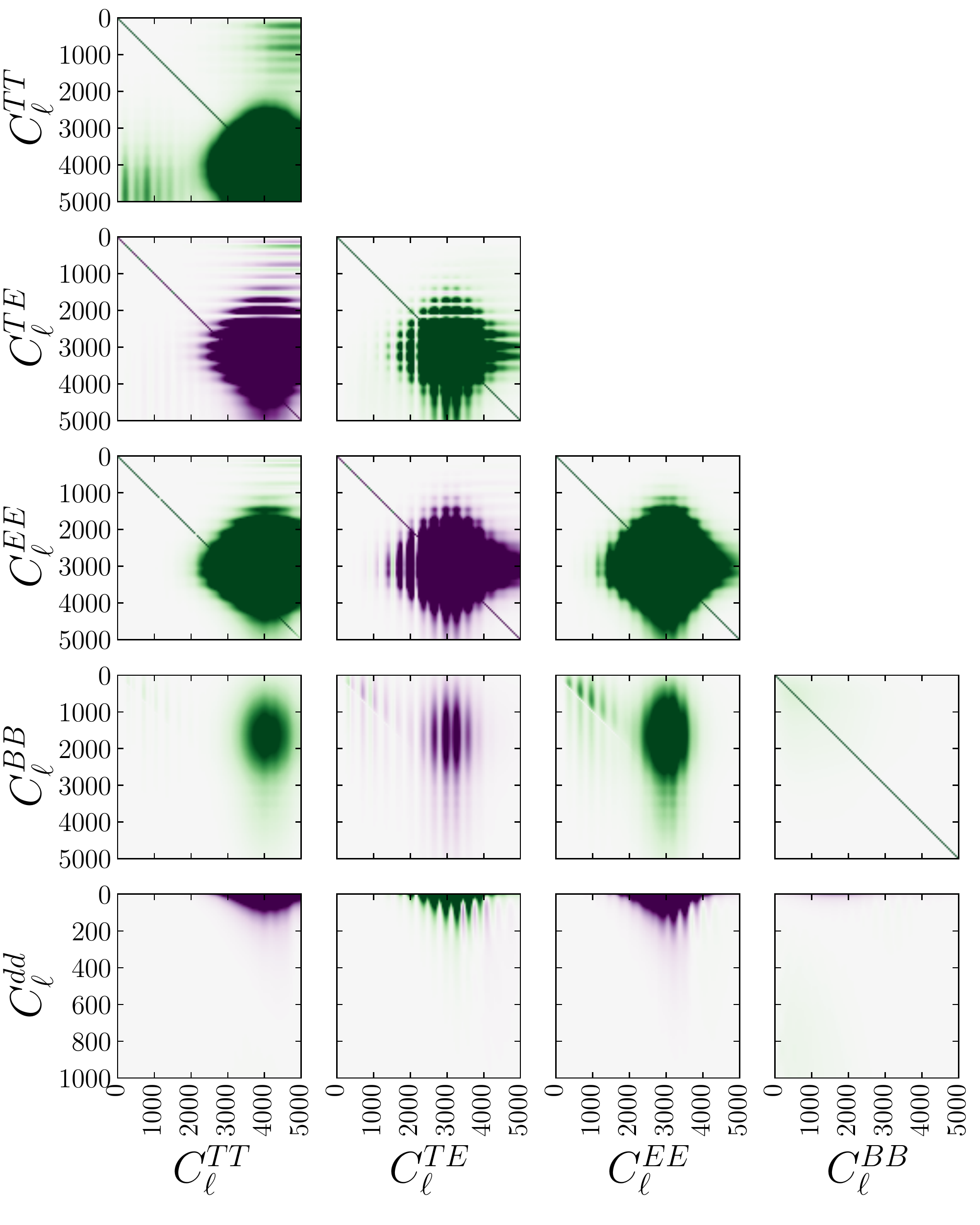}}
    \hfill
    \subfloat{\includegraphics[width=0.45\columnwidth]{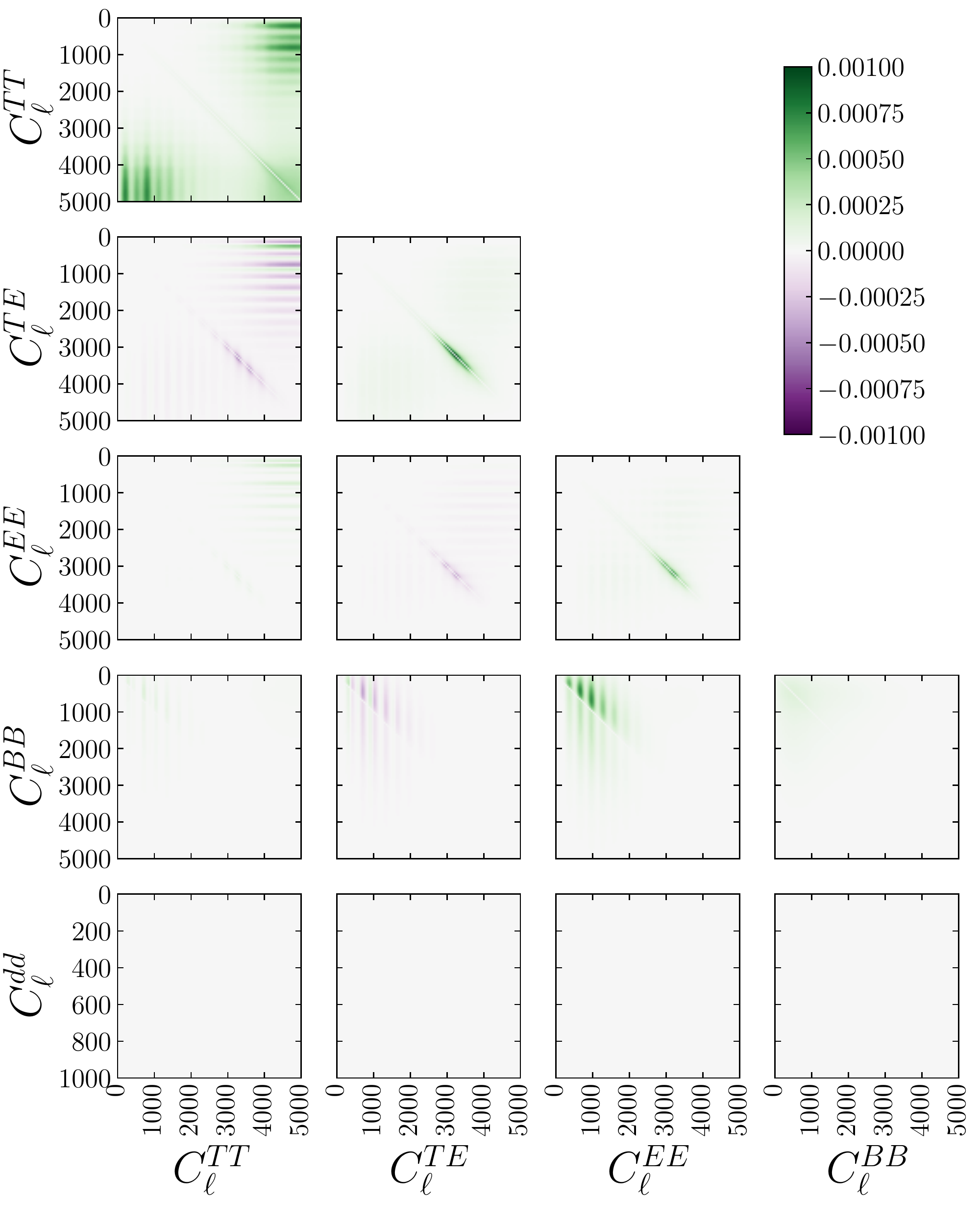}}
    \caption{
    (\textit{Top Left:}) Lensed power spectrum correlation (see Eq.~\eqref{eq:NGCorr}) for Experiment B defined in Table~\ref{table:experiments}. 
    (\textit{Top Right:}) Contribution to the lensed power spectrum correlation for Experiment B coming from the derivatives of the lensed spectra with respect to the unlensed spectra, shown in the second line of Eq.~\eqref{eq:covarianceNG}.
    (\textit{Bottom Left:}) Delensed power spectrum correlation for Experiment B.
    (\textit{Bottom Right:}) Contribution to the delensed spectrum correlation for Experiment B from the second line of Eq.~\eqref{eq:covarianceNG}.
    }
    \label{fig:S4_Lensed_Correlation_Combined}
\end{figure}

Since delensing reverses the effects of lensing, it also reduces the non-Gaussian power spectrum covariance for modes that are well measured~\cite{Green:2016cjr,Peloton:2016kbw}.
In order to visualize the non-Gaussian covariance, we define the power spectrum correlation matrix
\begin{equation}
    R_{\ell_1 \ell_2}^{XY,WZ} \equiv \frac{ \mathrm{Cov}_{\ell_1\ell_2}^{XY,WZ} } {\sqrt{\mathrm{Cov}_{\ell_1\ell_1}^{XY,XY} \mathrm{Cov}_{\ell_2\ell_2}^{WZ,WZ}} } \, .
    \label{eq:NGCorr}
\end{equation}
In Fig.~\ref{fig:S4_Lensed_Correlation_Combined} we show the lensed and delensed power spectrum correlation matrices for Experiment B defined in Table~\ref{table:experiments}.
We also show the contribution to the covariance coming from the derivatives of the lensed and delensed spectra with respect to the unlensed CMB spectra (the second line of Eq.~\eqref{eq:covarianceNG}), which was not included in the analysis of Ref.~\cite{Green:2016cjr}.
This contribution is particularly important to accurately model the non-Gaussian covariance of the $BB$ spectrum, since $B$ modes result entirely from the conversion of $E$ modes to $B$ modes by lensing in the cosmology we consider.
We find that this contribution has non-trivial impact on the covariance of the $TT$, $TE$, and $EE$ spectra as well, especially near the diagonal and on small scales.
It can be seen from Fig.~\ref{fig:S4_Lensed_Correlation_Combined} that delensing significantly reduces non-Gaussian covariances for modes that are measured with high signal-to-noise ratio.
Delensed covariance matrices, using the methods and code from Ref.~\cite{Green:2016cjr}, were used with data from the Atacama Cosmology Telescope by Ref.~\cite{ACT:2020goa}.

Including the non-Gaussian covariances for lensed and delensed spectra in parameter forecasts is important to avoid double-counting information and to make the forecasts internally consistent.
When the non-Gaussian covariances are properly included, delensing always improves parameter constraints compared to using lensed spectra~\cite{Green:2016cjr}.  
We find that the contributions from the second line of Eq.~\eqref{eq:covarianceNG} are particularly important for accurate forecasts involving parameters whose constraints are driven measuring the effects of lensing, such as $A_s$ and $\tau$.
This is especially true when including information from the $TT$ spectrum at $\ell>3000$, in which case the second and third lines of Eq.~\eqref{eq:covarianceNG} have nearly equal impact on the constraints of $A_s$ and $\tau$ in a $\Lambda$CDM cosmology, for example.

The lensing-induced non-Gaussian covariances will become the dominant source of uncertainty in searches for the effects of local-type primordial non-Gaussianity in future CMB experiments.
Reducing off-diagonal covariance by delensing has a direct and significant impact on improving the constraints on primordial non-Gaussianity~\cite{Coulton:2019odk}.


\section{Cosmological Parameter Forecasts}
\label{sec:forecasts}

In this section, we provide results from forecasts for upcoming CMB experiments, demonstrating how delensing can lead to tighter constraints on cosmological parameters.
We first give the details of our forecasting framework.
Then we show improvements for parameter constraints within the $\Lambda$CDM model before moving on to demonstrate improvements for extended models of cosmology.
We then show how delensing can mitigate the biases that result from incorrect modeling of the lensing power spectrum.

\subsection{Forecasting Formalism}
\label{subsec:fisher_formalism}

The unlensed, lensed, and lensing deflection power spectra used in these forecasts are all computed using the \texttt{CLASS} Boltzmann code~\cite{Blas:2011rf}. 
The delensed spectra are computed using \texttt{CLASS\_delens}, which implements the formalism described in Appendix~\ref{sec:correlation_functions}. 
We assume the noise in the CMB survey is Gaussian, with temperature noise spectrum
\begin{equation}
    N_{\ell}^{TT} = \Delta_T^2 \, \mathrm{exp} \left( \ell(\ell+1) \frac{\theta_{\mathrm{FWHM}}^2}{8\log{2}} \right) \, ,
    \label{eq:TT_noise}
\end{equation}
where $\Delta_T$ is the instrumental noise in $\mu$K-rad, and $\theta_{\mathrm{FWHM}}$ is the full-width at half-maximum beam size in radians. 
We take the polarization noise spectra to be $N_{\ell}^{EE} = N_{\ell}^{BB} = 2N_{\ell}^{TT}$, as is expected with fully polarized detectors. 
The lensing reconstruction noise is calculated according to the iterative delensing procedure outlined in Appendix~\ref{sec:lensing_reconstruction}.
In all forecasts presented here, we use the same iterative lensing reconstruction procedure such that the forecasts using lensed, delensed, and unlensed spectra have the same lensing reconstruction noise.
This choice is made to highlight the differences that arise from delensing the CMB spectra, rather than improvements that come from reduced lensing reconstruction noise.
As we showed in the previous section, delensing also leads to lower lensing reconstruction noise, and so in that sense the forecasts shown here underestimate the improvements of delensing compared to analyses with lensed spectra.

To forecast constraints on cosmological parameters, $\lambda^i$, we compute the Fisher matrix, with elements given by
\begin{equation}
    F_{ij} = \sum\limits_{\ell_1, \ell_2} \ \sum\limits_{W X Y Z} 
    \frac{\partial C_{\ell_1}^{XY}}{\partial \lambda^i} 
    \left[ \mathrm{Cov}_{\ell_1\ell_2}^{XY,WZ} \right]^{-1}
    \frac{\partial C_{\ell_2}^{WZ}}{\partial \lambda^j} \, .
    \label{eq:FisherMatrix}
\end{equation}
Unless otherwise specified, in each forecast, the sums over $\ell$ are taken from $\ell_\mathrm{min} = 30$ to $\ell_\mathrm{max} = 5000$, with the exception of the lensing spectrum, for which we use $\ell_\mathrm{min}=2$, and the $TT$ spectrum, for which we take $\ell_\mathrm{max} = 3000$, including when performing lensing reconstruction. 
We make this choice to avoid contamination from extragalactic foregrounds on small scales in $TT$, including radio point sources, the cosmic infrared background, and the kinetic Sunyaev-Zel'dovich effect. 
We use a beam size of 1.4~arcmin and a sky fraction of $f_\mathrm{sky} = 0.5$. 
All forecasts include $TT$, $TE$, $EE$, and $dd$ spectra, where $C_\ell^{dd} = \ell(\ell+1)C_\ell^{\phi\phi}$.
The covariances include the lensing-induced non-Gaussian contributions as given in Eq.~\eqref{eq:covarianceNG}. 
We include a prior on $\tau$ with $\sigma_\tau = 0.007$; however, we do not include external data, such as information from observations of baryon acoustic oscillations, in order to more transparently highlight the benefits of the delensing procedure and the effects of the non-Gaussian covariances.
Table~\ref{table:cosmo_fiducial} lists the cosmological parameters included in our forecasts, along with their fiducial values and the step sizes used to calculate numerical derivatives. We fix the sum of the neutrino masses at 0.06~eV and set the primordial helium abundance to be consistent with the predictions of standard Big Bang nucleosynthesis.
Our Fisher forecasting code, \texttt{FisherLens}, is publicly available\footnote{\url{https://github.com/ctrendafilova/FisherLens}}.

\begin{table}[t!]
\begin{center}
 \begin{tabular}{l@{\hskip 12pt}l @{\hskip 12pt}c@{\hskip 12pt}c} 
 \toprule
   Parameter & Symbol    &   Fiducial Value      & Step Size     \\ [0.5ex] 
 \hline
Physical cold dark matter density &   $\Omega_c h^2$ &   0.1197 	            & 0.0030 	    \\ 
Physical baryon density &   $\Omega_b h^2$ &   0.0222 	            & $8.0\times10^{-4}$ 	    \\
Angle subtended by acoustic scale &   $\theta_s$     &   0.010409 	            & $5.0\times10^{-5}$ 	    \\
Thomson optical depth to recombination &   $\tau$         &   0.060 	            & 0.020 	    \\
Primordial scalar fluctuation amplitude &   $A_s$          &   $2.196\times10^{-9}$  & $0.1\times10^{-9}$ 	    \\
Primordial scalar fluctuation slope &   $n_s$          &   0.9655 	            & 0.010 	    \\
   \hline
Effective number of neutrino species & $N_\mathrm{eff}$ & 3.046 & 0.080 \\
  \hline
\end{tabular}
    \caption{
    Fiducial cosmological parameters and step sizes for numerical derivatives used in forecasts, taken from \cite{Allison:2015qca}.
    The basic model we consider is the 6-parameter $\Lambda$CDM model. 
    The forecasts in Sec.~\ref{sub:bias} additionally include $N_\mathrm{eff}$.
    }
\label{table:cosmo_fiducial}
\end{center}
\end{table}

\subsection{Forecasts for \texorpdfstring{$\Lambda$}{L}CDM Parameters}
\label{sub:LCDM_forecasts}

\begin{figure}[t!]
    \vspace{-0.25cm}
    \includegraphics[width=\columnwidth]{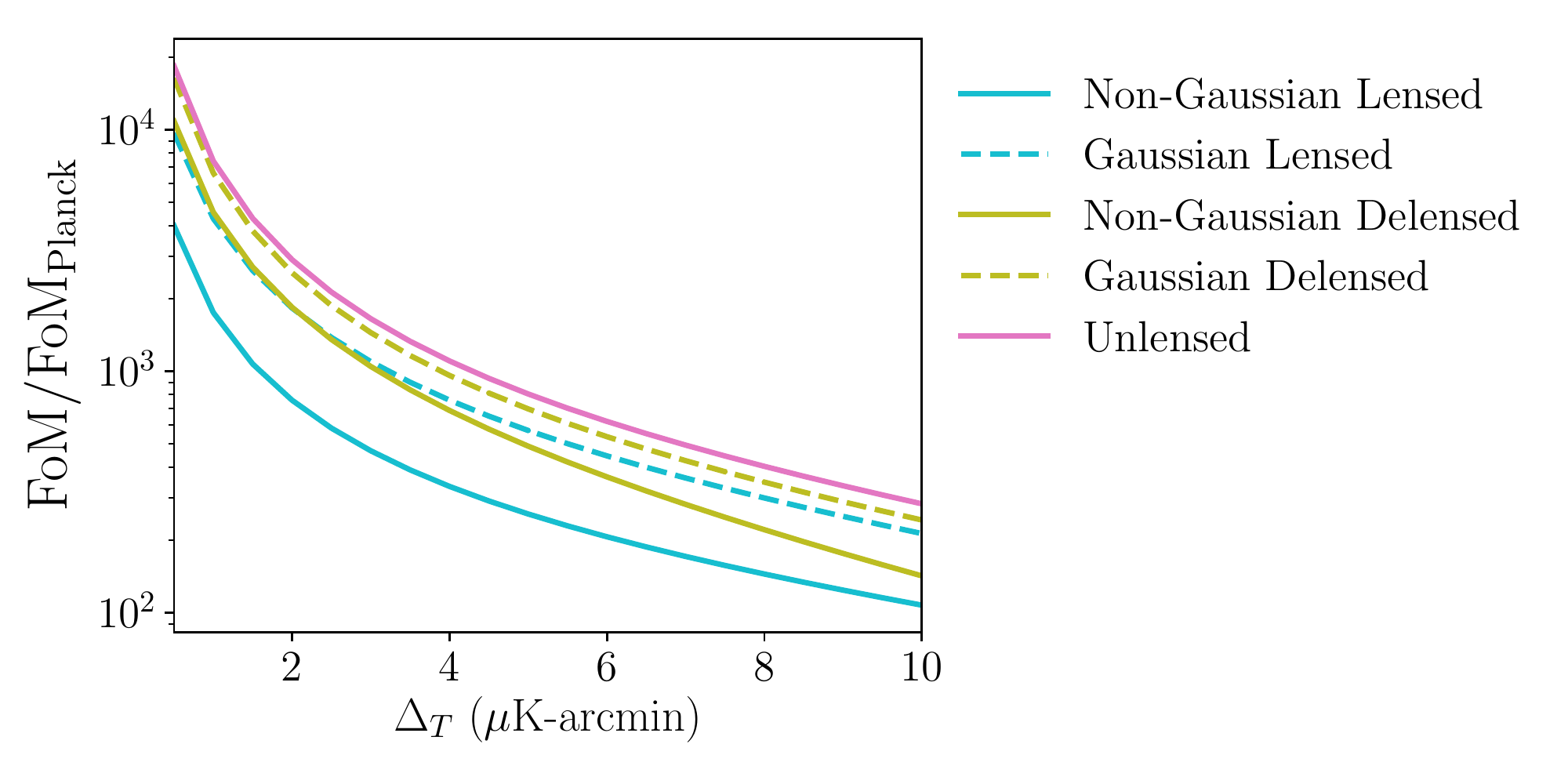}
    \vspace{-0.75cm}
    \caption{Figure of Merit for a 6-parameter $\Lambda$CDM forecast, calculated as $\mathrm{FoM} = \left[\mathrm{det}\left(F_{ij}^{-1}\right)\right]^{-1/2}$, and scaled relative to the Figure of Merit for Planck. 
    Delensing increases the Figure of Merit for all noise levels shown here, suggesting that delensing will provide a valuable improvement to constraining power for near-future CMB surveys.
    }
    \label{fig:FoM}
\end{figure}

In Fig.~\ref{fig:FoM}, we plot the Figure of Merit for a 6-parameter $\Lambda$CDM forecast, which we calculate as $\mathrm{FoM} = \left[\mathrm{det}\left(F_{ij}^{-1}\right)\right]^{-1/2}$~\cite{Wang:2008zh}, for a range of noise levels using lensed, unlensed, and delensed spectra. 
In each case the values are scaled to the Figure of Merit for Planck, which is calculated from a forecast using the same method described in Sec.~\ref{subsec:fisher_formalism}, and with noise specifications consistent with the treatment of Planck in Ref.~\cite{Allison:2015qca}.
While the Figure of Merit does not distinguish among the uncertainties on individual parameters, it provides a convenient summary of how well a given model will be constrained by various experiments, and we see here that constraints improve significantly with delensing.
We also show forecasts where non-Gaussian covariances are neglected, highlighting the overly optimistic nature of these results in comparison to forecasts where non-Gaussian covariances are included.

\begin{figure}[t!]
    \vspace{-0.25cm}
    \includegraphics[width=\columnwidth]{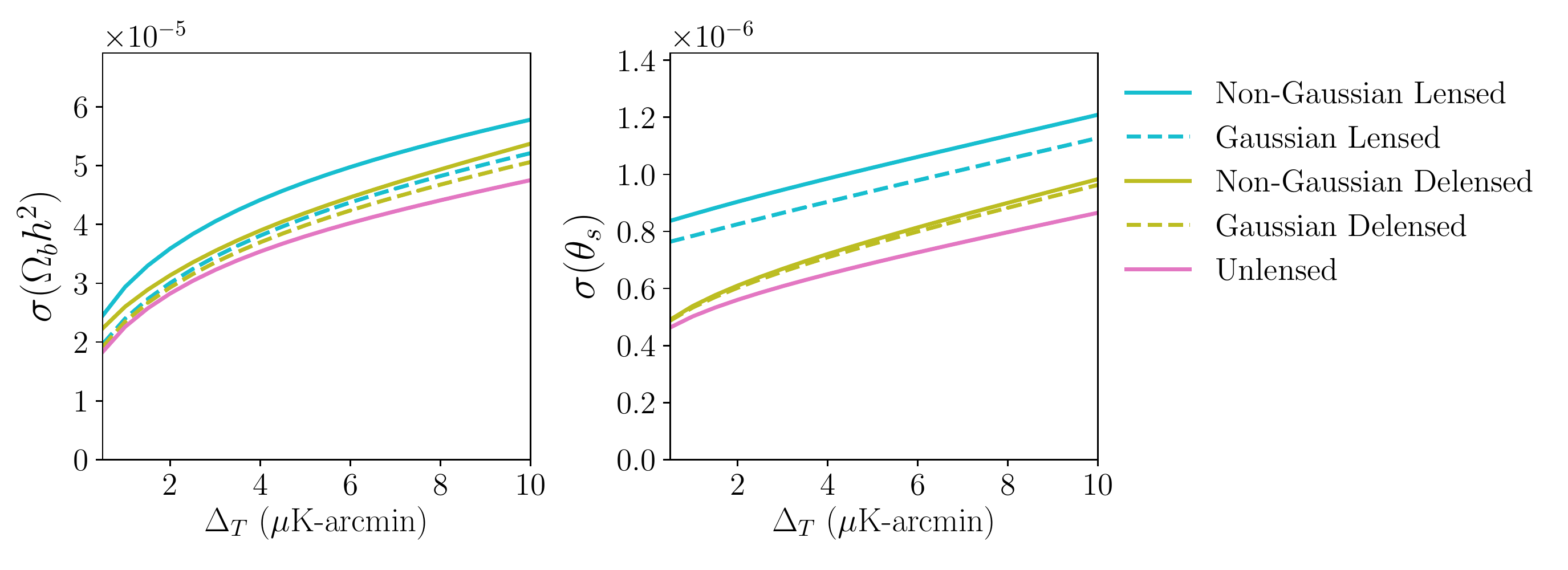}
    \vspace{-0.75cm}
    \caption{
    Forecasted constraints on $\Omega_b h^2$ and $\theta_s$ in a 6-parameter $\Lambda$CDM cosmology.
    Delensing significantly improves constraints on these parameters, as expected from the improvements on the measurement of peak positions discussed in Sec.~\ref{sub:peaklocations}.
    For $\theta_s$ in particular, the improvement from delensing an experiment with a noise level of $8~\mu$K-arcmin is equivalent to reducing the noise by a factor of 4.
    These forecasts also demonstrate the importance of including non-Gaussian covariances; the forecasted errors are too optimistic by more than 10\% if one uses Gaussian covariances.
    }
    \label{fig:omegab_thetas}
\end{figure}

In Fig.~\ref{fig:omegab_thetas}, we present forecasted constraints on the parameters of $\Lambda$CDM for the same range of noise levels. 
In particular, we show the constraints for $\Omega_b h^2$ and $\theta_s$, parameters which impact the inference of $H_0$ from CMB measurements.
It is clear that delensing leads to significant improvements on parameter constraints compared to using lensed spectra, as anticipated from the discussion of measuring peak positions discussed in Sec.~\ref{sub:peaklocations}. 
For an experiment with a noise level of $8~\mu$K-arcmin, delensing improves constraints on $\theta_s$ by an amount equivalent to the improvement that would come from reducing the noise level by a factor of 4, which would require 16 times more detector-years of observing. 
If non-Gaussian covariances are neglected, forecasted errors are too optimistic by more than 10\% for these parameters.

\subsection{Isocurvature Forecasts}
\label{sub:isocurvature_forecasts}

Next, we consider the impact of delensing on the measurement of isocurvature perturbations. 
Current CMB analysis from Planck~\citep{Planck:2018jri} provides strong constraints on generic forms of isocurvature on large-scales, suggesting isocurvature fluctuations are subdominant compared to the adiabatic fluctuations predicted by single field inflationary models and assumed in standard $\Lambda$CDM cosmology. 
Nevertheless, on small-scales, constraints are weaker and large amplitudes of isocurvature with blue-tilted spectra are allowed by current data. 
A detection of isocurvature fluctuations would rule out all single-field inflation models and provide valuable insight into the early universe~\cite{Polarski:1994rz,Gordon:2000hv}.

Isocurvature fluctuations predict spectra with peaks at different positions and heights compared to adiabatic fluctuations~\cite{Bucher:2000hy}.
Since delensing improves our ability to measure these aspects of the spectra, we expect that delensing will improve our constraints on isocurvature fluctuations.
We consider two models of cold dark matter isocurvature, one model in which the isocurvature fluctuations are fully correlated with the adiabatic fluctuations, and another in which the isocurvature fluctuations are uncorrelated and have a blue-tilted spectrum.
The parameters defining these two models are shown in Table~\ref{table:iso_fiducial}.
Models of blue-tilted cold dark matter isocurvature predict CMB spectra with peaks and troughs smoothed out, leading to a degeneracy with the effects of lensing on the spectra~\cite{Valiviita:2012ub}.
Delensing reverses the peak smoothing effect due to lensing, but does not alter any smoothing that comes from isocurvature, and we therefore expect that delensing should help to break the degeneracy between lensing and isocurvature effects.

\begin{table}[t!]
    \begin{center}
    \begin{tabular}{l@{\hskip 12pt}c@{\hskip 12pt}c@{\hskip 12pt}c@{\hskip 12pt}c}
        \toprule
        CDI Model 
        & Fiducial value of $f_\mathrm{cdi}$ at $k = 0.05~\mathrm{Mpc}^{-1}$  & $n_\mathrm{cdi}$ & $\alpha_\mathrm{cdi}$ & $c_\mathrm{cdi}$  \\ [0.5ex] 
        \hline
        Fully Correlated & 0.001 & 0.9655 & 0 & 1 \\   
        Uncorrelated, Blue-Tilted & 0.03  & 3 & 0 & 0 \\  
        \hline
    \end{tabular}
    \caption{
    Definitions of cold dark matter isocurvature (CDI) models under consideration.
    For both models, numerical derivatives with respect to $f_\mathrm{cdi}$ are computed with a step size of $0.0005$.
    }
    \label{table:iso_fiducial}
    \end{center}
\end{table}

\begin{figure}[t!]
    \vspace{-0.25cm}
    \includegraphics[width=\columnwidth]{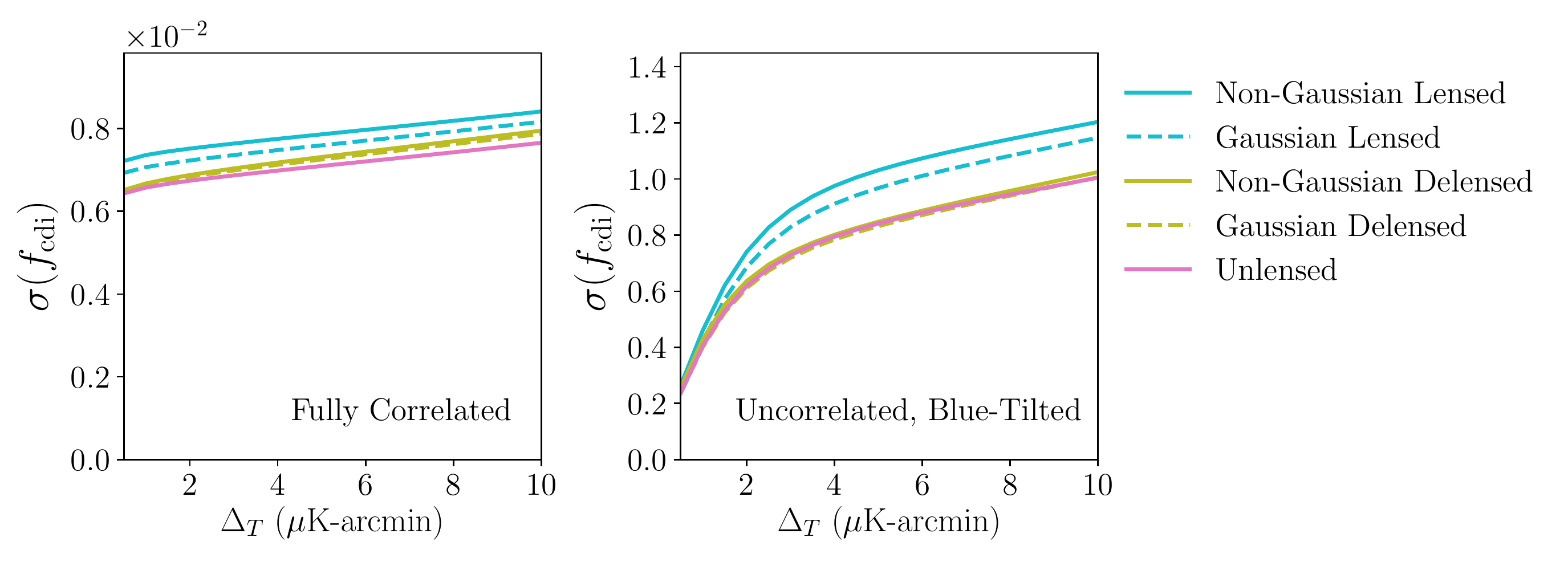}
    \vspace{-0.75cm}
    \caption{
    Forecasts showing the benefits of delensing in constraining the entropy-to-curvature ratio, $f_\mathrm{cdi}$, for two models of cold dark matter isocurvature, in a 7-parameter forecast of $\Lambda$CDM~+~$f_\mathrm{cdi}$. 
    Parameters defining these models are given in Table~\ref{table:iso_fiducial}.
    Delensing provides significantly improved constraints for both models.
    }
    \label{fig:cdm_iso}
\end{figure}

In Fig.~\ref{fig:cdm_iso} we show results for 7-parameter forecasts where we constrain the cold dark matter isocurvature entropy-to-curvature ratio $f_\mathrm{cdi} \equiv \left[\mathcal{S}_\mathrm{cdi} / \mathcal{R}\right](k = 0.05~\mathrm{Mpc}^{-1})$ in addition to $\Lambda$CDM parameters. 
The left panel shows constraints for the fully correlated model (the results for an anti-correlated case are very similar).
The effects of nearly scale-invariant cold dark matter isocurvature on the CMB spectra are restricted to large angular scales, and thus there is little improvement on the constraints with decreasing noise level.
However, delensing provides a non-trivial improvement in the constraints.
For the models we consider here, the improvement from delensing is comparable to the improvement in constraints when decreasing experimental noise from 10~$\mu$K-arcmin to 1~$\mu$K-arcmin when using lensed spectra.
The right panel shows results for the uncorrelated, blue-tilted isocurvature model.
In this case, reversing the peak smoothing due to lensing leads to a significant improvement in the constraints, nearly matching the constraints expected from unlensed spectra.

\subsection{Mitigation of Parameter Bias}
\label{sub:bias}

An additional benefit of delensing is that delensed spectra are less susceptible to biases that may result from incorrectly modeling the lensing spectrum.
The distribution of matter on small scales at late times is significantly impacted by the nonlinear growth of structure.
The CMB lensing potential is determined by the integrated mass density out to the surface of last scattering, and thus receives contributions from nonlinear fluctuations in the matter density.
Fluctuations in mildly nonlinear regime can be reliably calculated with perturbation theory~\cite{Bernardeau:2001qr}, but simulations are necessary to accurately model the scales where nonlinear effects are more significant.
Software such as \texttt{HALOFIT}~\cite{Smith:2002dz,Takahashi:2012em,Bird:2011rb} can be used to compute the nonlinear matter power spectrum and CMB lensing power spectrum using fits to numerical simulations inspired by the analytic halo model~\cite{Peacock:2000qk,Seljak:2000gq,Cooray:2002dia}.

Baryonic effects including supernovae, gas cooling, and feedback from active galactic nuclei also have a significant impact on the distribution of matter on small scales.  
Hydrodynamical simulations are used to model these effects, though not all simulations agree, meaning that there is a theoretical uncertainty in the small scale matter distribution and thus in the CMB lensing power spectrum~\cite{White:2004kv,Zhan:2004wq,Jing:2005gm,Rudd:2007zx,Semboloni:2011fe,Natarajan:2014xba,Copeland:2019bho,Schneider:2019xpf,Chung:2019bsk,McCarthy:2020dgq,McCarthy:2021lfp}.
Baryonic effects on the CMB lensing power spectrum impact the lensed CMB spectra, and failing to properly account for these feedback effects can lead to biased inferences of cosmological parameters from the lensed CMB~\cite{McCarthy:2021lfp}.

As discussed in Ref.~\cite{McCarthy:2021lfp}, delensing can serve to mitigate biases from incorrect modeling of the CMB lensing power spectrum.
Here we show quantitatively the degree to which delensing reduces biases that result from failing to account for baryonic feedback effects.
The bias on a given parameter $B(\lambda^i)$ can be computed in the Fisher formalism as~\cite{Huterer:2004tr,LoVerde:2006cj,Amara:2007as}
\begin{equation}
    B(\lambda^i) = \sum_j F_{ij}^{-1} \sum\limits_{\ell_1, \ell_2} \ \sum\limits_{W X Y Z} 
    \frac{\partial C_{\ell_1}^{XY}}{\partial \lambda^j} 
    \left[ \mathrm{Cov}_{\ell_1\ell_2}^{XY,WZ} \right]^{-1}
    \Delta C_{\ell_2}^{WZ} \, ,
    \label{eq:BiasVector}
\end{equation}
with the definition
\begin{equation}
    \Delta C_\ell^{XY} = C_\ell^{XY,\mathrm{true}} - C_\ell^{XY,\mathrm{fiducial}} \, .
\end{equation}
We use \texttt{HMcode}~\cite{Mead:2015yca} as implemented in \texttt{CLASS} with the dark matter only emulator model~\cite{Heitmann:2013bra} serving as the fiducial model and the OWLS AGN model~\cite{Schaye:2009bt,vanDaalen:2011xb} as the true model.
These choices are included in \texttt{HMcode} as specific values of the minimum concentration parameter $c_\mathrm{min}$ and halo bloating parameter $\eta_0$ that have been fit to each model.
Note that Refs.~\cite{Chung:2019bsk,McCarthy:2020dgq,McCarthy:2021lfp} used the matter power spectra directly from simulations and therefore may differ somewhat from the parametric fits of \texttt{HMcode} that we employ here.

\begin{figure}[t!]
    \vspace{-0.25cm}
    \includegraphics[width=\columnwidth]{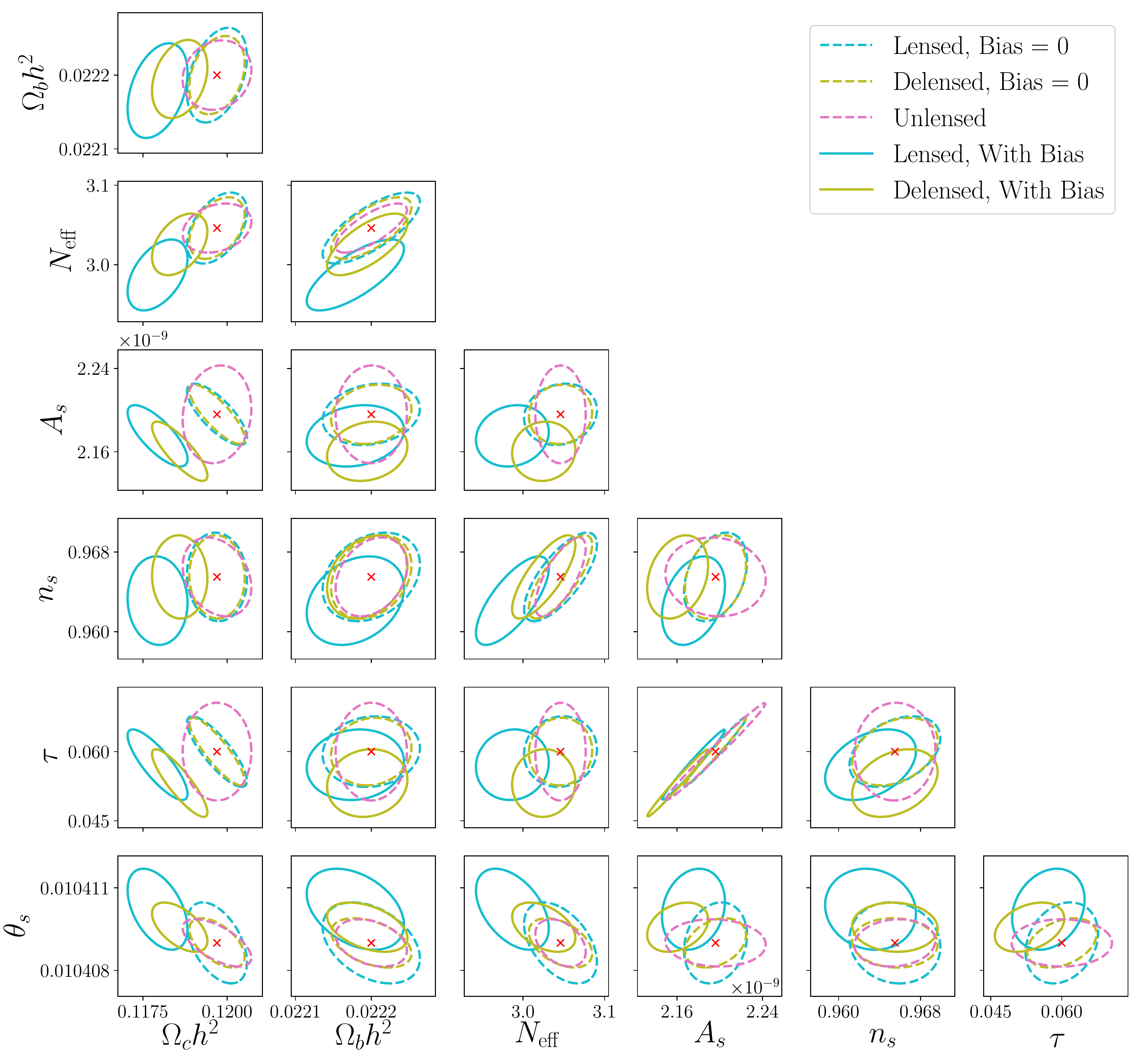}
    \vspace{-0.75cm}
    \caption{Forecasted errors and bias on the parameters of $\Lambda$CDM~+~$N_\mathrm{eff}$ cosmology for Experiment~B defined in Table~\ref{table:experiments} using only $TT$, $TE$, and $EE$ spectra, assuming that the true lensing spectrum is described by the OWLS AGN model while the fiducial model was taken from a dark matter only emulator, with both models implemented using \texttt{HMcode}.
    Delensing mitigates the bias for all parameters except $A_s$ and $\tau$ whose constraints are primarily driven by lensing. The largest residual effects of lensing appear in the high-$\ell$ $TT$ spectrum (see Fig.~\ref{fig:dampingtail}) and correspond to the most poorly modeled part of the lensing power spectrum.  In realistic surveys, this same region of the $TT$ power spectrum cannot be measured due to astrophysical foregrounds.
    }
    \label{fig:baryonbias}
\end{figure}

In Fig.~\ref{fig:baryonbias} we show the forecasted bias and uncertainty on the 7 parameters of the $\Lambda$CDM~+~$N_\mathrm{eff}$ cosmology before and after delensing.
To match with the setup described in Ref.~\cite{McCarthy:2021lfp} for these forecasts, we use $\ell_\mathrm{max} = 5000$ for all spectra, and we do not include lensing reconstruction information in the Fisher matrix.
Delensing provides a significant reduction in the bias for all parameters, with the exception of $A_s$ and $\tau$.
For the situation considered here, the degeneracy between $A_s$ and $\tau$ is primarily broken by the lensing effects on the CMB spectra. 

When considering the improvement from delensing alone, the residual effects of the (incorrectly modeled) lensing are responsible for the bias that remains after delensing, and the bias is exacerbated for the parameters with lensing-driven constraints. 
This is somewhat counter-intuitive, as one might expect that we can accurately reconstruct and remove the lensing modes that are biasing our data.  
However, as shown in Fig.~\ref{fig:dampingtail}, the lensing effect on the damping tail in $TT$ and $EE$ at high $\ell$ is not fully removed by delensing.  
Particularly in $TT$, this residual lensing is well above the noise and contributes significantly to parameter constraints.  
In addition, these contributions arise in the high-$L$ region of the lensing power spectrum where the baryonic effects are largest.  
The power spectrum constraints are then sensitive to the modeling of the lensing power spectrum, and the suppressed small scale lensing power in the OWLS AGN model leads to an inference of $A_s$ that is biased low due to the lower than expected lensing power, and this also pushes $\tau$ to smaller values since the combination $A_s e^{-2\tau}$ is tightly constrained by the temperature and $E$-mode power.

In a realistic survey, the $\ell > 3000$ region of the $TT$ power spectrum is limited by astrophysical foregrounds and will not contribute to the constraints on cosmological parameters.  
As a result, we anticipate these biases would be much smaller when excluding the small scale $TT$ power as the effect of the residual lensing is mostly limited by noise in the $E$-modes.
Furthermore, we find that marginalizing over the parameters $\eta_0$ and $c_\mathrm{min}$ that specify the baryonic feedback model in \texttt{HMcode} reduces biases of all parameters to less than 1\% of the $1\sigma$ error bars, in agreement with the results of Ref.~\cite{McCarthy:2021lfp}. 
After marginalizing over baryonic feedback parameters, delensing provides little additional improvement for biases, but delensing still reduces errors as described above.

\section{Conclusion}
\label{sec:conclusion}

Gravitational lensing of the CMB distorts our view of the primary anisotropies, injecting information from the low-redshift universe into these relics of recombination.  
Our ability to reconstruct the map of the lensing potential from data allows us to isolate these late time effects and probe the growth of structure.  
We can additionally use this map to remove the effects of lensing from the primary temperature and polarization maps (i.e.~delensing), in order to isolate the information from the recombination era.  
In future surveys, this procedure will enable us to clearly separate the primary CMB from the lensing potential for most of the modes observed with high signal-to-noise ratio.

In this paper, we explored several benefits provided by CMB delensing, extending the results of Ref.~\cite{Green:2016cjr} to include curved-sky effects and iterated delensing on all spectra. 
Delensed spectra have sharper acoustic peaks and more pronounced damping tails.
The observed $B$-mode power is reduced by delensing, thereby aiding in the search for primordial gravitational waves as well as sources of secondary $B$ modes.
Delensing reduces the variance on the reconstruction of lensing and other sources of non-stationary CMB statistics.
Non-Gaussian off-diagonal power spectrum covariance is reduced by delensing, simplifying analyses and improving constraints on primordial non-Gaussianity.
We showed how delensing results in tighter constraints on cosmological parameters in $\Lambda$CDM and in extended cosmological models.

One might reasonably wonder how it can be that delensing can increase cosmological constraining power, especially when the estimate of the lensing potential is derived from observations of the CMB itself.
We are not adding external data, so it must be that all of the information is already contained in the CMB maps that we observe.
Lensing has the effect of moving information from the two-point statistics into higher-order statistics of the CMB maps.
While it is possible in principle to forward-model any system we wish including the effects of lensing and to treat the whole analysis at the level of maps, this procedure would be intractable both for forecasting and for inferences from real data.
Delensing moves the information back into the two-point statistics where things are more easily calculated and where physical insights can more straightforwardly guide our understanding.
Furthermore, the procedure we have described here achieves this self-consistently.

It may be tempting to simply use lensed spectra along with the reconstructed lensing power spectrum, but this analysis is overly optimistic if lensing-induced covariances are ignored, and is sub-optimal when treated properly.
This is due to the fact that the lensing potential exhibits cosmic variance, and therefore the lensed CMB scatters more than would a Gaussian field with the same spectrum.
Delensing removes the actual realization of lensing on the sky (as opposed to simply deconvolving the lensing spectrum), thereby removing the extra scatter that comes from lensing variance.
This fact is well-understood for primordial gravitational wave searches, where simply subtracting the lensing $B$-mode power would provide weaker constraints than delensing due to the variance of the lensing $B$ modes.

Similarly, using unlensed spectra along with the reconstructed lensing power spectrum produces overly optimistic forecasts.
We do not actually observe the unlensed CMB, though parameters that mostly impact the primary anisotropies (such as $\theta_s$ and $N_\mathrm{eff}$) would be more tightly constrained if we could.

We discussed how delensing helps to mitigate biases that arise from incorrectly modeling the lensing power spectrum.
A related benefit of delensing is that by reducing lensing-induced covariance, delensing can help to identify whether such biases are present in the data.
Unless a given source of bias is perfectly degenerate with one or more cosmological parameters, incorrect modeling of a particular effect will tend to show up as an internal tension between various parts of the data set.
Delensing makes the temperature and polarization spectra more independent of the lensing spectrum (and of the temperature and polarization spectra on different scales), thereby making any disagreement between the cosmological parameters inferred from subsets of the data easier to identify.
To make this somewhat more clear, different combinations of $C_\ell^{\phi\phi}$ and $C_\ell^{TT,\unlensed}$ can lead to the same $C_\ell^{TT,\lensed}$ for some range of scales; however, delensing removes this ambiguity for all modes that are well-measured, leaving a temperature spectrum that is less sensitive to the modeling of lensing.

Delensing is made more valuable by the fact that it is possible to achieve at no additional experimental cost.
CMB surveys will collect all of the data that is necessary to delens the temperature and polarization maps using an internal estimate of the lensing potential.
Given the wide array of benefits provided by CMB delensing, it is a procedure worth applying wherever possible, especially with the high fidelity maps expected from future surveys.

\section*{Acknowledgments}

We thank Colin Hill, Marilena LoVerde, Srini Raghunathan, Kendrick Smith, and Ben Wallisch for helpful discussions.
DG is supported by the US~Department of Energy under Grants~\mbox{DE-SC0019035} and~\mbox{DE-SC0009919}.
JM and CT are supported by the US~Department of Energy under Grant~\mbox{DE-SC0010129}.
This work was completed in part at the Aspen Center for Physics, which is supported by National Science Foundation grant PHY-1607611.
SCH is supported by the Horizon Fellowship from Johns Hopkins University.
SCH also acknowledges the support of a grant from the Simons Foundation at the Aspen Center for Physics, Imperial College President's Fellowship and a postdoctoral fellowship from Imperial College London.
SCH would like to thank Imperial College High Performance Computing Service at Imperial College London (UK) for providing computational resources at various early stages of this project.
JM and CT carried out computations on ManeFrame~II, a shared high-performance computing cluster at Southern Methodist University. 
We acknowledge the use of \texttt{CLASS}~\cite{Blas:2011rf}, \texttt{IPython}~\cite{Perez:2007ipy}, and the Python packages \texttt{Matplotlib}~\cite{Hunter:2007mat}, \texttt{NumPy}~\cite{Harris:2020xlr}, and~\texttt{SciPy}~\cite{Virtanen:2019joe}.

\appendix 

\section{Delensed correlation functions}
\label{sec:correlation_functions}

The lensed all-sky correlation functions have been calculated in Ref.~\cite{Challinor:2005jy} and capture the lensing effect accurately on all scales.
We will apply the same methods to calculate the delensed correlation functions on the curved sky.
The expressions for the delensed correlation functions as well as the lensing reconstruction covariance described in Appendix~\ref{sec:lensing_reconstruction} have been shown in part in Ref.~\citep{Hotinli:2020adc} which appeared during the early stages of this work.
The lensed and delensed correlation functions depend only on the separation between two points on the sky, and hence are invariant under displacements, describing the correlation that is insensitive the bulk unobservable shifts of the unlensed CMB.  
The lensed correlation functions are~\cite{Challinor:2005jy}
\be
    {\xi}^{TT,\lensed}\!\!\simeq\!\!\sum\limits_{\ell m m'}\!\! \frac{2\ell+1}{4\pi}C_\ell^{TT, \unlensed}\!\Bigg\{X_{000}^2d^\ell_{00}\!+\!\frac{8}{\ell(\ell+1)}C_{\gl,2}X_{000}'^2d^{\ell}_{1-1}\!+\!C_{\gl,2}^2\left(X'^2_{000}d_{00}^\ell\!+\!X_{220}^2d_{2-2}^\ell\right)\!\Bigg\},\ \ \ \ \ \ 
    \label{eq:Lensed_T_corr}
\ee
\begin{align}
    \label{eq:Lensed_TE_corr}
    {\xi}^{TE,\lensed}\simeq\sum\limits_{\ell m m'} \frac{2\ell+1}{4\pi} C^{TE, \unlensed}_\ell\Bigg\{ X_{022}X_{000}d_{20}^\ell+&\frac{2C_{\gl,2}X_{000}'}{\sqrt{\ell(\ell+1)}}(X_{121}d_{11}^\ell+X_{132}d_{3-1}^\ell)  \\ +&\frac{1}{2}C_{\gl,2}^2\left[(2X_{022}'X_{000}'+X_{220}^2)d_{20}^\ell+X_{220}X_{242}d_{4-2}^\ell\right]\Bigg\}\,, \nonumber
\end{align}
\be
    {\xi}^{-,\lensed}\simeq\sum\limits_{\ell m m'}\frac{2\ell+1}{4\pi}(C_\ell^{EE, \unlensed}-C_\ell^{BB, \unlensed})\Bigg\{X_{022}^2&&d_{2-2}^\ell+C_{\gl,2}(X_{121}^2d_{1-1}^\ell+X_{132}^2d_{3-3}^\ell)\\ 
    +&&\frac{1}{2}C_{\gl,2}^2\left[2(X_{022}')^2d_{2-2}^\ell+X_{220}^2d_{00}^\ell+X_{242}^2d_{4-4}^\ell\right]\Bigg\} \, , \nonumber
    \label{eq:Lensed_minus_corr}
\ee
and
\be
    \xi^{+,\lensed}\!\simeq\!\sum\limits_{\ell m m'}\!\frac{2\ell+1}{4\pi}(C_\ell^{EE, \unlensed}\!+\!C_\ell^{BB, \unlensed})\Bigg\{X_{022}^2&&d_{22}^\ell\!+\!2C_{\gl,2}X_{132}X_{121}d_{31}^\ell \\
    +&&C_{\gl,2}^2\left[(X_{022}')^2d_{22}^\ell\!+\!X_{242}X_{220}d_{40}^\ell\right]\Bigg\}\,, \nonumber
    \label{eq:Lensed_plus_corr}
\ee
where $d^\ell_{mm'}$ are Wigner $d$-matrices,
\be
    X_{000}=e^{-\ell(\ell+1)\sigma^2/4}\,,
\ee
\be
    X_{220}=\frac{1}{4}\sqrt{(\ell+2)(\ell-1)\ell(\ell+1)}e^{-(\ell(\ell+1)-2)\sigma^2/4}\,,
\ee
and $\sigma^2(\beta)=C_{\gl}(0)-C_{\gl}(\beta)$, where
\be
    C_\gl(\beta)=\sum\limits_{\ell}\frac{2\ell+1}{4\pi}\ell(\ell+1)C_\ell^{\phi\phi}d^\ell_{11}(\beta)\,,
\ee
and
\be
    C_{\gl,2}(\beta)=\sum\limits_{\ell}\frac{2\ell+1}{4\pi}\ell(\ell+1)C_\ell^{\phi\phi}d_{-11}^\ell(\beta)\,,
\ee
and primes on the $X_{kmn}$ denote differentiation with respect to $\sigma^2$.
Similarly, the delensed correlation functions can be defined as
\be
    \xi^{TT,\delensed}= &&\sum\limits_\ell \frac{2\ell+1}{4\pi} \Bigg( |\bar{h}_\ell|^2 {C}_\ell^{TT,\lensed} \\
    +&&2\bar{h}_\ell h_\ell C_{\ell}^{TT,\unlensed} \left[ (X_{000}^{\bar{h}h})^2 d_{00}^\ell +\frac{8}{\ell(\ell+1)} C_{\gl,2}^{\bar{h}h} (X_{000}^{\bar{h}h}{}')^2 d_{1-1}^\ell +(C_{\gl,2}^{h^2})^2 \left( (X^{\bar{h}h}_{000}{}')^2 d_{000}^\ell +(X_{220}^{\bar{h}h})^2 d_{2-2}^\ell \right) \right] \nonumber\\
    \ \ \ +&& |h_\ell|^2 C_{\ell}^{TT,\unlensed} \left[ (X_{000}^{h^2})^2 d_{00}^\ell +\frac{8}{\ell(\ell+1)} C_{\gl,2}^{h^2} (X_{000}^{h^2}{}')^2 d_{1-1}^\ell +(C_{\gl,2}^{h^2})^2 \left( (X_{000}^{h^2}{}')^2  d_{00}^\ell +(X_{220}^{h^2})^2  d_{2-2}^\ell \right) \right] \Bigg)\nonumber
    \label{eq:Delensed_T_corr}
\ee
\be
    {\xi}^{TE,\delensed}=&&\sum\limits_{\ell}\frac{2\ell+1}{4\pi}\Bigg(\bar{h}_\ell \bar{h}_\ell^P {C}^{TE,\lensed}_\ell\\
    &&+({h}_\ell \bar{h}_\ell^P+\bar{h}{h}^P){C}^{TE,\unlensed}_\ell\Bigg[X^{h\bar{h}}_{022}X^{h\bar{h}}_{000}d_{20}^\ell+\frac{2C_{\gl,2}^{h\bar{h}}X_{000}^{h\bar{h}}{}'}{\sqrt{\ell(\ell+1)}}(X_{121}^{h\bar{h}}d_{11}^\ell+X_{132}^{h\bar{h}}d_{3-1}^\ell)\nonumber\\
    &&\ \ \ \ \ \ \ \ \ \ \ \ \ \ \ \ \ \ \ \ \ \ \ \ \ \
    +\frac{1}{2}(C_{\gl,2}^{h\bar{h}})^2\left[(2X_{022}^{h\bar{h}}{}'X_{000}^{h\bar{h}}{}'+(X_{220}^{h\bar{h}})^2)d_{20}^\ell+X_{220}^{h\bar{h}}X_{242}^{h\bar{h}}d_{4-2}^\ell\right]\Bigg]\nonumber\\
    &&+(h_\ell h_\ell^P){C}^{TE,\unlensed}_\ell\Bigg\{X^{h^2}_{022}X^{h^2}_{000}d_{20}^\ell+\frac{2C_{\gl,2}^{h^2}X_{000}^{h^2}{}'}{\sqrt{\ell(\ell+1)}}(X_{121}^{h^2}d_{11}^\ell+X_{132}^{h^2}d_{3-1}^\ell)\nonumber\\
    &&\ \ \ \ \ \ \ \ \ \ \ \ \ \ \ \ \ \ \ \ \ \ \ \ \ \
    +\frac{1}{2}(C_{\gl,2}^{h^2})^2\left[(2X_{022}^{h^2}{}'X_{000}^{h^2}{}'+(X_{220}^{h^2})^2)d_{20}^\ell+X_{220}^{h^2}X_{242}^{h^2}d_{4-2}^\ell\right]\Bigg\}\Bigg)\nonumber
    \label{eq:Delensed_TE_corr}
\ee
\be
    {\xi}^{-,\delensed}=&&\sum\limits_{\ell}\frac{2\ell+1}{4\pi}\Bigg(\left| \bar{h}_\ell^P \right|^2 ({C}_\ell^{EE,\lensed}-{C}_\ell^{BB,\lensed})\\
    &&+2{h}_\ell^P\bar{h}_\ell^P({C}^{EE,\unlensed}_\ell-{C}^{BB,\unlensed}_\ell)\Bigg\{(X_{022}^{h\bar{h}})^2d_{2-2}^\ell+C_{\gl,2}^{h\bar{h}}((X_{121}^{h\bar{h}})^2d_{1-1}^\ell+(X_{132}^{h\bar{h}})^2d_{3-3}^\ell)\nonumber\\
    &&\ \ \ \ \ \ \ \ \ \ \ \ \ \ \ \ \ \ \ \ \ \ \ \ \ \ \ \ \ \ \ \ \ \ \ \
    +\frac{1}{2}(C_{\gl,2}^{h\bar{h}})^2\left[2(X_{022}^{h\bar{h}}{}')^2d_{2-2}^\ell+(X_{220}^{h\bar{h}})^2d_{00}^\ell+(X_{242}^{h\bar{h}})^2d_{4-4}^\ell\right]\Bigg\}\nonumber\\
    &&+ \left| h^P_\ell \right| ^2 ({C}^{EE,\unlensed}_\ell-{C}^{BB,\unlensed}_\ell)\Bigg\{(X_{022}^{h^2})^2d_{2-2}^\ell+C_{\gl,2}^{h^2}((X_{121}^{h^2})^2d_{1-1}^\ell+(X_{132}^{h^2})^2d_{3-3}^\ell)\nonumber\\
    &&\ \ \ \ \ \ \ \ \ \ \ \ \ \ \ \ \ \ \ \ \ \ \ \ \ \ \ \ \ \ \ \ \ \ \ \
    +\frac{1}{2}(C_{\gl,2}^{h^2})^2\left[2(X_{022}^{h^2}{}')^2d_{2-2}^\ell+(X_{220}^{h^2})^2d_{00}^\ell+(X_{242}^{h^2})^2d_{4-4}^\ell\right]\Bigg\}\Bigg)\nonumber
    \label{eq:Delensed_minus_corr}
\ee
\be
    {\xi}^{+,\delensed}=&&\sum\limits_{\ell}\frac{2\ell+1}{4\pi}\Bigg(\left| \bar{h}_\ell^P \right|^2 ({C}_\ell^{EE,\lensed}+{C}_\ell^{BB,\lensed})\\
    &&+2{h}_\ell^P\bar{h}_\ell^P({C}^{EE,\unlensed}_\ell+{C}^{BB,\unlensed}_\ell)\Bigg\{(X_{022}^{h\bar{h}})^2d_{22}^\ell+2C_{\gl,2}^{h\bar{h}} X_{132}^{h\bar{h}}X_{121}^{h\bar{h}}d^\ell_{31} \nonumber\\
    &&\ \ \ \ \ \ \ \ \ \ \ \ \ \ \ \ \ \ \ \ \ \ \ \ \ \ \ \ \ \ \ \ \ \ \ \
    +(C_{\gl,2}^{h\bar{h}})^2\Big[(X_{022}^{h\bar{h}}{}')^2d_{22}^\ell+X_{242}^{h\bar{h}}X_{220}^{h\bar{h}}d^\ell_{40}\Big]\Bigg\} \nonumber
    \\
    &&+\left| h^P_\ell \right| ^2 ({C}^{EE,\unlensed}_\ell+{C}^{BB,\unlensed}_\ell)\Bigg\{(X_{022}^{h^2})^2d_{22}^\ell+2C_{\gl,2}^{h^2} X_{132}^{h^2}X_{121}^{h^2}d^\ell_{31} \nonumber\\
    &&\ \ \ \ \ \ \ \ \ \ \ \ \ \ \ \ \ \ \ \ \ \ \ \ \ \ \ \ \ \ \ \ \ \ \ \
    +(C_{\gl,2}^{h^2})^2\Big[(X_{022}^{h^2}{}')^2d_{22}^\ell+X_{242}^{h^2}X_{220}^{h^2}d^\ell_{40}\Big]\Bigg\}\Bigg)\nonumber
    \label{eq:Delensed_plus_corr}
\ee
where $h_\ell$ is a filter applied to the temperature
\be
    h_\ell=\frac{C_\ell^{TT}}{C_{\ell}^{TT}+N_{\ell}^{TT}}\,,
    \label{eq:h_filter}
\ee
with $\bar{h}_\ell$ defined to conserve total temperature power
\be
    \bar{h}_\ell=\sqrt{1-h_\ell^2(1-e^{-\ell(\ell+1)C_{\gl}^{\rm obs}(0)/2})}-h_\ell e^{-\ell(\ell+1)C_\gl^{\rm obs}(0)/4}\,,
    \label{eq:hbar_filter}
\ee
while $h_\ell^P$ is the polarization filter
\be
    h_\ell^P=\frac{C_\ell^{EE}}{C_{\ell}^{EE}+N_{\ell}^{EE}}\,,
    \label{eq:hP_filter}
\ee
with $\bar{h}_\ell^P$ defined similarly to $\bar{h}_\ell$ to conserve total polarization power
\be
    \bar{h}_\ell^P=\sqrt{1-(h_\ell^P)^2(1-e^{-\ell(\ell+1)C_{\gl}^{\rm obs}(0)/2})}-h_\ell^P e^{-\ell(\ell+1)C_\gl^{\rm obs}(0)/4}\,,
    \label{eq:hbarP_filter}
\ee
and
\be
    X^{\bar{h}h}_{000}=e^{-\ell(\ell+1)[\sigma^2(\beta)]^{\bar{h}h}/4}\,,
\ee
\be
    X^{h^2}_{000}=e^{-\ell(\ell+1)[\sigma^2(\beta)]^{h^2}/4}\,,
\ee
\be
    X^{\bar{h}h}_{210}=\frac{1}{4}\sqrt{(\ell+2)(\ell+1)\ell(\ell-1)}e^{-(\ell(\ell+1)-2)[\sigma^2(\beta)]^{\bar{h}h}/4}\,,
\ee
\be
    X^{h^2}_{220}=\frac{1}{4}\sqrt{(\ell+2)(\ell+1)\ell(\ell-1)}e^{-(\ell(\ell+1)-2)[\sigma^2(\beta)]^{h^2}/4}\,,
\ee
\be
    [\sigma^2(\beta)]^{\bar{h}h}=(C_\gl(0)-C_\gl(\beta))-(C_\gl^{\rm cross}(0)-C_{\gl}^{\rm cross}(\beta))+\frac{1}{2}C_{\gl}^{\rm obs}(0)\,,
\ee
\be
    [\sigma^2(\beta)]^{h^2}=(C_\gl(0)-C_\gl(\beta))-2(C_\gl^{\rm cross}(0)-C_{\gl}^{\rm cross}(\beta))+(C_{\gl}^{\rm obs}(\beta)-C_{\gl}^{\rm obs}(\beta))\,,
\ee
\be
    C_{\gl,2}^{\bar{h}h}(\beta)=C_{\gl,2}(\beta)-C_{\gl,2}^{\rm cross}(\beta)\,,
\ee
\be
    C_{\gl,2}^{h^2}=C_{\gl,2}(\beta)-2C_{\gl,2}^{\rm cross}(\beta)+C_{\gl,2}^{\rm obs}(\beta)\,,
\ee
\be
    C^{\rm obs}_\gl(\beta)=\sum\limits_\ell\frac{2\ell+1}{4\pi}\ell(\ell+1)|g_\ell|^2C_\ell^{\phi\phi,{\rm obs}}d_{11}^\ell(\beta)\,,
\ee
\be
    C^{\rm obs}_{\gl,2}(\beta)=\sum\limits_\ell\frac{2\ell+1}{4\pi}\ell(\ell+1)|g_\ell|^2C_{\ell}^{\phi\phi,{\rm obs}}d_{-11}^\ell(\beta)\,,
\ee
\be
    C_{\gl}^{\rm cross}(\beta)=\sum\limits_\ell\frac{2\ell+1}{4\pi}\ell(\ell+1)g_\ell C_{\ell}^{\phi\phi}d_{11}^\ell(\beta)\,,
\ee
\be
    C_{\gl,2}^{\rm cross}(\beta)=\sum\limits_\ell\frac{2\ell+1}{4\pi}\ell(\ell+1)g_\ell C_\ell^{\phi\phi}d_{-11}^\ell(\beta)\,,
\ee
with 
\be
    C_\ell^{\phi\phi,{\rm obs}}=C_{\ell}^{\phi\phi}+N_{\ell}^{\phi\phi}\,,
\ee
and 
\be
    g_\ell=\frac{{C_\ell}^{\phi\phi}}{C_{\ell}^{\phi\phi,{\rm obs}}}\,.
    \label{eq:g_filter}
\ee

The delensed power spectra can be calculated from the correlation functions 
\begin{align}
    \label{eq:delensed_powspec}
    C^{TT,\delensed}_\ell & =2\pi\int_{-1}^{1}\xi^{TT,\delensed}(\beta)d^\ell_{00}(\beta) \, \dd\cos{\beta}\,, \nonumber \\
    C^{TE,\delensed}_\ell & = 2\pi\int_{-1}^{1}\xi^{TE,\delensed}(\beta)d^\ell_{20}(\beta) \, \dd\cos{\beta}\,, \nonumber \\
    C^{EE,\delensed}_\ell - C^{BB,\delensed}_\ell & = 2\pi\int_{-1}^{1}\xi^{-,\delensed}(\beta)d^\ell_{22}(\beta) \, \dd\cos{\beta}\,, \nonumber \\
    C^{EE,\delensed}_\ell + C^{BB,\delensed}_\ell & = 2\pi\int_{-1}^{1}\xi^{+,\delensed}(\beta)d^\ell_{22}(\beta) \, \dd\cos{\beta}\, , 
\end{align}
and similarly for the lensed spectra.

\section{Full-sky lensing reconstruction with estimator covariances}
\label{sec:lensing_reconstruction}

In this section we calculate the full-sky lensing-reconstruction noise estimators. The full covariance matrix for the lensing-reconstruction noise takes the form 
\be
    \boldsymbol{N}_\ell=
    \begin{bmatrix}
    N_{\ell}^{TT;TT} & N_{\ell}^{TT;TE} & N_{\ell}^{TT;EE} & N_{\ell}^{TT;TB} & N_{\ell}^{TT;EB} & N_{\ell}^{TT;BB} \\
    N_{\ell}^{TE;TT} & N_{\ell}^{TE;TE} & N_{\ell}^{TE;EE} & N_{\ell}^{TE;TB} & N_{\ell}^{TE;EB} & N_{\ell}^{TE;BB} \\
    N_{\ell}^{EE;TT} & N_{\ell}^{EE;TE} & N_{\ell}^{EE;EE} & N_{\ell}^{EE;TB} & N_{\ell}^{EE;EB} & N_{\ell}^{EE;BB}\\
    N_{\ell}^{TB;TT} & N_{\ell}^{TB;TE} & N_{\ell}^{TB;EE} & N_{\ell}^{TB;TB} & N_{\ell}^{TB;EB} & N_{\ell}^{TB;BB}\\
    N_{\ell}^{EB;TT} & N_{\ell}^{EB;TE} & N_{\ell}^{EB;EE} & N_{\ell}^{EB;TB} & N_{\ell}^{EB;EB} & N_{\ell}^{EB;BB}\\
    N_{\ell}^{BB;TT} & N_{\ell}^{BB;TE} & N_{\ell}^{BB;EE} & N_{\ell}^{BB;TB} & N_{\ell}^{BB;EB} & N_{\ell}^{BB;BB}
    \end{bmatrix}
\ee
where the minimum variance estimator is given by
\be
    N_\ell^{\rm mv}=\frac{1}{\sum_{\alpha\beta}(\boldsymbol{N}_\ell^{-1})^{\alpha\beta}}\,. 
    \label{eq:Nlmv}
\ee 
Before calculating each of these elements, it is useful to define $F^{s}_{\ell_1\ell_2\ell_3}$ as in Ref.~\citep{Smith:2010gu} where 
\begin{equation}\label{eq:fdef}
    \begin{split}
    F^{s}_{\ell_1\ell_2\ell_3} = & - \sqrt{\frac{(2\ell_1+1)(2\ell_2+1)(2\ell_3+1)\ell_3(\ell_3+1)}{16\pi}} \\ 
    & \times\left[ \sqrt{(\ell_2-s)(\ell_2+s+1)}
    \begin{pmatrix} 
    \ell_1 & \ell_2 & \ell_3 \\ 
    -s & s+1 & -1 \\ 
    \end{pmatrix}+\sqrt{(\ell_2+s)(\ell_2-s+1)}
    \begin{pmatrix} 
    \ell_1 & \ell_2 & \ell_3 \\ 
    -s & s-1 & +1 \\ 
    \end{pmatrix} \right]\,. \\ 
    \end{split} 
\end{equation}
Here, we used the identities
\begin{equation}\label{eq:fdef2}
    \int_{-1}^{1}\dd(\cos{\theta})\ d^{\ell_1}_{s_1s'_1}\!(\theta)d^{\ell_2}_{s_2s'_2}\!(\theta)d^{\ell_3}_{s_3s'_3}\!(\theta)=2
    \begin{pmatrix}
    \ell_1 & \ell_2 & \ell_3 \\ 
    s_1 & s_2 & s_3 \\ 
    \end{pmatrix} 
    \begin{pmatrix} 
    \ell_1 & \ell_2 & \ell_3 \\ 
    s'_1 & s'_2 & s'_3 \\ 
    \end{pmatrix}\,,
\end{equation}
and
\begin{equation}
    \begin{split}
    &\begin{pmatrix} 
    \ell_2 & \ell_3 & \ell_1\\ 
    s_2 & s_3 & s_1 \\ 
    \end{pmatrix}\!=\!
    \begin{pmatrix} 
    \ell_1 & \ell_2 & \ell_3\\ 
    s_1 & s_2 & s_3 \\ 
    \end{pmatrix}\,,\\ 
    &\begin{pmatrix} 
    \ell_2 & \ell_1 & \ell_3\\ 
    s_2 & s_1 & s_3 \\ 
    \end{pmatrix}\!=\!
    (-1)^{\ell_1+\ell_2+\ell_3}\!\!\begin{pmatrix} 
    \ell_1 & \ell_2 & \ell_3\\ 
    s_1 & s_2 & s_3 \\ 
    \end{pmatrix}\,,\\
    &\begin{pmatrix} 
    \ell_1 & \ell_2 & \ell_3\\ 
    -s_1 & -s_2 & -s_3 \\ 
    \end{pmatrix}\!=\!
    (-1)^{\ell_1+\ell_2+\ell_3}\!\!\begin{pmatrix} 
    \ell_1 & \ell_2 & \ell_3\\ 
    s_1 & s_2 & s_3 \\ 
    \end{pmatrix}\,.
    \end{split}
\end{equation}
Also note
\begin{equation}
    F^{s}_{\ell_2\ell\ell_1}=F^{s}_{\ell_1\ell_2\ell} \ \ \ \ \mathrm{and} \ \ \ \  d^\ell_{m,m'}=d^\ell_{-m',-m}=(-1)^{m'-m}d^\ell_{m',m}\ .
\end{equation}
The components of the lensing-reconstruction noise covariance satisfy
\be\label{eq:offd}
    N_{\ell}^{XY;WZ} =&& \frac{N^{XY;XY*}_\ell N^{WZ;WZ}_\ell}{
    \ell(\ell+1)(2\ell+1)}  \\
    &&\times \sum\limits_{\ell_1\ell_2}\left\{g_{\ell_1\ell\ell_2}^{XY*}\left[C_{\ell_1}^{XW,\observed}C_{\ell_2}^{YZ,\observed}g_{\ell_1\ell\ell_2}^{WZ}+(-1)^{\ell_1+\ell_2+\ell}C_{\ell_1}^{XZ,\observed}C_{\ell_2}^{YW,\observed}g_{\ell_2\ell\ell_1}^{WZ}\right]\right\} \, , \nonumber
\ee
where we use the definition \citep{Okamoto:2003zw}
\be
    g_{\ell_1\ell\ell_2}^{{XY}}=\frac{C_{\ell_2}^{XX,\observed}C_{\ell_1}^{YY,\observed}f_{\ell_1\ell\ell_2}^{XY*}-(-1)^{\ell_1+\ell_2+\ell}C_{\ell_1}^{XY,\observed}C_{\ell_2}^{XY,\observed}f_{\ell_2\ell\ell_1}^{XY*}}{C^{XX,\observed}_{\ell_1}C^{XX,\observed}_{\ell_2}C^{YY,\observed}_{\ell_1}C^{YY,\observed}_{\ell_2}-(C^{XY,\observed}_{\ell_1}C^{XY,\observed}_{\ell_2})^2}\,.
\ee
Note that
\be
    g_{\ell_1\ell\ell_2}^{{XY}}=\frac{f_{\ell_1\ell\ell_2}^{{XY}}}{2 C_{\ell_1}^{{XX,\observed}}C_{\ell_2}^{{YY,\observed}}}\,,
\ee
for $X=Y$, and
\be
    g_{\ell_1\ell\ell_2}^{{XY}}=\frac{f_{\ell_1\ell\ell_2}^{{XY}}}{C_{\ell_1}^{{XX,\observed}}C_{\ell_2}^{{YY,\observed}}}\,,
\ee
for $XY\in\{EB,TB\}$. Finally, in what follows, we use the approximation 
\be
    g_{\ell_1\ell\ell_2}^{{TE}}\simeq\frac{f_{\ell_1\ell\ell_2}^{{TE}}}{C_{\ell_1}^{{TT,\observed}}C_{\ell_2}^{{EE,\observed}}}\,,
\ee
as done in Ref.~\citep{Okamoto:2003zw}. The quadratic estimator reconstruction noise for the diagonal covariances hence has the form 
\begin{equation}
    \label{eq:normalization}
    N^{XY;XY}_{\ell} = (2\ell+1)\left[\sum\limits_{\ell_1\ell_2}\frac{|f^{XY}_{\ell_1\ell\ell_2}|^2}{(1+\delta_{XY}){C}_{\ell_1}^{XX, {\rm obs}}{C}_{\ell_2}^{YY,\observed}}\right]^{-1}
\end{equation} 
for $XY \in \{ TT, \, TE, \, EE, \, EB, \, TB, \, BB \}$. 
Also note ${C}^{aa,\rm obs}_\ell=C^{aa}_\ell+N^{aa}_\ell$ where $C^{aa}_\ell$ is the theory prediction for the spectra, either consistently all lensed or all delensed in the expressions below, with one exception for the $BB$ estimator.
Using these equalities, we calculate the elements of reconstruction noise covariance below.

\begin{table}[h]
{\renewcommand{\arraystretch}{1.25} 
\begin{center}
\begin{tabular}{  | c | c | c | }
\hline
 $\alpha$ & $f^{\alpha}_{\ell_1\!\ell\ell_2}$   & Parity\\ 
 \hline 
 $TT$ & $C_{\ell_1}^{{TT}}F^{0}_{\ell_2\ell\ell_1}+C_{\ell_2}^{{TT}}F^{0}_{\ell_1\ell\ell_2}$  & -\\  
 \hline 
 $TE$ & $C_{\ell_1}^{{TE}}F^{2}_{\ell_2\ell\ell_1}+C_{\ell_2}^{{TE}}F^{0}_{\ell_1\ell\ell_2}$  & even\\  
 \hline  
 $EE$ & $C_{\ell_1}^{{EE}}F^{2}_{\ell_2\ell\ell_1}+C_{\ell_2}^{{EE}}F^{2}_{\ell_1\ell\ell_2}$  & even\\  
 \hline  
 $TB$ & $iC_{\ell_1}^{{TE}}F^{2}_{\ell_2\ell\ell_1}$  & odd\\  
 \hline  
 $EB$ & $i\left(C_{\ell_1}^{{EE}}F^{2}_{\ell_2\ell\ell_1}-C_{\ell_2}^{{BB}}F^{2}_{\ell_1\ell\ell_2}\right)$ & odd \\  
 \hline  
 $BB$ & $C_{\ell_1}^{{BB}}F^{2}_{\ell_2\ell\ell_1}+C_{\ell_2}^{{BB}}F^{0}_{\ell_1\ell\ell_2}$ & even\\
 \hline
\end{tabular}
\end{center}
}
\caption{The optimal filters for the full-sky CMB lensing quadratic estimator from \citep{Okamoto:2003zw}.}
    \label{tab:optimal_filters}
\end{table}

\subsection{Expressions for \texorpdfstring{$N^{TT;TT}_\ell$}{Nl{TT;TT}}}
First, note that
\begin{equation}
    C_{\ell_1}^{{TT}}F^{0}_{\ell_2\ell\ell_1}+C_{\ell_2}^{{TT}}F^{0}_{\ell_1\ell\ell_2}=C_{\ell_1}^{{TT}}F^{0}_{\ell_1\ell_2\ell}+C_{\ell_2}^{{TT}}F^{0}_{\ell_2\ell_1\ell}\ , 
\end{equation}
and
\begin{equation}
    \begin{split}
    \left[C_{\ell_1}^{{TT}}F^{0}_{{\ell_2\ell_1}\ell}+C_{\ell_2}^{{TT}}F^{0}_{{\ell_1\ell_2}\ell}\right]^2 = & \  (C_{\ell_1}^{{TT}}F_{{\ell_2\ell_1}\ell}^0)^2+2C_{\ell_1}^{{TT}}C_{\ell_2}^{{TT}}F_{{\ell_2\ell_1}\ell}^0F_{{\ell_1\ell_2}\ell}^0+(C_{\ell_2}^{{TT}}F_{{\ell_1\ell_2}\ell}^0)^2\,,
    \end{split}
\end{equation}
such that
\be
    \label{eq:TTTTnoise}
    |f_{\ell_1\ell\ell_2}^{TT}|^2=&&\Big[C_{\ell_1}^{{TT}}F^{0}_{{\ell_2\ell_1}\ell}+C_{\ell_2}^{{TT}}F^{0}_{{\ell_1\ell_2}\ell}\Big]^2 \\  \ =&&\frac{(2\ell_1+1)(2\ell_2+1)\ell(\ell+1)(2\ell+1)}{32\pi}\nonumber\\
    && \times \left(\ 2\,\ell_1(\ell_1+1)(C_{\ell_1}^{{TT}})^2\left[\int\!\dd\cos{\theta}\left( d_{00}^{\ell_2}d_{11}^{\ell_1}d_{11}^{\ell} + d_{00}^{\ell_1}d_{1-1}^{\ell_2}d_{1-1}^{\ell}\right)\right]\right.\nonumber\\ 
    &&\ \ \ \, +2\,\ell_2(\ell_2+1)(C_{\ell_2}^{{TT}})^2\left[\int\!\dd\cos{\theta}\left( d_{00}^{\ell_1}d_{11}^{\ell_2}d_{11}^{\ell} + d_{00}^{\ell_2}d_{1-1}^{\ell_1}d_{1-1}^{\ell}\right)\right]\nonumber\\
     &&+ \left.2\,\sqrt{\ell_1(\ell_1+1)}\sqrt{\ell_2(\ell_2+1)}C_{\ell_1}^{{TT}}C_{\ell_2}^{{TT}}\left[\int\!\dd\cos{\theta}\left(d_{01}^{\ell_1}d_{01}^{\ell_2}d_{11}^{\ell}-d_{01}^{\ell_1}d_{01}^{\ell_2}d_{1-1}^{\ell}\right)\right]\right)\,.\nonumber
\ee
The covariance element $N^{TT;TT}_\ell$ can then be found by using Eq.~\eqref{eq:TTTTnoise} and Eq.~\eqref{eq:normalization}. 

\subsection{Expressions for \texorpdfstring{$N^{EE;EE}_\ell$}{Nl{EE;EE}}}
Following on similar lines as before, $f_{\ell_1\ell\ell_2}^{{EE}}$ satisfies
\be
    |f_{\ell_1\ell\ell_2}^{{EE}}|^2 = && \left[C_{\ell_1}^{{{EE}}}F^{2}_{\ell_2\ell_1\ell}+C_{\ell_2}^{{EE}}F^{2}_{\ell_1\ell_2\ell}\right]^2 \nonumber \\
    = && \left[C_{\ell_1}^{{{EE}}}\left(\frac{F^{2}_{\ell_2\ell_1\ell}+F^{-2}_{\ell_2\ell_1\ell}}{2}\right)+C_{\ell_2}^{{EE}}\left(\frac{F^{2}_{\ell_1\ell_2\ell}+F^{-2}_{\ell_1\ell_2\ell}}{2}\right)\right]^2 \nonumber \\ 
    = && \  \ \ \ \ \frac{1}{4}\Big[(C_{\ell_1}^{{{EE}}}F^{2}_{\ell_2\ell_1\ell})^2+(C_{\ell_1}^{{{EE}}}F^{-2}_{\ell_2\ell_1\ell})^2+2(C_{\ell_1}^{{{EE}}}C_{\ell_1}^{{{EE}}}F^{2}_{\ell_2\ell_1\ell}F^{-2}_{\ell_2\ell_1\ell}) \nonumber\\ 
    && \ \ \ \ \ \ \ \ \ + (C_{\ell_2}^{{{EE}}}F^{2}_{\ell_1\ell_2\ell})^2+(C_{\ell_2}^{{{EE}}}F^{-2}_{\ell_1\ell_2\ell})^2+2(C_{\ell_2}^{{{EE}}}C_{\ell_2}^{{{EE}}}F^{2}_{\ell_1\ell_2\ell}F^{-2}_{\ell_1\ell_2\ell})\nonumber \\ 
    && \ \ \ \ \ \ \ \ \ 
    + 2\,(C_{\ell_1}^{{{EE}}}C_{\ell_2}^{{{EE}}}F^{2}_{\ell_2\ell_1\ell}F^{2}_{\ell_1\ell_2\ell})+2(C_{\ell_1}^{{{EE}}}C_{\ell_2}^{{{EE}}}F^{2}_{\ell_2\ell_1\ell}F^{-2}_{\ell_1\ell_2\ell})\nonumber \\ 
    && \ \ \ \ \ \ \ \ \ + 2\,(C_{\ell_1}^{{{EE}}}C_{\ell_2}^{{{EE}}}F^{-2}_{\ell_2\ell_1\ell}F^{2}_{\ell_1\ell_2\ell}) + 2(C_{\ell_1}^{{{EE}}}C_{\ell_2}^{{{EE}}}F^{-2}_{\ell_2\ell_1\ell}F^{-2}_{\ell_1\ell_2\ell})\Big]\,.
    \label{eq:BB_filter}
\ee
Applying this expression, the variance of the $EE$ lensing estimator takes the form
\begin{equation}
    \begin{split}
    N^{{EE;EE}}_\ell=\Big(\frac{\pi\ell(\ell+1)}{16}\int_{-1}^{+1}\dd\cos{\theta}\ &\left[\ 
    \ 4\left( \zeta_{22}^{{EE}}\zeta_{33}^{{EE}}+\zeta_{22}^{{EE}}\zeta_{11}^{{EE}}+2\zeta_{2-2}^{{EE}}\zeta_{3-1}^{{EE}}\right)d_{11}^\ell\right. \\ 
    &\left.\ + 4\left(\zeta_{2-2}^{{EE}}\zeta_{3-3}^{{EE}}+\zeta_{2-2}^{{EE}}\zeta_{1-1}^{{EE}}+2\zeta_{22}^{{EE}}\zeta_{31}^{{EE}}\right)d_{1-1}^\ell\right.\\
    &\left.\ -4\left(\zeta_{3-2}^{{EE}}\zeta_{3-2}^{{EE}}+\zeta_{2-1}^{{EE}}\zeta_{2-1}^{{EE}}+2\zeta_{32}^{{EE}}\zeta_{21}^{{EE}}\right)d_{11}^\ell\right.\\
    &\left.\ +4\left(\zeta_{32}^{{EE}}\zeta_{32}^{{EE}}+\zeta_{21}^{{EE}}\zeta_{21}^{{EE}}-2\zeta_{3-2}^{{EE}}\zeta_{2-1}^{{EE}}\right)d_{1-1}^\ell\right]\Big)^{-1}\,,
    \label{eq:NlEEEE}
    \end{split}
\end{equation}
where 
\be
    \zeta_{3\pm1}^{\alpha\beta} && =\sum\limits_\ell\frac{2\ell+1}{4\pi}\frac{{\left({C}_\ell^{\alpha\beta}\right)^2}}{C_\ell^{\alpha\beta,\observed}}\sqrt{(\ell+2)(\ell-1)(\ell-2)(\ell+3)}d_{3\pm1}^{\ell} \ee\be
    \zeta_{3\pm3}^{\alpha\beta} && =\sum\limits_\ell\frac{2\ell+1}{4\pi}\frac{{\left({C}_\ell^{\alpha\beta}\right)^2}}{C_\ell^{\alpha\beta,\observed}}(\ell-2)(\ell+3)d_{3\pm3}^{\ell} \ee\be 
    \zeta_{1\pm1}^{\alpha\beta} && =\sum\limits_\ell\frac{2\ell+1}{4\pi}\frac{{\left({C}_\ell^{\alpha\beta}\right)^2}}{C_\ell^{\alpha\beta,\observed}}(\ell+2)(\ell-1)d_{1\pm1}^{\ell} \ee\be 
    \zeta_{2\pm2}^{\alpha\beta} && = \sum\limits_\ell\frac{2\ell+1}{4\pi}\frac{1}{C_\ell^{\alpha\beta,\observed}}d_{2\pm2}^{\ell}\ee\be
    \zeta_{3\pm2}^{\alpha\beta} && = \sum\limits_\ell\frac{2\ell+1}{4\pi}\frac{{C}_\ell^{\alpha\beta}}{C_\ell^{\alpha\beta,\observed}}\sqrt{(\ell_2-2)(\ell_2+3)}d_{3\pm2}^{\ell}\ee\be
    \zeta_{2\pm1}^{\alpha\beta} && = \sum\limits_\ell\frac{2\ell+1}{4\pi}\frac{{C}_\ell^{\alpha\beta}}{C_\ell^{\alpha\beta,\observed}}\sqrt{(\ell_2+2)(\ell_2-1)}d_{2\pm1}^{\ell} \ .
\ee

\subsection{Expressions for \texorpdfstring{$N^{BB;BB}_\ell$}{Nl{BB;BB}}}
The expressions for $N^{BB;BB}_\ell$, are the same as for $N^{EE;EE}_\ell$, but with $C_\ell^{EE}$ replaced by $C_\ell^{BB,\unlensed}$ in Eq.~\eqref{eq:BB_filter}.
Notice that the variance on the $BB$ estimator is infinite when $C_\ell^{BB,\unlensed}=0$.

\subsection{Expressions for \texorpdfstring{$N^{EB;EB}_\ell$}{Nl{EB;EB}}}
 The function $f_{\ell_1\ell\ell_2}^{{EB}}$ satisfies
\be
    |f_{\ell_1\ell\ell_2}^{{EB}}|^2 = && \left[C_{\ell_1}^{{{EE}}}F^{2}_{{\ell_2\ell_1}\ell}-C_{\ell_2}^{{BB}}F^{2}_{{\ell_1\ell_2}\ell}\right]^2\ \ ,\ \\
    = && \left[C_{\ell_1}^{{{EE}}}\left(\frac{F^{2}_{{\ell_2\ell_1}\ell}-F^{-2}_{{\ell_2\ell_1}\ell}}{2}\right)-C_{\ell_2}^{{BB}}\left(\frac{F^{2}_{{\ell_1\ell_2}\ell}-F^{-2}_{{\ell_1\ell_2}\ell}}{2}\right)\right]^2\\ 
    = && \  \ \ \ \ \frac{1}{4}\left[(C_{\ell_1}^{{{EE}}}F^{2}_{{\ell_2\ell_1}\ell})^2+(C_{\ell_1}^{{{EE}}}F^{-2}_{{\ell_2\ell_1}\ell})^2-2(C_{\ell_1}^{{{EE}}}C_{\ell_1}^{{{EE}}}F^{2}_{{\ell_2\ell_1}\ell}F^{-2}_{{\ell_2\ell_1}\ell})\right. \\ 
    && \  \ \ \ \ \ + (C_{\ell_2}^{{{BB}}}F^{2}_{{\ell_1\ell_2}\ell})^2+(C_{\ell_2}^{{{BB}}}F^{-2}_{{\ell_1\ell_2}\ell})^2-2(C_{\ell_2}^{{{BB}}}C_{\ell_2}^{{{BB}}}F^{2}_{{\ell_1\ell_2}\ell}F^{-2}_{{\ell_1\ell_2}\ell}) \nonumber\\ 
    && \ \ \ \ \ \ -C_{\ell_1}^{{{EE}}}C_{\ell_2}^{{{BB}}}(F^{2}_{{\ell_2\ell_1}\ell}F^{2}_{{\ell_1\ell_2}\ell}+F^{2}_{{\ell_1\ell_2}\ell}F^{2}_{{\ell_2\ell_1}\ell})+C_{\ell_1}^{{{EE}}}C_{\ell_2}^{{{BB}}}(F^{2}_{{\ell_2\ell_1}\ell}F^{-2}_{{\ell_1\ell_2}\ell}+F^{-2}_{{\ell_1\ell_2}\ell}F^{2}_{{\ell_2\ell_1}\ell}) \nonumber\\ 
    && \ \ \ \ \ \  + \left.C_{\ell_1}^{{{EE}}}C_{\ell_2}^{{{BB}}}(F^{-2}_{{\ell_2\ell_1}\ell}F^{2}_{{\ell_1\ell_2}\ell}+F^{2}_{{\ell_1\ell_2}\ell}F^{-2}_{{\ell_2\ell_1}\ell}) - C_{\ell_1}^{{{EE}}}C_{\ell_2}^{{{BB}}}(F^{-2}_{{\ell_2\ell_1}\ell}F^{-2}_{{\ell_1\ell_2}\ell}+F^{-2}_{{\ell_1\ell_2}\ell}F^{-2}_{{\ell_2\ell_1}\ell})\right]\ .\nonumber
\ee
Note that in the case $C_{\ell}^{{BB}}=0$ only the first line above is non-zero which gives the expression found in \citep{Smith:2010gu}. Using the equalities in Eq.~\eqref{eq:zetas} we get 
\be
    N^{{EB}}_\ell=&&\Bigg(\frac{\pi\ell(\ell+1)}{4}\int_{-1}^{+1}\dd\cos{\theta}\\
    &&\Big[ 
    \left( \zeta^{BB}_{22}\zeta_{33}^{{EE}}+\zeta^{BB}_{22}\zeta_{11}^{{EE}}-2\zeta^{BB}_{2-2}\zeta_{3-1}^{{EE}}+ \zeta_{22}^{{EE}}\zeta_{33}^{{BB}}+\zeta_{22}^{{EE}}\zeta_{11}^{{BB}}-2\zeta_{2-2}^{{EE}}\zeta_{3-1}^{{BB}}\right)d_{11}^\ell \nonumber\\ 
    && \! -\left( \zeta_{2-2}^{{EE}}\zeta_{3-3}^{{BB}}+\zeta_{2-2}^{{EE}}\zeta_{1-1}^{{BB}}-2{\zeta_{22}^{{EE}}}\zeta_{31}^{{BB}}+\zeta^{BB}_{2-2}\zeta_{3-3}^{{EE}}+\zeta^{BB}_{2-2}\zeta_{1-1}^{{EE}}-2\zeta^{BB}_{22}\zeta_{31}^{{EE}}\right)d_{1-1}^\ell \nonumber\\
    &&\left.\ \!+\left(\zeta_{3-2}^{{EE}/{{BB}}}\zeta_{3-2}^{{BB}/{EE}}+\zeta_{2-1}^{{EE}/{{BB}}}\zeta_{2-1}^{{BB}/{EE}}-\zeta_{32}^{{EE}/{{BB}}}\zeta_{21}^{{BB}/{EE}}-\zeta_{32}^{{BB}/{EE}}\zeta_{21}^{{EE}/{{BB}}}\right)d_{11}^\ell\right.\nonumber\\
    && +\left(\zeta_{32}^{{EE}/{{BB}}}\zeta_{32}^{{BB}/{EE}}+\zeta_{21}^{{EE}/{{BB}}}\zeta_{21}^{{BB}/{EE}}+\zeta_{3-2}^{{EE}/{{BB}}}\zeta_{2-1}^{{BB}/{EE}}+\zeta_{3-2}^{{BB}}\zeta_{2-1}^{{EE}/{{BB}}}\right)d_{1-1}^\ell{\Big]}\Bigg)^{-1}\,,\nonumber
\ee
where
\be
    \label{eq:zetas}
    \zeta_{3\pm2}^{\alpha\alpha/\beta\beta} && = \sum\limits_\ell\frac{2\ell+1}{4\pi}\frac{{C}_\ell^{\alpha\alpha}}{C_\ell^{\beta\beta,\observed}}\sqrt{(\ell-2)(\ell+3)}d_{3\pm2}^{\ell}
\ee
\be
    \zeta_{2\pm1}^{\alpha\alpha/\beta\beta} && = \sum\limits_\ell\frac{2\ell+1}{4\pi}\frac{{C}_\ell^{\alpha\alpha}}{C_\ell^{\beta\beta,\observed}}\sqrt{(\ell+2)(\ell-1)}d_{2\pm1}^{\ell} \ .
\ee

\subsection{Expressions for \texorpdfstring{$N^{TE;TE}_\ell$}{Nl{TE;TE}}}
 The term $f^{{TE}}_{\ell_1\ell\ell_2}$ satisfies 
\be
    |f^{{TE}}_{\ell_1\ell\ell_2}|^2  && = \left[C_{\ell_1}^{{TE}}F^{2}_{\ell_2\ell\ell_1}+C_{\ell_2}^{{TE}}F^{0}_{\ell_1\ell\ell_2}\right]^2\\
    && = \left[C_{\ell_1}^{{TE}}\left(\frac{F^2_{\ell_1\ell_2\ell}+F^{-2}_{\ell_1\ell_2\ell}}{2}\right)+C_{\ell_2}^{{TE}}F^{0}_{\ell_2\ell_1\ell}\right]^2\\ 
    && = \frac{1}{4}\left[(C_{\ell_1}^{{{TE}}}F^{2}_{{\ell_2\ell_1}\ell})^2+(C_{\ell_1}^{{{TE}}}F^{-2}_{{\ell_2\ell_1}\ell})^2+2(C_{\ell_1}^{{{TE}}}C_{\ell_1}^{{{TE}}}F^{2}_{{\ell_2\ell_1}\ell}F^{-2}_{{\ell_2\ell_1}\ell})\right.\nonumber\\ 
    &&  + 2C_{\ell_1}^{{{TE}}}C_{\ell_2}^{{{TE}}}(F^{2}_{{\ell_2\ell_1}\ell}F^{0}_{{\ell_1\ell_2}\ell}+F^{0}_{{\ell_1\ell_2}\ell}F^{2}_{{\ell_2\ell_1}\ell})\nonumber\\
    &&\left.+2C_{\ell_1}^{{{TE}}}C_{\ell_2}^{{{TE}}}(F^{-2}_{{\ell_2\ell_1}\ell}F^{0}_{{\ell_1\ell_2}\ell}+F^{0}_{{\ell_1\ell_2}\ell}F^{-2}_{{\ell_2\ell_1}\ell})+4(C_{\ell_2}^{{{TE}}}F^{0}_{{\ell_1\ell_2}\ell})^2\right]\,. 
\ee
The expression for $N_\ell^{{TE;TE}}$ becomes 
\begin{equation}
    \begin{split}
    N^{{TE;TE}}_\ell=\Big(\frac{\pi\ell(\ell+1)}{8}\int_{-1}^{+1}\dd\cos{\theta}\ &\left [
     \ +2 \left(\zeta_{22}^{{TE}}\zeta_{11}^{{TE}/{TT}}+\zeta_{22}^{{TE}}\zeta_{33}^{{TE}}+2\zeta_{2-2}^{{TE}}\zeta_{3-1}^{{TE}}\right)d_{11}^\ell\right.\\
     & \ \ +2 \left(\zeta_{2-2}^{{TE}}\zeta_{1-1}^{{TE}/{TT}}+\zeta_{2-2}^{{TE}}\zeta_{3-3}^{{TE}}+2\zeta_{22}^{{TE}}\zeta_{31}^{{TE}}\right)d_{1-1}^\ell\\
     &  \ \ +8\left(\zeta_{21}^{{TE}}\zeta_{01}^{{TE}}-\zeta_{2-1}^{{TE}}\zeta_{30}^{{TE}}\right)d_{11}^\ell \\ 
     &  \ \ +8\left(\zeta_{2-1}^{{TE}}\zeta_{01}^{{TE}}-\zeta_{21}^{{TE}}\zeta_{30}^{{TE}}\right)d_{1-1}^\ell \\ 
     &  \ \ \left.+\,8\left(\zeta_{00}^{{TE}}\zeta_{11}^{{TE}/{TT}}d_{11}^\ell+\zeta_{00}^{{TE}}\zeta_{1-1}^{{TE}/{TT}}d_{1-1}^\ell\right)\right]\Big)^{-1}\,,
    \end{split}
\end{equation}
where for convenience we defined
\be
    \label{eq:zetas2}
    \zeta_{3\pm1}^{{TE}} && =\sum\limits_{\ell}\frac{2\ell+1}{4\pi}\sqrt{(\ell-1)(\ell+2)(\ell-2)(\ell+3)}\frac{{({C}_\ell^{{TE}})^2}}{C_\ell^{{TT,\observed}}}d_{3\pm1}^{\ell} 
\ee
\be 
    \zeta_{1\pm1}^{{TE}/\alpha}  && =\sum\limits_{\ell}\frac{2\ell+1}{4\pi}(\ell+2)(\ell-1)\frac{{({C}_\ell^{{TE}})^2}}{C_\ell^{\alpha,\observed}}d_{1\pm1}^{\ell} 
\ee
\be
    \zeta_{2\pm2}^{{TE}}  && =\sum\limits_{\ell}\frac{2\ell+1}{4\pi}\frac{1}{C_\ell^{{EE,\observed}}}d_{2\pm2}^{\ell} 
\ee
\be 
    \zeta_{00}^{{TE}}  && =\sum\limits_{\ell}\frac{2\ell+1}{4\pi}\frac{1}{C_\ell^{{TT,\observed}}}d_{00}^{\ell} 
\ee
\be 
    \zeta_{3\pm3}^{{TE}}  && =\sum\limits_{\ell}\frac{2\ell+1}{4\pi}(\ell-2)(\ell+3)\frac{{{({C}_\ell^{{TE}})^2}}}{C_\ell^{{TT,\observed}}}d_{3\pm3}^{\ell} 
\ee
\be
    \zeta_{\pm10}^{{TE}}  && =\sum\limits_{\ell}\frac{2\ell+1}{4\pi}\sqrt{(\ell+2)(\ell-1)}\frac{{C}_\ell^{{TE}}}{C_\ell^{{TT,\observed}}}d_{\pm10}^{\ell} 
\ee
\be 
    \zeta_{2\pm1}^{{TE},}  && =\sum\limits_{\ell}\frac{2\ell+1}{4\pi}\sqrt{\ell(\ell+1)}\frac{{C}_\ell^{{TE}}}{C_\ell^{{EE},\observed}}d_{2\pm1}^{\ell} 
\ee
\be 
    \zeta_{30}^{{TE}}  && = \sum\limits_\ell\frac{2\ell+1}{4\pi}C_{\ell}^{{TE}}\sqrt{(\ell-2)(\ell+3)}\frac{{C}_\ell^{{TE}}}{C_\ell^{{TT,\observed}}}d_{30}^{\ell} 
\ee

\subsection{Expressions for \texorpdfstring{$N^{TB;TB}_\ell$}{Nl{TB;TB}}}
 We begin by defining
\hspace*{-2cm}
\begin{equation}
    g^{{TB}}_{\ell_1\ell\ell_2}f^{{TB}}_{\ell_1\ell\ell_2}=\frac{1}{4}\frac{({C}_{\ell_1}^{{TE}})^2}{C_{\ell_1}^{{TT,\observed}}C_{\ell_2}^{{BB,\observed}}}\left[\left(F^{2}_{{\ell_2\ell_1}\ell}-F^{-2}_{{\ell_2\ell_1}\ell}\right)\left(F^{2}_{{\ell_2\ell_1}\ell}-F^{-2}_{{\ell_2\ell_1}\ell}\right)\right]\,,
\end{equation}
and
\be
    \zeta^{{TB}}_{2\pm2}= && \sum\limits_{\ell}\frac{2\ell+1}{4\pi}\frac{1}{C_\ell^{{BB,\observed}}}d_{2\pm2}^\ell\,, 
\ee
\be
    \zeta^{{TB}}_{3\pm1}= && \sum\limits_{\ell}\frac{2\ell+1}{4\pi}\sqrt{(\ell+2)(\ell-2)(\ell-1)(\ell+3)}\frac{({C}_\ell^{{TE}})^2}{C_\ell^{{TT,\observed}}}d_{3\pm1}^\ell \,,
\ee
\be
    \zeta^{{TB}}_{3\pm3}= && \sum\limits_{\ell}\frac{2\ell+1}{4\pi}(\ell-2)(\ell+3)\frac{{({C}_\ell^{{TE}})^2}}{C_\ell^{{TT,\observed}}}d_{3\pm3}^\ell \,,
\ee
\be
    \zeta^{{TB}}_{1\pm1}= && \sum\limits_{\ell}\frac{2\ell+1}{4\pi}(\ell-1)(\ell+2)\frac{{({C}_\ell^{{TE}})^2}}{C_\ell^{{TT,\observed}}}d_{1\pm1}^\ell\,,
\ee
for convenience. The expression for $N_\ell^{{TB;TB}}$ becomes
\begin{equation}
    \begin{split}
    N^{{TB;TB}}_\ell=\Big(\frac{\pi\ell(\ell+1)}{8}\int_{-1}^{+1}\dd\cos{\theta}\ &\left [\right. + 2\left( \zeta_{22}^{{TB}} \zeta_{11}^{{TB}}+ \zeta_{22}^{{TB}} \zeta_{33}^{{TB}} -  2\zeta_{2-2}^{{TB}} \zeta_{3-1}^{{TB}}\right)d_{11}^\ell\\
    & \ \  - 2\left( \zeta_{2-2}^{{TB}} \zeta_{1-1}^{{TB}}+ \zeta_{2-2}^{{TB}} \zeta_{3-3}^{{TB}} -  2\zeta_{22}^{{TB}} \zeta_{31}^{{TB}}\right)d_{1-1}^\ell\left.\right]\Big)^{-1}\,.
    \end{split}
\end{equation}

\subsection{Expressions for \texorpdfstring{$N^{TE;EE}_\ell$}{Nl{TE;EE}}}
 We use the equality
\be
    &&g^{{TE}*}_{\ell_1\ell\ell_2}[{C_{\ell_1}^{{TE,\observed}}C_{\ell_2}^{{EE,\observed}}}\left(g^{{EE}}_{\ell_1\ell\ell_2}+g^{{EE}}_{\ell_2\ell\ell_1}\right)]\\
    &&= \frac{C_{\ell_1}^{TE,\observed}}{8C_{\ell_1}^{{TT,\observed}}C_{\ell_2}^{{EE,\observed}}C_{\ell_1}^{EE,\observed}}\left({C}_{\ell_1}^{{EE}}\left[2\,{C}_{\ell_1}^{{TE}}{\left(F_{\ell_2\ell_1\ell}^{2}+F_{\ell_2\ell_1\ell}^{-2}\right)^2}+4{C}_{\ell_2}^{{TE}}{\left(F_{\ell_1\ell_2\ell}^{0}F_{\ell_2\ell_1\ell}^{2}+F_{\ell_1\ell_2\ell}^{0}F_{\ell_2\ell_1\ell}^{-2}\right)}\right]\right) 
    \nonumber\\ 
    && 
    \ \ \ \ \ \ \ \ \ \ \ \ \ \ \ \ \ \
    \ \ \ \ \ \ \ \ \ \ \ \ \ \ \ \ \ \ +\Big({C}_{\ell_2}^{{EE}}\Big[2\,{C}_{\ell_1}^{{TE}}{\Big(
    F_{\ell_2\ell_1\ell}^{2}F_{\ell_1\ell_2\ell}^{2}+F_{\ell_2\ell_1\ell}^{2}F_{\ell_1\ell_2\ell}^{-2}+F_{\ell_2\ell_1\ell}^{-2}F_{\ell_1\ell_2\ell}^{2}+F_{\ell_2\ell_1\ell}^{-2}F_{\ell_1\ell_2\ell}^{-2}\Big)}
    \nonumber\\ 
    && 
    \ \ \ \ \ \ \ \ \ \ \ \ \ \ \ \ \ \
    \ \ \ \ \ \ \ \ \ \ \ \ \ \ \ \ \ \ 
    \ \ \ \ \ \ \ \ \ \ \ 
    \left.+4\,{C}_{\ell_2}^{{TE}}{\left(F_{\ell_1\ell_2\ell}^{0}F_{\ell_1\ell_2\ell}^{2}+F_{\ell_1\ell_2\ell}^{0}F_{\ell_1\ell_2\ell}^{-2}\right)}\Big]\right) \nonumber
\ee
to get
\be
    N^{TE;EE}_\ell && = \!\!=\!\frac{N^{TE;TE*}_\ell N^{EE;EE}_\ell}{4\ell(\ell+1)(2\ell+1)} \int_{-1}^{+1}\!\!\!\!\!\dd\cos{\theta}\!\left[\left(\zeta^{{TE;EE}}_{22}\zeta^{{TE;EE}}_{33}+\zeta^{{TE;EE}}_{22}\zeta^{{TE;EE}}_{11,(1)}+2\zeta^{{TE;EE}}_{2-2}\zeta^{{TE;EE}}_{3-1,(1)}\right)d_{11}^\ell \right. \nonumber\\ 
    &&  \ \ \ \ \ \ \ \ \ \ \  +\left(\zeta^{{TE;EE}}_{2-2}\zeta^{{TE;EE}}_{3-3}+\zeta^{{TE;EE}}_{2-2}\zeta^{{TE;EE}}_{1-1,(1)}+2\zeta^{{TE;EE}}_{22}\zeta^{{TE;EE}}_{31,(1)}\right)d_{1-1}^\ell  \nonumber\\ 
    &&  \ \ \ \ \ \ \ \ \ \ \  +2\left(\zeta^{{TE;EE}}_{01}\zeta^{{TE;EE}}_{21}-\zeta^{{TE;EE}}_{30}\zeta^{{TE;EE}}_{2-1}\right)d_{11}^\ell \nonumber\\ 
    &&  \ \ \ \ \ \ \ \ \ \ \  +2\left(\zeta^{{TE;EE}}_{01}\zeta^{{TE;EE}}_{2-1}-\zeta^{{TE;EE}}_{30}\zeta^{{TE;EE}}_{21}\right)d_{1-1}^\ell\\
    &&  \ \ \ \ \ \ \ \ \ \ \  -\left(\zeta^{{TE;EE}}_{3-2,(\alpha)}\zeta^{{TE;EE}}_{3-2,(\beta)}+\zeta^{{TE;EE}}_{2-1,(\alpha)}\zeta^{{TE;EE}}_{2-1,(\beta)}
        +\zeta^{{TE;EE}}_{32,(\alpha)}\zeta^{{TE;EE}}_{21,(\alpha)}+\zeta^{{TE;EE}}_{21,(\beta)}\zeta^{{TE;EE}}_{32,(\beta)}\right)d_{11}^\ell \nonumber\\
    &&  \ \ \ \ \ \ \ \ \ \ \  +\left(\zeta^{{TE;EE}}_{32,(\alpha)}\zeta^{{TE;EE}}_{32,(\beta)}+\zeta^{{TE;EE}}_{21,(\alpha)}\zeta^{{TE;EE}}_{21,(\beta)}
        -\zeta^{{TE;EE}}_{3-2,(\alpha)}\zeta^{{TE;EE}}_{2-1,(\alpha)}-\zeta^{{TE;EE}}_{2-1,(\beta)}\zeta^{{TE;EE}}_{3-2,(\beta)}\right)d_{1-1}^\ell \nonumber\\
    &&  \ \ \ \ \ \ \ \ \ \ \  +2\left(\zeta^{{TE;EE}}_{20}\zeta^{{TE;EE}}_{1-1,(2)}+\zeta^{{TE;EE}}_{20}\zeta^{{TE;EE}}_{31,(2)}\right)d_{11}^\ell \nonumber\\
    &&  \ \ \ \ \ \ \ \ \ \ \  +2\left(\zeta^{{TE;EE}}_{20}\zeta^{{TE;EE}}_{11,(2)}+\zeta^{{TE;EE}}_{20}\zeta^{{TE;EE}}_{3-1,(2)}\right)d_{1-1}^\ell\,, \nonumber
\ee
where
\be
    \zeta^{{TE;EE}}_{2\pm2}  &&=  \sum\limits_\ell\frac{2\ell+1}{4\pi}\frac{1}{C_{\ell}^{{EE},\observed}}d_{2\pm2}^\ell
\ee
\be
    \zeta^{{TE;EE}}_{3\pm3} &&=   \sum\limits_\ell\frac{(2\ell+1)(\ell-2)(\ell+3)}{4\pi} \frac{{C}_{\ell}^{{TE,\observed}}{C}_{\ell}^{{EE}}{C}_{\ell}^{{TE}}}{C_{\ell}^{{TT,\observed}}C_{\ell}^{{EE,\observed}}}d_{3\pm3}^\ell
\ee
\be
    \zeta^{{TE;EE}}_{3\pm1,(1)} &&=   \sum\limits_\ell\frac{(2\ell+1)\sqrt{(\ell+2)(\ell-1)(\ell-2)(\ell+3)}}{4\pi} \frac{{C}_{\ell}^{{TE},\observed}{C}_{\ell}^{{EE}}{C}_{\ell}^{{TE}}}{C_{\ell}^{{TT,\observed}}C_{\ell}^{{EE,\observed}}}  d_{3\pm1}^\ell
\ee
\be
    \zeta^{{TE;EE}}_{1\pm1,(1)} &&=     \sum\limits_\ell  \frac{(2\ell+1)(\ell+2)(\ell-1)}{4\pi} \frac{{C}_{\ell}^{{TE},\observed}{C}_{\ell}^{{EE}}{C}_{\ell}^{{TE}}}{C_{\ell}^{{TT,\observed}}C_{\ell}^{{EE,\observed}}}  d_{1\pm1}^\ell  
\ee
\be 
    \zeta^{{TE;EE}}_{01} &&=  \sum\limits_\ell\frac{(2\ell+1)\sqrt{(\ell+2)\ell}}{4\pi}\frac{{C}_{\ell}^{{TE,\observed}}{C}_{\ell}^{{EE}}}{C_{\ell}^{{EE,\observed}}C_{\ell}^{{TT,\observed}}}d_{01}^\ell 
\ee
\be
    \zeta^{{TE;EE}}_{30} &&=   \sum\limits_\ell\frac{(2\ell+1)\sqrt{(\ell-2)(\ell+3)}}{4\pi}\frac{{C}_{\ell}^{{TE,\observed}}{C}_{\ell}^{{EE}}}{C_{\ell}^{{EE,\observed}}C_{\ell}^{{TT,\observed}}}d_{30}^\ell 
\ee
\be
    \zeta^{{TE;EE}}_{2\pm1} &&=    \sum\limits_\ell\frac{(2\ell+1)\ell(\ell+1)}{4\pi}\frac{{C}_{\ell}^{{TE}}}{{C}_{\ell}^{{EE,\observed}}}d_{2\pm1}^\ell 
\ee
\be 
    { \zeta^{{TE;EE}}_{2\pm1,(\alpha)}} &&=   \sum\limits_\ell\frac{(2\ell+1)\sqrt{\ell+2(\ell-1)}}{4\pi}\frac{{C}_{\ell}^{{EE}}}{{C}_{\ell}^{{EE,\observed}}}d_{2\pm1}^\ell
\ee
\be
    { \zeta^{{TE;EE}}_{3\pm2,(\alpha)}}
    &&=    \sum\limits_\ell\frac{(2\ell+1)\sqrt{(\ell-2)(\ell+3)}}{4\pi}\frac{{C}_{\ell}^{{TE,\observed}}{C}_{\ell}^{{TE}}}{{C}_{\ell}^{{EE,\observed}}{C}_{\ell}^{{TT,\observed}}}d_{3\pm2}^\ell  
\ee
\be
    { \zeta^{{TE;EE}}_{2\pm1,(\beta)} } 
    &&=    \sum\limits_\ell\frac{(2\ell+1)(\ell+2)(\ell-1)}{4\pi}\frac{{C}_{\ell}^{{TE,\observed}}{C}_{\ell}^{{TE}}}{{C}_{\ell}^{{EE,\observed}}{C}_{\ell}^{{TT,\observed}}}d_{2\pm1}^\ell 
\ee
\be
    { \zeta^{{TE;EE}}_{3\pm2,(\beta)}} &&=    \sum\limits_\ell\frac{(2\ell+1)\sqrt{(\ell-2)(\ell+3)}}{4\pi}\frac{{C}_{\ell}^{{EE}}}{{C}_{\ell}^{{EE}}}d_{3\pm2}^\ell  
\ee
\be
    { \zeta^{{TE;EE}}_{20}  } && =     \sum\limits_\ell\frac{2\ell+1}{4\pi}\frac{{C}_{\ell}^{{TE,\observed}}{{C}_{\ell}^{{EE}}}}{{C}_{\ell}^{{TT,\observed}}{C}_{\ell}^{{EE,\observed}}} d_{20}^\ell
\ee
\be
    { \zeta^{{TE;EE}}_{1\pm1,(2)}}  &&=   \sum\limits_\ell\frac{(2\ell+1)\sqrt{\ell(\ell+1)(\ell+2)(\ell-1)}}{4\pi} \frac{{C}_{\ell}^{{TE}}}{C_{\ell}^{{EE,\observed}}} d_{1\pm1}^\ell  
\ee
\be
    { \zeta^{{TE;EE}}_{3\pm1,(2)}}  &&=   \sum\limits_\ell\frac{(2\ell+1)\sqrt{\ell(\ell+1)(\ell-2)(\ell+3)}}{4\pi} \frac{{C}_{\ell}^{{TE}}}{C_{\ell}^{{EE,\observed}}}d_{3\pm1}^\ell     
\ee

\subsection{Expressions for \texorpdfstring{$N^{TT;TE}_\ell$}{Nl{TT;TE}}}
We require an additional term of the form
\be
    \label{eq:TTTEg}
    &&g^{{TT}*}_{\ell_1\ell\ell_2}(C_{\ell_1}^{{TT,\observed}}C_{\ell_2}^{{TE,\observed}}g^{{TE}}_{\ell_1\ell\ell_2}+C_{\ell_1}^{{TE,\observed}}C_{\ell_2}^{{TT,\observed}}g^{{TE}}_{\ell_2\ell\ell_1})\\  
    &&=\frac{1}{2C_{\ell_1}^{{TT,\observed}}C_{\ell_2}^{{TT,\observed}}}\left[\frac{C_{\ell_2}^{{TE,\observed}}}{C_{\ell_2}^{{EE,\observed}}}\left({C}_{\ell_1}^{TT}\left[\frac{{C}_{\ell_1}^{{TE}}}{2}{\left(F^{0}_{\ell_2\ell_1\ell}F^{2}_{\ell_2\ell_1\ell}+F^{0}_{\ell_2\ell_1\ell}F^{-2}_{\ell_2\ell_1\ell}\right)}+{C}_{\ell_2}^{{TE}}{\left(F^0_{\ell_2\ell_1\ell}F^0_{\ell_1\ell_2\ell}\right)}\right]\right.\right. \nonumber\\ 
    &&\ \ \ \ \ \ \ \ \ \ \ \ \ \ \ \ \ \ \ \ \ \ \ \ \ \ \ \ \ \ \ \ \ \left.{C}_{\ell_2}^{TT}\left[\frac{{C}_{\ell_1}^{{TE}}}{2}{\left(F^{0}_{\ell_1\ell_2\ell}F^{2}_{\ell_2\ell_1\ell}+F^{0}_{\ell_1\ell_2\ell}F^{-2}_{\ell_2\ell_1\ell}\right)}+{C}_{\ell_2}^{{TE}}{\left(F^0_{\ell_1\ell_2\ell}F^0_{\ell_1\ell_2\ell}\right)}\right]\right) \nonumber\\
    &&\ \ \ \ \ \ \ \ \ \ \ \ \ \ \ \ \ \ \ \ +  \frac{C_{\ell_1}^{{TE,\observed}}}{C_{\ell_1}^{{EE,\observed}}}\left({C}_{\ell_1}^{TT}\left[\frac{{C}_{\ell_2}^{{TE}}}{2}{\left(F^{0}_{\ell_2\ell_1\ell}F^{2}_{\ell_1\ell_2\ell}+F^{0}_{\ell_2\ell_1\ell}F^{-2}_{\ell_1\ell_2\ell}\right)}+{C}_{\ell_1}^{{TE}}{\left(F^0_{\ell_2\ell_1\ell}F^0_{\ell_2\ell_1\ell}\right)}\right]\right.\nonumber\\
    &&\ \ \ \ \ \ \ \ \ \ \ \ \ \ \ \ \ \ \ \ \ \ \ \ \ \ \ \ \ \ \ \ \ \left.\left.{C}_{\ell_2}^{TT}\left[\frac{{C}_{\ell_2}^{{TE}}}{2}{\left(F^{0}_{\ell_1\ell_2\ell}F^{2}_{\ell_1\ell_2\ell}+F^{0}_{\ell_1\ell_2\ell}F^{-2}_{\ell_1\ell_2\ell}\right)}+{C}_{\ell_1}^{{TE}}{\left(F^0_{\ell_1\ell_2\ell}F^0_{\ell_2\ell_1\ell}\right)}\right]\right)\right]\,. \nonumber
\ee
Using Eq.~\eqref{eq:TTTEg} in  Eq.~\eqref{eq:offd}, we get 
\be
    {N^{TT;TE}_\ell}\!\!=\!\frac{N^{TT;TT*}_\ell N^{TE;TE}_\ell}{2\ell(\ell+1)(2\ell+1)}&&\int_{-1}^{+1}\!\!\!\!\!\dd\!\cos{\theta}\!\!\left[
    \left({\zeta^{{TT;TE}}_{20}\zeta^{{TT;TE}}_{1-1,(1)}+\zeta^{{TT;TE}}_{20}\zeta^{{TT;TE}}_{31}}{-2\zeta^{{TT;TE}}_{01,(1)}\zeta^{{TT;TE}}_{01,(2)}}\right)\!d_{11}^\ell\hspace*{1cm}\right.\\
    && \ \ \ \ \ \ \ \ \ \ \ \   +\!\left({\zeta^{{TT;TE}}_{20}\zeta^{{TT;TE}}_{11,(1)}+\zeta^{{TT;TE}}_{20}\zeta^{{TT;TE}}_{3-1}}
    {+2\zeta^{{TT;TE}}_{01,(1)}\zeta^{{TT;TE}}_{01,(2)}}\right)d_{1-1}^\ell \nonumber\\
    && \ \ \ \ \ \ \ \ \ \ \ \  +\!\left({\zeta^{{TT;TE}}_{01,(3)}\zeta^{{TT;TE}}_{21}-\zeta^{{TT;TE}}_{30}\zeta^{{TT;TE}}_{2-1}}{ +2\zeta^{{TT;TE}}_{00,(1)}\zeta^{{TT;TE}}_{11,(2)}}\right)d^\ell_{11} \nonumber\\ 
    && \ \ \ \ \ \ \ \ \ \ \ \   +\!\left({\zeta^{{TT;TE}}_{01,(3)}\zeta^{{TT;TE}}_{2-1}-\zeta^{{TT;TE}}_{30}\zeta^{{TT;TE}}_{21}}{ +2\zeta^{{TT;TE}}_{00,(1)}\zeta^{{TT;TE}}_{1-1,(2)}}\right)d^\ell_{1-1} \nonumber 
\ee
where
\be
    {\zeta^{{TT;TE}}_{20}}&&=     \sum\limits_\ell\frac{2\ell+1}{4\pi}\frac{C_\ell^{{TE}}}{C_\ell^{{EE,\observed}}C_\ell^{{TT,\observed}}} d_{20}^\ell  
\ee
\be
    { \zeta^{{TT;TE}}_{1\pm1,(1)}} && =   \sum\limits_\ell\frac{(2\ell+1)\sqrt{\ell(\ell+1)(\ell+2)(\ell-1)}}{4\pi} \frac{{C}_{\ell}^{{TT}}{C}_{\ell}^{{TE}}}{C_{\ell}^{{TT,\observed}}} d_{1\pm1}^\ell 
\ee
\be
    { \zeta^{{TT;TE}}_{3\pm1}} && =   \sum\limits_\ell\frac{(2\ell+1)\sqrt{\ell(\ell+1)(\ell-2)(\ell+3)}}\ \frac{{C}_{\ell}^{{TT}}{C}_{\ell}^{{TE}}}{C_{\ell}^{{TT,\observed}}}d_{3\pm1}^\ell 
\ee
\be
    { \zeta^{{TT;TE}}_{01,(3)}  }&&=     \sum\limits_\ell\frac{(2\ell+1)\sqrt{\ell(\ell+2)}}{4\pi} \frac{{C}_\ell^{{TE}}}{{C}_\ell^{{TT,\observed}}}  d_{01}^\ell \ee\be
    { \zeta^{{TT;TE}}_{2\pm1}} &&=   \sum\limits_\ell\frac{(2\ell+1)(\ell+2)(\ell-1)}{4\pi} \frac{C_\ell^{{TE}}{C}_\ell^{{TT}}}{{C}_\ell^{{TT,\observed}}{C}_\ell^{{EE,\observed}}} d_{2\pm1}^\ell
\ee
\be
    { \zeta^{{TT;TE}}_{30}}&&=   \sum\limits_\ell\frac{(2\ell+1)\sqrt{(\ell-2)(\ell+3)}}{4\pi} \frac{{C}_\ell^{{TE}}}{{C}_\ell^{{TT,\observed}}}d_{30}^\ell
\ee
\be
    { \zeta^{{TT;TE}}_{01,(1)}}&&=     \sum\limits_\ell\frac{(2\ell+1)\sqrt{\ell(\ell+1)}}{4\pi} \frac{{C}_\ell^{{TE,\observed}}{C}_\ell^{{TE}}}{{C}_\ell^{{TT,\observed}}{C}_\ell^{{EE,\observed}}}  d_{01}^\ell 
\ee
\be
{ \zeta^{{TT;TE}}_{01,(2)}}&&=     \sum\limits_\ell\frac{(2\ell+1)\sqrt{\ell(\ell+1)}}{4\pi} \frac{{C}_\ell^{{TT}}}{{C}_\ell^{{TT},\observed}} d_{01}^\ell \ee\be
    { \zeta^{{TT;TE}}_{00,(1)}}&&=     \sum\limits_\ell\frac{2\ell+1}{4\pi} \frac{1}{{C}_\ell^{{TT,\observed}}}  d_{01}^\ell \ee\be
    { \zeta^{{TT;TE}}_{1\pm1,(2)}}&&=     \sum\limits_\ell\frac{(2\ell+1)\ell(\ell+1)}{4\pi} \frac{{C}_\ell^{{TE,\observed}}{C}_\ell^{{TT}}{C}_\ell^{{TE}}}{{C}_\ell^{{TT,\observed}}{{C}_\ell^{{EE,\observed}}}}  d_{01}^\ell
\ee

\subsection{Expressions for \texorpdfstring{$N^{TT;EE}_\ell$}{Nl{TT;EE}}}
This element requires that we calculate
\be
    \sum\limits_{\ell_1\ell_2}&&\{g^{{TT}*}_{\ell_1\ell\ell_2}C_{\ell_1}^{{TE,\observed}}C_{\ell_2}^{{TE,\observed}}\left(g^{{EE}}_{\ell_1\ell\ell_2}+g^{{EE}}_{\ell_2\ell\ell_1}\right)\}\nonumber\\  
    &&=\frac{C_{\ell_1}^{{TE,\observed}}C_{\ell_2}^{{TE,\observed}}}{2C_{\ell_1}^{{TT,\observed}}C_{\ell_2}^{{TT,\observed}}C_{\ell_1}^{{EE,\observed}}C_{\ell_2}^{{EE,\observed}}}\left[{C}_{\ell_1}^{{TT}}{C}_{\ell_1}^{{EE}}{\left(F^0_{\ell_2\ell_1\ell}F^2_{\ell_2\ell_1\ell}+F^0_{\ell_2\ell_1\ell}F^{-2}_{\ell_2\ell_1\ell}\right)}\right.\\ 
    && \ \ \ \ \ \ \ \ \ \ \ \ \ \ \ \ \ \ \ \ \ \ \ \ \ \ \ \ \ \ \ \ \ \ \ \ \ \ \ \ \ \ \ \  \left.+{C}_{\ell_1}^{{TT}}{C}_{\ell_2}^{{EE}}{\left(F^0_{\ell_2\ell_1\ell}F^2_{\ell_1\ell_2\ell}+F^0_{\ell_2\ell_1\ell}F^{-2}_{\ell_1\ell_2\ell}\right)}\right]\nonumber
\ee
such that
\be
    {N^{TT;EE}_\ell}\!\!=\!\frac{N^{TT;TT*}_\ell N^{EE;EE}_\ell}{2\ell(\ell+1)(2\ell+1)}&&\int_{-1}^{+1}\dd\cos{\theta}\ \left[
    \left({\zeta^{{TT;TE}}_{20}\zeta^{{TT;TE}}_{1-1}+\zeta^{{TT;TE}}_{20}\zeta^{{TT;TE}}_{31}}\right)d_{11}^\ell\right. \\
    && \ \ \ \ \ \ \ \ \ \ \ \ \ \ \ \ \ \  +\left({\zeta^{{TT;TE}}_{20}\zeta^{{TT;TE}}_{11}+\zeta^{{TT;TE}}_{20}\zeta^{{TT;TE}}_{3-1}}
    \right)d_{1-1}^\ell \nonumber\\
    && \ \ \ \ \ \ \ \ \ \ \ \ \ \ \ \ \ \  +\left({\zeta^{{TT;TE}}_{01}\zeta^{{TT;TE}}_{21}-\zeta^{{TT;TE}}_{30}\zeta^{{TT;TE}}_{2-1}}\right)d^\ell_{11} \nonumber\\ 
    && \ \ \ \ \ \ \ \ \ \ \ \ \ \ \ \ \ \  +\left.\left({\zeta^{{TT;TE}}_{01}\zeta^{{TT;TE}}_{2-1}-\zeta^{{TT;TE}}_{30}\zeta^{{TT;TE}}_{21}}\right)d^\ell_{1-1}\right] \nonumber
\ee
where
\be
    { \zeta^{{TT;EE}}_{20}  }&&=     \sum\limits_\ell\frac{2\ell+1}{4\pi}\frac{C_\ell^{{TE}}}{C_\ell^{{EE,\observed}}C_\ell^{{TT,\observed}}} d_{20}^\ell 
\ee
\be
    { \zeta^{{TT;EE}}_{1\pm1}}&&=   \sum\limits_\ell\frac{(2\ell+1)\sqrt{\ell(\ell+1)(\ell+2)(\ell-1)}}{4\pi} \frac{{C}_{\ell}^{{TE,\observed}}{C}_{\ell}^{{TT}}{C}_{\ell}^{{EE}}}{C_{\ell}^{{TT,\observed}}{C}_{\ell}^{{EE,\observed}}} d_{1\pm1}^\ell 
\ee
\be
    { \zeta^{{TT;EE}}_{3\pm1}}&&=   \sum\limits_\ell\frac{(2\ell+1)\sqrt{\ell(\ell+1)(\ell-2)(\ell+3)}}{4\pi} \frac{{C}_{\ell}^{{TE}}{C}_{\ell}^{{TT}}{C}_{\ell}^{{EE}}}{C_{\ell}^{{TT,\observed}}{C}_{\ell}^{{EE,\observed}}}d_{3\pm1}^\ell   
\ee
\be
    { \zeta^{{TT;EE}}_{01}  }&&=     \sum\limits_\ell\frac{(2\ell+1)\sqrt{\ell(\ell+2)}}{4\pi} \frac{{C}_{\ell}^{{TE,\observed}}{C}_{\ell}^{{EE}}}{C_{\ell}^{{TT,\observed}}{C}_{\ell}^{{EE,\observed}}}  d_{01}^\ell 
    \ee\be
    { \zeta^{{TT;EE}}_{2\pm1}}&&=   \sum\limits_\ell\frac{(2\ell+1)(\ell+2)(\ell-1)}{4\pi} \frac{C_\ell^{{TE,\observed}}{C}_\ell^{{TT}}}{{C}_\ell^{{TT,\observed}}{C}_\ell^{{EE,\observed}}} d_{2\pm1}^\ell\ee\be
    { \zeta^{{TT;EE}}_{30}}&&=   \sum\limits_\ell\frac{(2\ell+1)\sqrt{(\ell-2)(\ell+3)}}{4\pi}\frac{C_\ell^{{TE,\observed}}{C}_\ell^{{TT}}}{{C}_\ell^{{TT,\observed}}{C}_\ell^{{EE,\observed}}}d_{30}^\ell     
\ee
\subsection{Expressions for \texorpdfstring{$N^{TB;EB}_\ell$}{Nl{TB;EB}}}
Here we need
\be
    \sum\limits_{\ell_1\ell_2}&&g^{{TB}*}_{\ell_1\ell\ell_2} \left(C_{\ell_1}^{{TE,\observed}}C_{\ell_2}^{{BB,\observed}}g^{{EB}}_{\ell_1\ell\ell_2}\right) \\ = &&\sum\limits_{\ell_1\ell_2}\frac{C_{\ell_1}^{{TE,\observed}}{C}_{\ell_1}^{{TE}}}{C_{\ell_1}^{{TT,\observed}}C_{\ell_1}^{{EE,\observed}}C_{\ell_2}^{{BB,\observed}}}\left[\frac{{C}_{\ell_1}^{{EE}}}{4}{\left(F_{\ell_2\ell_1\ell}^2-F_{\ell_2\ell_1\ell}^{-2}\right)^2} \right.\nonumber \\ 
    &&\qquad\qquad\qquad\qquad\qquad\qquad \left. -\frac{{C}_{\ell_2}^{{BB}}}{4}{\left(F^{2}_{\ell_2\ell_1\ell}F^{2}_{\ell_1\ell_2\ell}-F^{2}_{\ell_2\ell_1\ell}F^{-2}_{\ell_1\ell_2\ell}-F^{-2}_{\ell_2\ell_1\ell}F^{2}_{\ell_1\ell_2\ell}+F^{-2}_{\ell_2\ell_1\ell}F^{2}_{\ell_1\ell_2\ell}\right)}\right] \nonumber
\ee
such that 
\be
    {N^{TB;EB}_\ell}\!\!=&&\frac{N^{TB;TB*}_\ell N^{EB;EB}_\ell}{4\ell(\ell+1)(2\ell+1)}\int_{-1}^{+1}\!\!\!\!\dd\cos{\theta}\!\left[{\left(\zeta^{{TB;EB}}_{22}\zeta^{{TB;EB}}_{33}+\zeta^{{TB;EB}}_{22}\zeta^{{TB;EB}}_{11,(1)}-2\zeta^{{TB;EB}}_{2-2}\zeta^{{TB;EB}}_{3-1,(}\right)d_{11}^\ell}\right.\nonumber \\ 
    \qquad\qquad&&\qquad\qquad\qquad \ \ \  \ {-\left(\zeta^{{TB;EB}}_{2-2}\zeta^{{TB;EB}}_{3-3}+\zeta^{{TB;EB}}_{2-2}\zeta^{{TB;EB}}_{1-1,(1)}-2\zeta^{{TB;EB}}_{22}\zeta^{{TB;EB}}_{31,(1)}\right)d_{1-1}^\ell} \\ 
    && \qquad -\left(\zeta^{{TB;EB}}_{3-2,(\alpha)}\zeta^{{TB;EB}}_{3-2,(\beta)}+\zeta^{{TB;EB}}_{2-1,(\alpha)}\zeta^{{TB;EB}}_{2-1,(\beta)}-\zeta^{{TB;EB}}_{32,(\alpha)}\zeta^{{TB;EB}}_{21,(\alpha)}-\zeta^{{TB;EB}}_{21,(\beta)}\zeta^{{TB;EB}}_{32,(\beta)}\right)d_{11}^\ell \nonumber\\ &&  \qquad -\left(\zeta^{{TB;EB}}_{32,(\alpha)}\zeta^{{TB;EB}}_{32,(\beta)}+\zeta^{{TB;EB}}_{21,(\alpha)}\zeta^{{TB;EB}}_{21,(\beta)}+\zeta^{{TB;EB}}_{3-2,(\alpha)}\zeta^{{TB;EB}}_{2-1,(\alpha)}+\zeta^{{TB;EB}}_{2-1,(\beta)}\zeta^{{TB;EB}}_{3-2,(\beta)}\right)d_{1-1}^\ell\Big] \nonumber
\ee
where 
\be
    \zeta^{{TB;EB}}_{2\pm2} && =  \sum\limits_\ell\frac{2\ell+1}{4\pi}\frac{1}{C_\ell^{{BB,\observed}}}d_{2\pm2}^\ell  \ee\be
    \zeta^{{TB;EB}}_{3\pm3} && =   \sum\limits_\ell\frac{(2\ell+1){(\ell-2)(\ell+3)}}{4\pi} \frac{C_\ell^{{TE,\observed}}{C}_\ell^{{TE}}{C}_\ell^{{EE}}}{C_\ell^{{TT,\observed}}C_\ell^{{EE,\observed}}}d_{3\pm3}^\ell 
\ee
\be
    \zeta^{{TB;EB}}_{3\pm1,(1)} && =   \sum\limits_\ell\frac{(2\ell+1)\sqrt{(\ell+2)(\ell-1)(\ell-2)(\ell+3)}}{4\pi}  \frac{C_\ell^{{TE,\observed}}{C}_\ell^{{TE}}{C}_\ell^{{EE}}}{C_\ell^{{TT,\observed}}C_\ell^{{EE,\observed}}}  d_{3\pm1}^\ell\ \ \ \ \ \  
\ee
\be
    \zeta^{{TB;EB}}_{1\pm1,(1)} && =     \sum\limits_\ell\frac{(2\ell+1)(\ell+2)(\ell-1)}{4\pi}  \frac{C_\ell^{{TE,\observed}}{C}_\ell^{{TE}}{C}_\ell^{{EE}}}{C_\ell^{{TT,\observed}}C_\ell^{{EE,\observed}}} d_{1\pm1}^\ell   
\ee
\be
    { \zeta^{{TB;EB}}_{2\pm1,(\alpha)}}
    &&=   \sum\limits_\ell\frac{(2\ell+1)(\ell+2)(\ell-1)}{4\pi}\frac{{C}_\ell^{{BB}}}{{C}_\ell^{{BB,\observed}}}d_{2\pm1}^\ell   
\ee
\be
    { \zeta^{{TB;EB}}_{3\pm2,(\alpha)}}
    &&=    \sum\limits_\ell\frac{(2\ell+1)\sqrt{(\ell-2)(\ell+3)}}{4\pi}\frac{C_\ell^{{TE,\observed}}{C}_\ell^{{TE}}}{C_\ell^{{TT,\observed}}C_\ell^{{EE,\observed}}}d_{3\pm2}^\ell  
\ee
\be
    { \zeta^{{TB;EB}}_{2\pm1,(\beta)} } 
    &&= \sum\limits_\ell\frac{(2\ell+1)(\ell+2)(\ell-1)}{4\pi} \frac{C_\ell^{{TE,\observed}}{C}_\ell^{{TE}}}{C_\ell^{{TT,\observed}}C_\ell^{{EE,\observed}}}d_{2\pm1}^\ell   
\ee
\be
    { \zeta^{{TB;EB}}_{3\pm2,(\beta)}} &&= \sum\limits_\ell\frac{(2\ell+1)\sqrt{(\ell-2)(\ell+3)}}{4\pi}\frac{{C}_\ell^{{BB}}}{{C}_\ell^{{BB,\observed}}}d_{3\pm2}^\ell 
\ee

\bibliographystyle{utphys}
\bibliography{delensing}

\providecommand{\href}[2]{#2}\begingroup\raggedright\begin{thebibliography}{100}

\bibitem{Lewis:2006fu}
A.~Lewis and A.~Challinor, ``{Weak gravitational lensing of the CMB},''
  \href{http://dx.doi.org/10.1016/j.physrep.2006.03.002}{{\em Phys. Rept.}
  {\bfseries 429} (2006) 1--65},
  \href{http://arxiv.org/abs/astro-ph/0601594}{{\ttfamily
  arXiv:astro-ph/0601594}}.

\bibitem{Hu:2001kj}
W.~Hu and T.~Okamoto, ``{Mass reconstruction with cmb polarization},''
  \href{http://dx.doi.org/10.1086/341110}{{\em Astrophys. J.} {\bfseries 574}
  (2002) 566--574}, \href{http://arxiv.org/abs/astro-ph/0111606}{{\ttfamily
  arXiv:astro-ph/0111606}}.

\bibitem{Okamoto:2003zw}
T.~Okamoto and W.~Hu, ``{CMB lensing reconstruction on the full sky},''
  \href{http://dx.doi.org/10.1103/PhysRevD.67.083002}{{\em Phys. Rev. D}
  {\bfseries 67} (2003) 083002},
  \href{http://arxiv.org/abs/astro-ph/0301031}{{\ttfamily
  arXiv:astro-ph/0301031}}.

\bibitem{Planck:2018lbu}
{\bfseries Planck} Collaboration, N.~Aghanim {\em et~al.}, ``{Planck 2018
  results. VIII. Gravitational lensing},''
  \href{http://dx.doi.org/10.1051/0004-6361/201833886}{{\em Astron. Astrophys.}
  {\bfseries 641} (2020) A8}, \href{http://arxiv.org/abs/1807.06210}{{\ttfamily
  arXiv:1807.06210 [astro-ph.CO]}}.

\bibitem{SimonsObservatory:2018koc}
{\bfseries Simons Observatory} Collaboration, P.~Ade {\em et~al.}, ``{The
  Simons Observatory: Science goals and forecasts},''
  \href{http://dx.doi.org/10.1088/1475-7516/2019/02/056}{{\em JCAP} {\bfseries
  02} (2019) 056}, \href{http://arxiv.org/abs/1808.07445}{{\ttfamily
  arXiv:1808.07445 [astro-ph.CO]}}.

\bibitem{CMB-S4:2016ple}
{\bfseries CMB-S4} Collaboration, K.~N. Abazajian {\em et~al.}, ``{CMB-S4
  Science Book, First Edition},''
  \href{http://arxiv.org/abs/1610.02743}{{\ttfamily arXiv:1610.02743
  [astro-ph.CO]}}.

\bibitem{NASAPICO:2019thw}
{\bfseries NASA PICO} Collaboration, S.~Hanany {\em et~al.}, ``{PICO: Probe of
  Inflation and Cosmic Origins},''
  \href{http://arxiv.org/abs/1902.10541}{{\ttfamily arXiv:1902.10541
  [astro-ph.IM]}}.

\bibitem{Sehgal:2019ewc}
N.~Sehgal {\em et~al.}, ``{CMB-HD: An Ultra-Deep, High-Resolution
  Millimeter-Wave Survey Over Half the Sky},''
  \href{http://arxiv.org/abs/1906.10134}{{\ttfamily arXiv:1906.10134
  [astro-ph.CO]}}.

\bibitem{Dolgov:2002wy}
A.~D. Dolgov, ``{Neutrinos in cosmology},''
  \href{http://dx.doi.org/10.1016/S0370-1573(02)00139-4}{{\em Phys. Rept.}
  {\bfseries 370} (2002) 333--535},
  \href{http://arxiv.org/abs/hep-ph/0202122}{{\ttfamily arXiv:hep-ph/0202122}}.

\bibitem{Kaplinghat:2003bh}
M.~Kaplinghat, L.~Knox, and Y.-S. Song, ``{Determining neutrino mass from the
  CMB alone},'' \href{http://dx.doi.org/10.1103/PhysRevLett.91.241301}{{\em
  Phys. Rev. Lett.} {\bfseries 91} (2003) 241301},
  \href{http://arxiv.org/abs/astro-ph/0303344}{{\ttfamily
  arXiv:astro-ph/0303344}}.

\bibitem{Lesgourgues:2006nd}
J.~Lesgourgues and S.~Pastor, ``{Massive neutrinos and cosmology},''
  \href{http://dx.doi.org/10.1016/j.physrep.2006.04.001}{{\em Phys. Rept.}
  {\bfseries 429} (2006) 307--379},
  \href{http://arxiv.org/abs/astro-ph/0603494}{{\ttfamily
  arXiv:astro-ph/0603494}}.

\bibitem{Dvorkin:2019jgs}
C.~Dvorkin {\em et~al.}, ``{Neutrino Mass from Cosmology: Probing Physics
  Beyond the Standard Model},''
  \href{http://arxiv.org/abs/1903.03689}{{\ttfamily arXiv:1903.03689
  [astro-ph.CO]}}.

\bibitem{Green:2021xzn}
D.~Green and J.~Meyers, ``{Cosmological Implications of a Neutrino Mass
  Detection},'' \href{http://arxiv.org/abs/2111.01096}{{\ttfamily
  arXiv:2111.01096 [astro-ph.CO]}}.

\bibitem{Frieman:2008sn}
J.~Frieman, M.~Turner, and D.~Huterer, ``{Dark Energy and the Accelerating
  Universe},''
  \href{http://dx.doi.org/10.1146/annurev.astro.46.060407.145243}{{\em Ann.
  Rev. Astron. Astrophys.} {\bfseries 46} (2008) 385--432},
  \href{http://arxiv.org/abs/0803.0982}{{\ttfamily arXiv:0803.0982
  [astro-ph]}}.

\bibitem{Seljak:2008xr}
U.~Seljak, ``{Extracting primordial non-gaussianity without cosmic variance},''
  \href{http://dx.doi.org/10.1103/PhysRevLett.102.021302}{{\em Phys. Rev.
  Lett.} {\bfseries 102} (2009) 021302},
\href{http://arxiv.org/abs/0807.1770}{{\ttfamily arXiv:0807.1770 [astro-ph]}}.

\bibitem{Schaan:2016ois}
E.~Schaan, E.~Krause, T.~Eifler, O.~Dor\'e, H.~Miyatake, J.~Rhodes, and D.~N.
  Spergel, ``{Looking through the same lens: Shear calibration for LSST,
  Euclid, and WFIRST with stage 4 CMB lensing},''
  \href{http://dx.doi.org/10.1103/PhysRevD.95.123512}{{\em Phys. Rev. D}
  {\bfseries 95} no.~12, (2017) 123512},
  \href{http://arxiv.org/abs/1607.01761}{{\ttfamily arXiv:1607.01761
  [astro-ph.CO]}}.

\bibitem{Schmittfull:2017ffw}
M.~Schmittfull and U.~Seljak, ``{Parameter constraints from cross-correlation
  of CMB lensing with galaxy clustering},''
  \href{http://dx.doi.org/10.1103/PhysRevD.97.123540}{{\em Phys. Rev. D}
  {\bfseries 97} no.~12, (2018) 123540},
  \href{http://arxiv.org/abs/1710.09465}{{\ttfamily arXiv:1710.09465
  [astro-ph.CO]}}.

\bibitem{Yu:2018tem}
B.~Yu, R.~Z. Knight, B.~D. Sherwin, S.~Ferraro, L.~Knox, and M.~Schmittfull,
  ``{Towards Neutrino Mass from Cosmology without Optical Depth Information},''
  \href{http://arxiv.org/abs/1809.02120}{{\ttfamily arXiv:1809.02120
  [astro-ph.CO]}}.

\bibitem{Yu:2021vce}
B.~Yu, S.~Ferraro, Z.~R. Knight, L.~Knox, and B.~D. Sherwin, ``{The Physical
  Origin of Dark Energy Constraints from Rubin Observatory and CMB-S4 Lensing
  Tomography},'' \href{http://arxiv.org/abs/2108.02801}{{\ttfamily
  arXiv:2108.02801 [astro-ph.CO]}}.

\bibitem{Kamionkowski:1996zd}
M.~Kamionkowski, A.~Kosowsky, and A.~Stebbins, ``{A Probe of primordial gravity
  waves and vorticity},''
  \href{http://dx.doi.org/10.1103/PhysRevLett.78.2058}{{\em Phys. Rev. Lett.}
  {\bfseries 78} (1997) 2058--2061},
  \href{http://arxiv.org/abs/astro-ph/9609132}{{\ttfamily
  arXiv:astro-ph/9609132}}.

\bibitem{Zaldarriaga:1996xe}
M.~Zaldarriaga and U.~Seljak, ``{An all sky analysis of polarization in the
  microwave background},''
  \href{http://dx.doi.org/10.1103/PhysRevD.55.1830}{{\em Phys. Rev. D}
  {\bfseries 55} (1997) 1830--1840},
  \href{http://arxiv.org/abs/astro-ph/9609170}{{\ttfamily
  arXiv:astro-ph/9609170}}.

\bibitem{Seljak:1996gy}
U.~Seljak and M.~Zaldarriaga, ``{Signature of gravity waves in polarization of
  the microwave background},''
  \href{http://dx.doi.org/10.1103/PhysRevLett.78.2054}{{\em Phys. Rev. Lett.}
  {\bfseries 78} (1997) 2054--2057},
  \href{http://arxiv.org/abs/astro-ph/9609169}{{\ttfamily
  arXiv:astro-ph/9609169}}.

\bibitem{Kamionkowski:1996ks}
M.~Kamionkowski, A.~Kosowsky, and A.~Stebbins, ``{Statistics of cosmic
  microwave background polarization},''
  \href{http://dx.doi.org/10.1103/PhysRevD.55.7368}{{\em Phys. Rev. D}
  {\bfseries 55} (1997) 7368--7388},
  \href{http://arxiv.org/abs/astro-ph/9611125}{{\ttfamily
  arXiv:astro-ph/9611125}}.

\bibitem{Knox:2002pe}
L.~Knox and Y.-S. Song, ``{A Limit on the detectability of the energy scale of
  inflation},'' \href{http://dx.doi.org/10.1103/PhysRevLett.89.011303}{{\em
  Phys. Rev. Lett.} {\bfseries 89} (2002) 011303},
  \href{http://arxiv.org/abs/astro-ph/0202286}{{\ttfamily
  arXiv:astro-ph/0202286}}.

\bibitem{Kesden:2002ku}
M.~Kesden, A.~Cooray, and M.~Kamionkowski, ``{Separation of gravitational wave
  and cosmic shear contributions to cosmic microwave background
  polarization},'' \href{http://dx.doi.org/10.1103/PhysRevLett.89.011304}{{\em
  Phys. Rev. Lett.} {\bfseries 89} (2002) 011304},
  \href{http://arxiv.org/abs/astro-ph/0202434}{{\ttfamily
  arXiv:astro-ph/0202434}}.

\bibitem{Seljak:2003pn}
U.~Seljak and C.~M. Hirata, ``{Gravitational lensing as a contaminant of the
  gravity wave signal in CMB},''
  \href{http://dx.doi.org/10.1103/PhysRevD.69.043005}{{\em Phys. Rev. D}
  {\bfseries 69} (2004) 043005},
  \href{http://arxiv.org/abs/astro-ph/0310163}{{\ttfamily
  arXiv:astro-ph/0310163}}.

\bibitem{Smith:2010gu}
K.~M. Smith, D.~Hanson, M.~LoVerde, C.~M. Hirata, and O.~Zahn, ``{Delensing CMB
  Polarization with External Datasets},''
  \href{http://dx.doi.org/10.1088/1475-7516/2012/06/014}{{\em JCAP} {\bfseries
  06} (2012) 014}, \href{http://arxiv.org/abs/1010.0048}{{\ttfamily
  arXiv:1010.0048 [astro-ph.CO]}}.

\bibitem{Abazajian:2016yjj}
{\bfseries CMB-S4} Collaboration, K.~N. Abazajian {\em et~al.}, ``{CMB-S4
  Science Book, First Edition},''
  \href{http://arxiv.org/abs/1610.02743}{{\ttfamily arXiv:1610.02743
  [astro-ph.CO]}}.

\bibitem{Abazajian:2019eic}
K.~Abazajian {\em et~al.}, ``{CMB-S4 Science Case, Reference Design, and
  Project Plan},'' \href{http://arxiv.org/abs/1907.04473}{{\ttfamily
  arXiv:1907.04473 [astro-ph.IM]}}.

\bibitem{CMB-S4:2020lpa}
{\bfseries CMB-S4} Collaboration, K.~Abazajian {\em et~al.}, ``{CMB-S4:
  Forecasting Constraints on Primordial Gravitational Waves},''
  \href{http://arxiv.org/abs/2008.12619}{{\ttfamily arXiv:2008.12619
  [astro-ph.CO]}}.

\bibitem{Green:2016cjr}
D.~Green, J.~Meyers, and A.~van Engelen, ``{CMB Delensing Beyond the B
  Modes},'' \href{http://dx.doi.org/10.1088/1475-7516/2017/12/005}{{\em JCAP}
  {\bfseries 12} (2017) 005}, \href{http://arxiv.org/abs/1609.08143}{{\ttfamily
  arXiv:1609.08143 [astro-ph.CO]}}.

\bibitem{Larsen:2016wpa}
P.~Larsen, A.~Challinor, B.~D. Sherwin, and D.~Mak, ``{Demonstration of cosmic
  microwave background delensing using the cosmic infrared background},''
  \href{http://dx.doi.org/10.1103/PhysRevLett.117.151102}{{\em Phys. Rev.
  Lett.} {\bfseries 117} no.~15, (2016) 151102},
  \href{http://arxiv.org/abs/1607.05733}{{\ttfamily arXiv:1607.05733
  [astro-ph.CO]}}.

\bibitem{Carron:2017vfg}
J.~Carron, A.~Lewis, and A.~Challinor, ``{Internal delensing of Planck CMB
  temperature and polarization},''
  \href{http://dx.doi.org/10.1088/1475-7516/2017/05/035}{{\em JCAP} {\bfseries
  05} (2017) 035}, \href{http://arxiv.org/abs/1701.01712}{{\ttfamily
  arXiv:1701.01712 [astro-ph.CO]}}.

\bibitem{ACT:2020goa}
{\bfseries ACT} Collaboration, D.~Han {\em et~al.}, ``{The Atacama Cosmology
  Telescope: delensed power spectra and parameters},''
  \href{http://dx.doi.org/10.1088/1475-7516/2021/01/031}{{\em JCAP} {\bfseries
  01} (2021) 031}, \href{http://arxiv.org/abs/2007.14405}{{\ttfamily
  arXiv:2007.14405 [astro-ph.CO]}}.

\bibitem{Millea:2020iuw}
M.~Millea {\em et~al.}, ``{Optimal CMB Lensing Reconstruction and Parameter
  Estimation with SPTpol Data},''
  \href{http://arxiv.org/abs/2012.01709}{{\ttfamily arXiv:2012.01709
  [astro-ph.CO]}}.

\bibitem{Blas:2011rf}
D.~Blas, J.~Lesgourgues, and T.~Tram, ``{The Cosmic Linear Anisotropy Solving
  System (CLASS) II: Approximation schemes},''
  \href{http://dx.doi.org/10.1088/1475-7516/2011/07/034}{{\em JCAP} {\bfseries
  07} (2011) 034}, \href{http://arxiv.org/abs/1104.2933}{{\ttfamily
  arXiv:1104.2933 [astro-ph.CO]}}.

\bibitem{Sherwin:2015baa}
B.~D. Sherwin and M.~Schmittfull, ``{Delensing the CMB with the Cosmic Infrared
  Background},'' \href{http://dx.doi.org/10.1103/PhysRevD.92.043005}{{\em Phys.
  Rev. D} {\bfseries 92} no.~4, (2015) 043005},
  \href{http://arxiv.org/abs/1502.05356}{{\ttfamily arXiv:1502.05356
  [astro-ph.CO]}}.

\bibitem{Lewis:1999bs}
A.~Lewis, A.~Challinor, and A.~Lasenby, ``{Efficient computation of CMB
  anisotropies in closed FRW models},''
  \href{http://dx.doi.org/10.1086/309179}{{\em \apj} {\bfseries 538} (2000)
  473--476},
\href{http://arxiv.org/abs/astro-ph/9911177}{{\ttfamily arXiv:astro-ph/9911177
  [astro-ph]}}.

\bibitem{Challinor:2005jy}
A.~Challinor and A.~Lewis, ``{Lensed CMB power spectra from all-sky correlation
  functions},'' \href{http://dx.doi.org/10.1103/PhysRevD.71.103010}{{\em Phys.
  Rev. D} {\bfseries 71} (2005) 103010},
  \href{http://arxiv.org/abs/astro-ph/0502425}{{\ttfamily
  arXiv:astro-ph/0502425}}.

\bibitem{Planck:2018jri}
{\bfseries Planck} Collaboration, Y.~Akrami {\em et~al.}, ``{Planck 2018
  results. X. Constraints on inflation},''
  \href{http://dx.doi.org/10.1051/0004-6361/201833887}{{\em Astron. Astrophys.}
  {\bfseries 641} (2020) A10},
  \href{http://arxiv.org/abs/1807.06211}{{\ttfamily arXiv:1807.06211
  [astro-ph.CO]}}.

\bibitem{Hirata:2002jy}
C.~M. Hirata and U.~Seljak, ``{Analyzing weak lensing of the cosmic microwave
  background using the likelihood function},''
  \href{http://dx.doi.org/10.1103/PhysRevD.67.043001}{{\em Phys. Rev. D}
  {\bfseries 67} (2003) 043001},
  \href{http://arxiv.org/abs/astro-ph/0209489}{{\ttfamily
  arXiv:astro-ph/0209489}}.

\bibitem{Millea:2017fyd}
M.~Millea, E.~Anderes, and B.~D. Wandelt, ``{Bayesian delensing of CMB
  temperature and polarization},''
  \href{http://dx.doi.org/10.1103/PhysRevD.100.023509}{{\em Phys. Rev. D}
  {\bfseries 100} no.~2, (2019) 023509},
  \href{http://arxiv.org/abs/1708.06753}{{\ttfamily arXiv:1708.06753
  [astro-ph.CO]}}.

\bibitem{Horowitz:2017iql}
B.~Horowitz, S.~Ferraro, and B.~D. Sherwin, ``{Reconstructing Small Scale
  Lenses from the Cosmic Microwave Background Temperature Fluctuations},''
  \href{http://dx.doi.org/10.1093/mnras/stz566}{{\em Mon. Not. Roy. Astron.
  Soc.} {\bfseries 485} no.~3, (2019) 3919--3929},
  \href{http://arxiv.org/abs/1710.10236}{{\ttfamily arXiv:1710.10236
  [astro-ph.CO]}}.

\bibitem{Caldeira:2018ojb}
J.~a. Caldeira, W.~L.~K. Wu, B.~Nord, C.~Avestruz, S.~Trivedi, and K.~T. Story,
  ``{DeepCMB: Lensing Reconstruction of the Cosmic Microwave Background with
  Deep Neural Networks},''
  \href{http://dx.doi.org/10.1016/j.ascom.2019.100307}{{\em Astron. Comput.}
  {\bfseries 28} (2019) 100307},
  \href{http://arxiv.org/abs/1810.01483}{{\ttfamily arXiv:1810.01483
  [astro-ph.CO]}}.

\bibitem{Hadzhiyska:2019cle}
B.~Hadzhiyska, B.~D. Sherwin, M.~Madhavacheril, and S.~Ferraro, ``{Improving
  Small-Scale CMB Lensing Reconstruction},''
  \href{http://dx.doi.org/10.1103/PhysRevD.100.023547}{{\em Phys. Rev. D}
  {\bfseries 100} no.~2, (2019) 023547},
  \href{http://arxiv.org/abs/1905.04217}{{\ttfamily arXiv:1905.04217
  [astro-ph.CO]}}.

\bibitem{Millea:2020cpw}
M.~Millea, E.~Anderes, and B.~D. Wandelt, ``{Sampling-based inference of the
  primordial CMB and gravitational lensing},''
  \href{http://dx.doi.org/10.1103/PhysRevD.102.123542}{{\em Phys. Rev. D}
  {\bfseries 102} no.~12, (2020) 123542},
  \href{http://arxiv.org/abs/2002.00965}{{\ttfamily arXiv:2002.00965
  [astro-ph.CO]}}.

\bibitem{Dvorkin:2009ah}
C.~Dvorkin, W.~Hu, and K.~M. Smith, ``{B-mode CMB Polarization from Patchy
  Screening during Reionization},''
  \href{http://dx.doi.org/10.1103/PhysRevD.79.107302}{{\em Phys. Rev. D}
  {\bfseries 79} (2009) 107302},
  \href{http://arxiv.org/abs/0902.4413}{{\ttfamily arXiv:0902.4413
  [astro-ph.CO]}}.

\bibitem{Peebles:1970ag}
P.~J.~E. Peebles and J.~T. Yu, ``{Primeval adiabatic perturbation in an
  expanding universe},'' \href{http://dx.doi.org/10.1086/150713}{{\em
  Astrophys. J.} {\bfseries 162} (1970) 815--836}.

\bibitem{Hu:1994jd}
W.~Hu and N.~Sugiyama, ``{Toward understanding CMB anisotropies and their
  implications},'' \href{http://dx.doi.org/10.1103/PhysRevD.51.2599}{{\em Phys.
  Rev. D} {\bfseries 51} (1995) 2599--2630},
  \href{http://arxiv.org/abs/astro-ph/9411008}{{\ttfamily
  arXiv:astro-ph/9411008}}.

\bibitem{Pan:2016zla}
Z.~Pan, L.~Knox, B.~Mulroe, and A.~Narimani, ``{Cosmic Microwave Background
  Acoustic Peak Locations},''
  \href{http://dx.doi.org/10.1093/mnras/stw833}{{\em Mon. Not. Roy. Astron.
  Soc.} {\bfseries 459} no.~3, (2016) 2513--2524},
  \href{http://arxiv.org/abs/1603.03091}{{\ttfamily arXiv:1603.03091
  [astro-ph.CO]}}.

\bibitem{Planck:2018nkj}
{\bfseries Planck} Collaboration, N.~Aghanim {\em et~al.}, ``{Planck 2018
  results. I. Overview and the cosmological legacy of Planck},''
  \href{http://dx.doi.org/10.1051/0004-6361/201833880}{{\em Astron. Astrophys.}
  {\bfseries 641} (2020) A1}, \href{http://arxiv.org/abs/1807.06205}{{\ttfamily
  arXiv:1807.06205 [astro-ph.CO]}}.

\bibitem{Planck:2018vyg}
{\bfseries Planck} Collaboration, N.~Aghanim {\em et~al.}, ``{Planck 2018
  results. VI. Cosmological parameters},''
  \href{http://dx.doi.org/10.1051/0004-6361/201833910}{{\em Astron. Astrophys.}
  {\bfseries 641} (2020) A6}, \href{http://arxiv.org/abs/1807.06209}{{\ttfamily
  arXiv:1807.06209 [astro-ph.CO]}}. [Erratum: Astron.Astrophys. 652, C4
  (2021)].

\bibitem{Knox:2019rjx}
L.~Knox and M.~Millea, ``{Hubble constant hunter\textquoteright{}s guide},''
  \href{http://dx.doi.org/10.1103/PhysRevD.101.043533}{{\em Phys. Rev. D}
  {\bfseries 101} no.~4, (2020) 043533},
  \href{http://arxiv.org/abs/1908.03663}{{\ttfamily arXiv:1908.03663
  [astro-ph.CO]}}.

\bibitem{Bashinsky:2003tk}
S.~Bashinsky and U.~Seljak, ``{Neutrino perturbations in CMB anisotropy and
  matter clustering},''
  \href{http://dx.doi.org/10.1103/PhysRevD.69.083002}{{\em Phys. Rev.}
  {\bfseries D69} (2004) 083002},
\href{http://arxiv.org/abs/astro-ph/0310198}{{\ttfamily arXiv:astro-ph/0310198
  [astro-ph]}}.

\bibitem{Follin:2015hya}
B.~Follin, L.~Knox, M.~Millea, and Z.~Pan, ``{First Detection of the Acoustic
  Oscillation Phase Shift Expected from the Cosmic Neutrino Background},''
  \href{http://dx.doi.org/10.1103/PhysRevLett.115.091301}{{\em Phys. Rev.
  Lett.} {\bfseries 115} no.~9, (2015) 091301},
  \href{http://arxiv.org/abs/1503.07863}{{\ttfamily arXiv:1503.07863
  [astro-ph.CO]}}.

\bibitem{Baumann:2015rya}
D.~Baumann, D.~Green, J.~Meyers, and B.~Wallisch, ``{Phases of New Physics in
  the CMB},'' \href{http://dx.doi.org/10.1088/1475-7516/2016/01/007}{{\em JCAP}
  {\bfseries 1601} (2016) 007},
\href{http://arxiv.org/abs/1508.06342}{{\ttfamily arXiv:1508.06342
  [astro-ph.CO]}}.

\bibitem{Hou:2011ec}
Z.~Hou, R.~Keisler, L.~Knox, M.~Millea, and C.~Reichardt, ``{How Massless
  Neutrinos Affect the Cosmic Microwave Background Damping Tail},''
  \href{http://dx.doi.org/10.1103/PhysRevD.87.083008}{{\em Phys. Rev. D}
  {\bfseries 87} (2013) 083008},
  \href{http://arxiv.org/abs/1104.2333}{{\ttfamily arXiv:1104.2333
  [astro-ph.CO]}}.

\bibitem{1967Natur.215.1155S}
J.~{Silk}, ``{Fluctuations in the Primordial Fireball},''
  \href{http://dx.doi.org/10.1038/2151155a0}{{\em \nat} {\bfseries 215}
  no.~5106, (Sept., 1967) 1155--1156}.

\bibitem{Weinberg:1971mx}
S.~Weinberg, ``{Entropy generation and the survival of protogalaxies in an
  expanding universe},'' \href{http://dx.doi.org/10.1086/151073}{{\em
  Astrophys. J.} {\bfseries 168} (1971) 175}.

\bibitem{1983MNRAS.202.1169K}
N.~{Kaiser}, ``{Small-angle anisotropy of the microwave background radiation in
  the adiabatic theory},''
  \href{http://dx.doi.org/10.1093/mnras/202.4.1169}{{\em \mnras} {\bfseries
  202} (Mar., 1983) 1169--1180}.

\bibitem{Hu:1996vq}
W.~Hu and M.~J. White, ``{Acoustic signatures in the cosmic microwave
  background},'' \href{http://dx.doi.org/10.1086/177951}{{\em Astrophys. J.}
  {\bfseries 471} (1996) 30--51},
  \href{http://arxiv.org/abs/astro-ph/9602019}{{\ttfamily
  arXiv:astro-ph/9602019}}.

\bibitem{Zaldarriaga:1995gi}
M.~Zaldarriaga and D.~D. Harari, ``{Analytic approach to the polarization of
  the cosmic microwave background in flat and open universes},''
  \href{http://dx.doi.org/10.1103/PhysRevD.52.3276}{{\em Phys. Rev. D}
  {\bfseries 52} (1995) 3276--3287},
  \href{http://arxiv.org/abs/astro-ph/9504085}{{\ttfamily
  arXiv:astro-ph/9504085}}.

\bibitem{Williams:2020hqk}
J.~Williams, A.~Rotti, and R.~Battye, ``{Constraining cosmic polarization
  rotation and implications for primordial B-modes},''
  \href{http://dx.doi.org/10.1088/1475-7516/2020/09/006}{{\em JCAP} {\bfseries
  09} (2020) 006}, \href{http://arxiv.org/abs/2006.04899}{{\ttfamily
  arXiv:2006.04899 [astro-ph.CO]}}.

\bibitem{Santos:2003jb}
M.~G. Santos, A.~Cooray, Z.~Haiman, L.~Knox, and C.-P. Ma, ``{Small - scale CMB
  temperature and polarization anisotropies due to patchy reionization},''
  \href{http://dx.doi.org/10.1086/378772}{{\em Astrophys. J.} {\bfseries 598}
  (2003) 756--766}, \href{http://arxiv.org/abs/astro-ph/0305471}{{\ttfamily
  arXiv:astro-ph/0305471}}.

\bibitem{Zahn:2005fn}
O.~Zahn, M.~Zaldarriaga, L.~Hernquist, and M.~McQuinn, ``{The Influence of
  non-uniform reionization on the CMB},''
  \href{http://dx.doi.org/10.1086/431947}{{\em Astrophys. J.} {\bfseries 630}
  (2005) 657--666}, \href{http://arxiv.org/abs/astro-ph/0503166}{{\ttfamily
  arXiv:astro-ph/0503166}}.

\bibitem{McQuinn:2005ce}
M.~McQuinn, S.~R. Furlanetto, L.~Hernquist, O.~Zahn, and M.~Zaldarriaga, ``{The
  Kinetic Sunyaev-Zel'dovich effect from reionization},''
  \href{http://dx.doi.org/10.1086/432049}{{\em Astrophys. J.} {\bfseries 630}
  (2005) 643--656}, \href{http://arxiv.org/abs/astro-ph/0504189}{{\ttfamily
  arXiv:astro-ph/0504189}}.

\bibitem{Dore:2007bz}
O.~Dore, G.~Holder, M.~Alvarez, I.~T. Iliev, G.~Mellema, U.-L. Pen, and P.~R.
  Shapiro, ``{The Signature of Patchy Reionization in the Polarization
  Anisotropy of the CMB},''
  \href{http://dx.doi.org/10.1103/PhysRevD.76.043002}{{\em Phys. Rev. D}
  {\bfseries 76} (2007) 043002},
  \href{http://arxiv.org/abs/astro-ph/0701784}{{\ttfamily
  arXiv:astro-ph/0701784}}.

\bibitem{Carroll:1989vb}
S.~M. Carroll, G.~B. Field, and R.~Jackiw, ``{Limits on a Lorentz and Parity
  Violating Modification of Electrodynamics},''
  \href{http://dx.doi.org/10.1103/PhysRevD.41.1231}{{\em Phys. Rev. D}
  {\bfseries 41} (1990) 1231}.

\bibitem{Harari:1992ea}
D.~Harari and P.~Sikivie, ``{Effects of a Nambu-Goldstone boson on the
  polarization of radio galaxies and the cosmic microwave background},''
  \href{http://dx.doi.org/10.1016/0370-2693(92)91363-E}{{\em Phys. Lett. B}
  {\bfseries 289} (1992) 67--72}.

\bibitem{Carroll:1998zi}
S.~M. Carroll, ``{Quintessence and the rest of the world},''
  \href{http://dx.doi.org/10.1103/PhysRevLett.81.3067}{{\em Phys. Rev. Lett.}
  {\bfseries 81} (1998) 3067--3070},
  \href{http://arxiv.org/abs/astro-ph/9806099}{{\ttfamily
  arXiv:astro-ph/9806099}}.

\bibitem{Lue:1998mq}
A.~Lue, L.-M. Wang, and M.~Kamionkowski, ``{Cosmological signature of new
  parity violating interactions},''
  \href{http://dx.doi.org/10.1103/PhysRevLett.83.1506}{{\em Phys. Rev. Lett.}
  {\bfseries 83} (1999) 1506--1509},
  \href{http://arxiv.org/abs/astro-ph/9812088}{{\ttfamily
  arXiv:astro-ph/9812088}}.

\bibitem{Kosowsky:1996yc}
A.~Kosowsky and A.~Loeb, ``{Faraday rotation of microwave background
  polarization by a primordial magnetic field},''
  \href{http://dx.doi.org/10.1086/177751}{{\em Astrophys. J.} {\bfseries 469}
  (1996) 1--6}, \href{http://arxiv.org/abs/astro-ph/9601055}{{\ttfamily
  arXiv:astro-ph/9601055}}.

\bibitem{Kamionkowski:1997na}
M.~Kamionkowski and A.~Loeb, ``{Getting around cosmic variance},''
  \href{http://dx.doi.org/10.1103/PhysRevD.56.4511}{{\em Phys. Rev. D}
  {\bfseries 56} (1997) 4511--4513},
  \href{http://arxiv.org/abs/astro-ph/9703118}{{\ttfamily
  arXiv:astro-ph/9703118}}.

\bibitem{Sazonov:1999zp}
S.~Y. Sazonov and R.~A. Sunyaev, ``{Microwave polarization in the direction of
  galaxy clusters induced by the CMB quadrupole anisotropy},''
  \href{http://dx.doi.org/10.1046/j.1365-8711.1999.02981.x}{{\em Mon. Not. Roy.
  Astron. Soc.} {\bfseries 310} (1999) 765--772},
  \href{http://arxiv.org/abs/astro-ph/9903287}{{\ttfamily
  arXiv:astro-ph/9903287}}.

\bibitem{1983Natur.302..315B}
M.~{Birkinshaw} and S.~F. {Gull}, ``{A test for transverse motions of clusters
  of galaxies},'' \href{http://dx.doi.org/10.1038/302315a0}{{\em \nat}
  {\bfseries 302} no.~5906, (Mar., 1983) 315--317}.

\bibitem{Dvorkin:2008tf}
C.~Dvorkin and K.~M. Smith, ``{Reconstructing Patchy Reionization from the
  Cosmic Microwave Background},''
  \href{http://dx.doi.org/10.1103/PhysRevD.79.043003}{{\em Phys. Rev. D}
  {\bfseries 79} (2009) 043003},
  \href{http://arxiv.org/abs/0812.1566}{{\ttfamily arXiv:0812.1566
  [astro-ph]}}.

\bibitem{Smith:2016lnt}
K.~M. Smith and S.~Ferraro, ``{Detecting Patchy Reionization in the Cosmic
  Microwave Background},''
  \href{http://dx.doi.org/10.1103/PhysRevLett.119.021301}{{\em Phys. Rev.
  Lett.} {\bfseries 119} no.~2, (2017) 021301},
  \href{http://arxiv.org/abs/1607.01769}{{\ttfamily arXiv:1607.01769
  [astro-ph.CO]}}.

\bibitem{Kamionkowski:2008fp}
M.~Kamionkowski, ``{How to De-Rotate the Cosmic Microwave Background
  Polarization},'' \href{http://dx.doi.org/10.1103/PhysRevLett.102.111302}{{\em
  Phys. Rev. Lett.} {\bfseries 102} (2009) 111302},
  \href{http://arxiv.org/abs/0810.1286}{{\ttfamily arXiv:0810.1286
  [astro-ph]}}.

\bibitem{Yadav:2009eb}
A.~P.~S. Yadav, R.~Biswas, M.~Su, and M.~Zaldarriaga, ``{Constraining a
  spatially dependent rotation of the Cosmic Microwave Background
  Polarization},'' \href{http://dx.doi.org/10.1103/PhysRevD.79.123009}{{\em
  Phys. Rev. D} {\bfseries 79} (2009) 123009},
  \href{http://arxiv.org/abs/0902.4466}{{\ttfamily arXiv:0902.4466
  [astro-ph.CO]}}.

\bibitem{Gluscevic:2009mm}
V.~Gluscevic, M.~Kamionkowski, and A.~Cooray, ``{De-Rotation of the Cosmic
  Microwave Background Polarization: Full-Sky Formalism},''
  \href{http://dx.doi.org/10.1103/PhysRevD.80.023510}{{\em Phys. Rev. D}
  {\bfseries 80} (2009) 023510},
  \href{http://arxiv.org/abs/0905.1687}{{\ttfamily arXiv:0905.1687
  [astro-ph.CO]}}.

\bibitem{Alizadeh:2012vy}
E.~Alizadeh and C.~M. Hirata, ``{How to detect gravitational waves through the
  cross-correlation of the galaxy distribution with the CMB polarization},''
  \href{http://dx.doi.org/10.1103/PhysRevD.85.123540}{{\em Phys. Rev. D}
  {\bfseries 85} (2012) 123540},
  \href{http://arxiv.org/abs/1201.5374}{{\ttfamily arXiv:1201.5374
  [astro-ph.CO]}}.

\bibitem{Deutsch:2017cja}
A.-S. Deutsch, M.~C. Johnson, M.~M\"unchmeyer, and A.~Terrana, ``{Polarized
  Sunyaev Zel'dovich tomography},''
  \href{http://dx.doi.org/10.1088/1475-7516/2018/04/034}{{\em JCAP} {\bfseries
  04} (2018) 034}, \href{http://arxiv.org/abs/1705.08907}{{\ttfamily
  arXiv:1705.08907 [astro-ph.CO]}}.

\bibitem{Deutsch:2017ybc}
A.-S. {Deutsch}, E.~{Dimastrogiovanni}, M.~C. {Johnson}, M.~{M{\"u}nchmeyer},
  and A.~{Terrana}, ``{Reconstruction of the remote dipole and quadrupole
  fields from the kinetic Sunyaev Zel'dovich and polarized Sunyaev Zel'dovich
  effects},'' \href{http://dx.doi.org/10.1103/PhysRevD.98.123501}{{\em \prd}
  {\bfseries 98} no.~12, (December, 2018) 123501},
  \href{http://arxiv.org/abs/1707.08129}{{\ttfamily arXiv:1707.08129
  [astro-ph.CO]}}.

\bibitem{Meyers:2017rtf}
J.~Meyers, P.~D. Meerburg, A.~van Engelen, and N.~Battaglia, ``{Beyond CMB
  cosmic variance limits on reionization with the polarized
  Sunyaev-Zel\textquoteright{}dovich effect},''
  \href{http://dx.doi.org/10.1103/PhysRevD.97.103505}{{\em Phys. Rev. D}
  {\bfseries 97} no.~10, (2018) 103505},
  \href{http://arxiv.org/abs/1710.01708}{{\ttfamily arXiv:1710.01708
  [astro-ph.CO]}}.

\bibitem{Hotinli:2018yyc}
S.~C. Hotinli, J.~Meyers, N.~Dalal, A.~H. Jaffe, M.~C. Johnson, J.~B. Mertens,
  M.~M\"unchmeyer, K.~M. Smith, and A.~van Engelen, ``{Transverse Velocities
  with the Moving Lens Effect},''
  \href{http://dx.doi.org/10.1103/PhysRevLett.123.061301}{{\em Phys. Rev.
  Lett.} {\bfseries 123} no.~6, (2019) 061301},
  \href{http://arxiv.org/abs/1812.03167}{{\ttfamily arXiv:1812.03167
  [astro-ph.CO]}}.

\bibitem{Hotinli:2020ntd}
S.~C. Hotinli, M.~C. Johnson, and J.~Meyers, ``{Optimal filters for the moving
  lens effect},'' \href{http://dx.doi.org/10.1103/PhysRevD.103.043536}{{\em
  Phys. Rev. D} {\bfseries 103} no.~4, (2021) 043536},
  \href{http://arxiv.org/abs/2006.03060}{{\ttfamily arXiv:2006.03060
  [astro-ph.CO]}}.

\bibitem{Hotinli:2021hih}
S.~C. Hotinli, K.~M. Smith, M.~S. Madhavacheril, and M.~Kamionkowski,
  ``{Cosmology with the moving lens effect},''
  \href{http://dx.doi.org/10.1103/PhysRevD.104.083529}{{\em Phys. Rev. D}
  {\bfseries 104} no.~8, (2021) 083529},
  \href{http://arxiv.org/abs/2108.02207}{{\ttfamily arXiv:2108.02207
  [astro-ph.CO]}}.

\bibitem{Hotinli:2020csk}
S.~C. Hotinli and M.~C. Johnson, ``{Reconstructing large scales at cosmic
  dawn},'' \href{http://arxiv.org/abs/2012.09851}{{\ttfamily arXiv:2012.09851
  [astro-ph.CO]}}.

\bibitem{Cayuso:2021ljq}
J.~Cayuso, R.~Bloch, S.~C. Hotinli, M.~C. Johnson, and F.~McCarthy, ``{Velocity
  reconstruction with the cosmic microwave background and galaxy surveys},''
  \href{http://arxiv.org/abs/2111.11526}{{\ttfamily arXiv:2111.11526
  [astro-ph.CO]}}.

\bibitem{Guzman:2021nfk}
E.~Guzman and J.~Meyers, ``{Reconstructing patchy reionization with deep
  learning},'' \href{http://dx.doi.org/10.1103/PhysRevD.104.043529}{{\em Phys.
  Rev. D} {\bfseries 104} no.~4, (2021) 043529},
  \href{http://arxiv.org/abs/2101.01214}{{\ttfamily arXiv:2101.01214
  [astro-ph.CO]}}.

\bibitem{Guzman:2021ygf}
E.~Guzman and J.~Meyers, ``{Reconstructing Cosmic Polarization Rotation with
  ResUNet-CMB},'' \href{http://arxiv.org/abs/2109.09715}{{\ttfamily
  arXiv:2109.09715 [astro-ph.CO]}}.

\bibitem{Su:2011ff}
M.~Su, A.~P.~S. Yadav, M.~McQuinn, J.~Yoo, and M.~Zaldarriaga, ``{An Improved
  Forecast of Patchy Reionization Reconstruction with CMB},''
  \href{http://arxiv.org/abs/1106.4313}{{\ttfamily arXiv:1106.4313
  [astro-ph.CO]}}.

\bibitem{Smith:2018bpn}
K.~M. {Smith}, M.~S. {Madhavacheril}, M.~{M{\"u}nchmeyer}, S.~{Ferraro},
  U.~{Giri}, and M.~C. {Johnson}, ``{KSZ tomography and the bispectrum},'' {\em
  arXiv e-prints} (October, 2018) arXiv:1810.13423,
  \href{http://arxiv.org/abs/1810.13423}{{\ttfamily arXiv:1810.13423
  [astro-ph.CO]}}.

\bibitem{Munchmeyer:2018eey}
M.~{M{\"u}nchmeyer}, M.~S. {Madhavacheril}, S.~{Ferraro}, M.~C. {Johnson}, and
  K.~M. {Smith}, ``{Constraining local non-Gaussianities with kinetic
  Sunyaev-Zel'dovich tomography},''
  \href{http://dx.doi.org/10.1103/PhysRevD.100.083508}{{\em \prd} {\bfseries
  100} no.~8, (October, 2019) 083508},
  \href{http://arxiv.org/abs/1810.13424}{{\ttfamily arXiv:1810.13424
  [astro-ph.CO]}}.

\bibitem{Zhang:2015uta}
P.~{Zhang} and M.~C. {Johnson}, ``{Testing eternal inflation with the kinetic
  Sunyaev Zel'dovich effect},''
  \href{http://dx.doi.org/10.1088/1475-7516/2015/06/046}{{\em \jcap} {\bfseries
  2015} no.~6, (June, 2015) 046},
  \href{http://arxiv.org/abs/1501.00511}{{\ttfamily arXiv:1501.00511
  [astro-ph.CO]}}.

\bibitem{Hotinli:2019wdp}
S.~C. Hotinli, J.~B. Mertens, M.~C. Johnson, and M.~Kamionkowski, ``{Probing
  correlated compensated isocurvature perturbations using scale-dependent
  galaxy bias},'' \href{http://dx.doi.org/10.1103/PhysRevD.100.103528}{{\em
  Phys. Rev.} {\bfseries D100} no.~10, (2019) 103528},
\href{http://arxiv.org/abs/1908.08953}{{\ttfamily arXiv:1908.08953
  [astro-ph.CO]}}.

\bibitem{Cayuso:2019hen}
J.~I. {Cayuso} and M.~C. {Johnson}, ``{Towards testing CMB anomalies using the
  kinetic and polarized Sunyaev-Zel'dovich effects},''
  \href{http://dx.doi.org/10.1103/PhysRevD.101.123508}{{\em \prd} {\bfseries
  101} no.~12, (June, 2020) 123508},
  \href{http://arxiv.org/abs/1904.10981}{{\ttfamily arXiv:1904.10981
  [astro-ph.CO]}}.

\bibitem{Alvarez:2020gvl}
M.~A. Alvarez, S.~Ferraro, J.~C. Hill, R.~Hlo\v{z}ek, and M.~Ikape,
  ``{Mitigating the optical depth degeneracy using the kinematic
  Sunyaev-Zel'dovich effect with CMB-S4},''
  \href{http://dx.doi.org/10.1103/PhysRevD.103.063518}{{\em Phys. Rev. D}
  {\bfseries 103} no.~6, (2021) 063518},
  \href{http://arxiv.org/abs/2006.06594}{{\ttfamily arXiv:2006.06594
  [astro-ph.CO]}}.

\bibitem{Deutsch:2018umo}
A.-S. Deutsch, E.~Dimastrogiovanni, M.~Fasiello, M.~C. Johnson, and
  M.~M\"unchmeyer, ``{Primordial gravitational wave phenomenology with
  polarized Sunyaev Zel\textquoteright{}dovich tomography},''
  \href{http://dx.doi.org/10.1103/PhysRevD.100.083538}{{\em Phys. Rev. D}
  {\bfseries 100} no.~8, (2019) 083538},
  \href{http://arxiv.org/abs/1810.09463}{{\ttfamily arXiv:1810.09463
  [astro-ph.CO]}}.

\bibitem{Zaldarriaga:2000ud}
M.~Zaldarriaga, ``{Lensing of the CMB: Non-Gaussian aspects},''
  \href{http://dx.doi.org/10.1103/PhysRevD.62.063510}{{\em Phys. Rev. D}
  {\bfseries 62} (2000) 063510},
  \href{http://arxiv.org/abs/astro-ph/9910498}{{\ttfamily
  arXiv:astro-ph/9910498}}.

\bibitem{Hu:2001fa}
W.~Hu, ``{Angular trispectrum of the CMB},''
  \href{http://dx.doi.org/10.1103/PhysRevD.64.083005}{{\em Phys. Rev. D}
  {\bfseries 64} (2001) 083005},
  \href{http://arxiv.org/abs/astro-ph/0105117}{{\ttfamily
  arXiv:astro-ph/0105117}}.

\bibitem{Hu:2001tn}
W.~Hu, ``{Mapping the dark matter through the cmb damping tail},''
  \href{http://dx.doi.org/10.1086/323253}{{\em Astrophys. J. Lett.} {\bfseries
  557} (2001) L79--L83},
  \href{http://arxiv.org/abs/astro-ph/0105424}{{\ttfamily
  arXiv:astro-ph/0105424}}.

\bibitem{Smith:2004up}
K.~M. Smith, W.~Hu, and M.~Kaplinghat, ``{Weak lensing of the CMB: Sampling
  errors on B-modes},''
  \href{http://dx.doi.org/10.1103/PhysRevD.70.043002}{{\em Phys. Rev. D}
  {\bfseries 70} (2004) 043002},
  \href{http://arxiv.org/abs/astro-ph/0402442}{{\ttfamily
  arXiv:astro-ph/0402442}}.

\bibitem{Smith:2005ue}
S.~Smith, A.~Challinor, and G.~Rocha, ``{What can be learned from the lensed
  cosmic microwave background b-mode polarization power spectrum?},''
  \href{http://dx.doi.org/10.1103/PhysRevD.73.023517}{{\em Phys. Rev. D}
  {\bfseries 73} (2006) 023517},
  \href{http://arxiv.org/abs/astro-ph/0511703}{{\ttfamily
  arXiv:astro-ph/0511703}}.

\bibitem{Smith:2006nk}
K.~M. Smith, W.~Hu, and M.~Kaplinghat, ``{Cosmological Information from Lensed
  CMB Power Spectra},''
  \href{http://dx.doi.org/10.1103/PhysRevD.74.123002}{{\em Phys. Rev. D}
  {\bfseries 74} (2006) 123002},
  \href{http://arxiv.org/abs/astro-ph/0607315}{{\ttfamily
  arXiv:astro-ph/0607315}}.

\bibitem{Li:2006pu}
C.~Li, T.~L. Smith, and A.~Cooray, ``{Non-Gaussian Covariance of CMB B-modes of
  Polarization and Parameter Degradation},''
  \href{http://dx.doi.org/10.1103/PhysRevD.75.083501}{{\em Phys. Rev. D}
  {\bfseries 75} (2007) 083501},
  \href{http://arxiv.org/abs/astro-ph/0607494}{{\ttfamily
  arXiv:astro-ph/0607494}}.

\bibitem{Benoit-Levy:2012dqi}
A.~Benoit-Levy, K.~M. Smith, and W.~Hu, ``{Non-Gaussian structure of the lensed
  CMB power spectra covariance matrix},''
  \href{http://dx.doi.org/10.1103/PhysRevD.86.123008}{{\em Phys. Rev. D}
  {\bfseries 86} (2012) 123008},
  \href{http://arxiv.org/abs/1205.0474}{{\ttfamily arXiv:1205.0474
  [astro-ph.CO]}}.

\bibitem{Schmittfull:2013uea}
M.~M. Schmittfull, A.~Challinor, D.~Hanson, and A.~Lewis, ``{Joint analysis of
  CMB temperature and lensing-reconstruction power spectra},''
  \href{http://dx.doi.org/10.1103/PhysRevD.88.063012}{{\em Phys. Rev. D}
  {\bfseries 88} no.~6, (2013) 063012},
  \href{http://arxiv.org/abs/1308.0286}{{\ttfamily arXiv:1308.0286
  [astro-ph.CO]}}.

\bibitem{Peloton:2016kbw}
J.~Peloton, M.~Schmittfull, A.~Lewis, J.~Carron, and O.~Zahn, ``{Full
  covariance of CMB and lensing reconstruction power spectra},''
  \href{http://dx.doi.org/10.1103/PhysRevD.95.043508}{{\em Phys. Rev. D}
  {\bfseries 95} no.~4, (2017) 043508},
  \href{http://arxiv.org/abs/1611.01446}{{\ttfamily arXiv:1611.01446
  [astro-ph.CO]}}.

\bibitem{Hu:2001fb}
W.~Hu, ``{Dark synergy: Gravitational lensing and the CMB},''
  \href{http://dx.doi.org/10.1103/PhysRevD.65.023003}{{\em Phys. Rev. D}
  {\bfseries 65} (2002) 023003},
  \href{http://arxiv.org/abs/astro-ph/0108090}{{\ttfamily
  arXiv:astro-ph/0108090}}.

\bibitem{Coulton:2019odk}
W.~R. Coulton, P.~D. Meerburg, D.~G. Baker, S.~Hotinli, A.~J. Duivenvoorden,
  and A.~van Engelen, ``{Minimizing gravitational lensing contributions to the
  primordial bispectrum covariance},''
  \href{http://dx.doi.org/10.1103/PhysRevD.101.123504}{{\em Phys. Rev. D}
  {\bfseries 101} no.~12, (2020) 123504},
  \href{http://arxiv.org/abs/1912.07619}{{\ttfamily arXiv:1912.07619
  [astro-ph.CO]}}.

\bibitem{Allison:2015qca}
R.~Allison, P.~Caucal, E.~Calabrese, J.~Dunkley, and T.~Louis, ``{Towards a
  cosmological neutrino mass detection},''
  \href{http://dx.doi.org/10.1103/PhysRevD.92.123535}{{\em Phys. Rev. D}
  {\bfseries 92} no.~12, (2015) 123535},
  \href{http://arxiv.org/abs/1509.07471}{{\ttfamily arXiv:1509.07471
  [astro-ph.CO]}}.

\bibitem{Wang:2008zh}
Y.~Wang, ``{Figure of Merit for Dark Energy Constraints from Current
  Observational Data},''
  \href{http://dx.doi.org/10.1103/PhysRevD.77.123525}{{\em Phys. Rev. D}
  {\bfseries 77} (2008) 123525},
  \href{http://arxiv.org/abs/0803.4295}{{\ttfamily arXiv:0803.4295
  [astro-ph]}}.

\bibitem{Polarski:1994rz}
D.~Polarski and A.~A. Starobinsky, ``{Isocurvature perturbations in multiple
  inflationary models},''
  \href{http://dx.doi.org/10.1103/PhysRevD.50.6123}{{\em Phys. Rev. D}
  {\bfseries 50} (1994) 6123--6129},
  \href{http://arxiv.org/abs/astro-ph/9404061}{{\ttfamily
  arXiv:astro-ph/9404061}}.

\bibitem{Gordon:2000hv}
C.~Gordon, D.~Wands, B.~A. Bassett, and R.~Maartens, ``{Adiabatic and entropy
  perturbations from inflation},''
  \href{http://dx.doi.org/10.1103/PhysRevD.63.023506}{{\em Phys. Rev. D}
  {\bfseries 63} (2000) 023506},
  \href{http://arxiv.org/abs/astro-ph/0009131}{{\ttfamily
  arXiv:astro-ph/0009131}}.

\bibitem{Bucher:2000hy}
M.~Bucher, K.~Moodley, and N.~Turok, ``{Constraining isocurvature perturbations
  with CMB polarization},''
  \href{http://dx.doi.org/10.1103/PhysRevLett.87.191301}{{\em Phys. Rev. Lett.}
  {\bfseries 87} (2001) 191301},
  \href{http://arxiv.org/abs/astro-ph/0012141}{{\ttfamily
  arXiv:astro-ph/0012141}}.

\bibitem{Valiviita:2012ub}
J.~Valiviita, M.~Savelainen, M.~Talvitie, H.~Kurki-Suonio, and S.~Rusak,
  ``{Constraints on scalar and tensor perturbations in phenomenological and
  two-field inflation models: Bayesian evidences for primordial isocurvature
  and tensor modes},''
  \href{http://dx.doi.org/10.1088/0004-637X/753/2/151}{{\em Astrophys. J.}
  {\bfseries 753} (2012) 151}, \href{http://arxiv.org/abs/1202.2852}{{\ttfamily
  arXiv:1202.2852 [astro-ph.CO]}}.

\bibitem{Bernardeau:2001qr}
F.~Bernardeau, S.~Colombi, E.~Gaztanaga, and R.~Scoccimarro, ``{Large scale
  structure of the universe and cosmological perturbation theory},''
  \href{http://dx.doi.org/10.1016/S0370-1573(02)00135-7}{{\em Phys. Rept.}
  {\bfseries 367} (2002) 1--248},
  \href{http://arxiv.org/abs/astro-ph/0112551}{{\ttfamily
  arXiv:astro-ph/0112551}}.

\bibitem{Smith:2002dz}
{\bfseries VIRGO Consortium} Collaboration, R.~E. Smith, J.~A. Peacock,
  A.~Jenkins, S.~D.~M. White, C.~S. Frenk, F.~R. Pearce, P.~A. Thomas,
  G.~Efstathiou, and H.~M.~P. Couchmann, ``{Stable clustering, the halo model
  and nonlinear cosmological power spectra},''
  \href{http://dx.doi.org/10.1046/j.1365-8711.2003.06503.x}{{\em Mon. Not. Roy.
  Astron. Soc.} {\bfseries 341} (2003) 1311},
  \href{http://arxiv.org/abs/astro-ph/0207664}{{\ttfamily
  arXiv:astro-ph/0207664}}.

\bibitem{Takahashi:2012em}
R.~Takahashi, M.~Sato, T.~Nishimichi, A.~Taruya, and M.~Oguri, ``{Revising the
  Halofit Model for the Nonlinear Matter Power Spectrum},''
  \href{http://dx.doi.org/10.1088/0004-637X/761/2/152}{{\em Astrophys. J.}
  {\bfseries 761} (2012) 152}, \href{http://arxiv.org/abs/1208.2701}{{\ttfamily
  arXiv:1208.2701 [astro-ph.CO]}}.

\bibitem{Bird:2011rb}
S.~Bird, M.~Viel, and M.~G. Haehnelt, ``{Massive Neutrinos and the Non-linear
  Matter Power Spectrum},''
  \href{http://dx.doi.org/10.1111/j.1365-2966.2011.20222.x}{{\em Mon. Not. Roy.
  Astron. Soc.} {\bfseries 420} (2012) 2551--2561},
  \href{http://arxiv.org/abs/1109.4416}{{\ttfamily arXiv:1109.4416
  [astro-ph.CO]}}.

\bibitem{Peacock:2000qk}
J.~A. Peacock and R.~E. Smith, ``{Halo occupation numbers and galaxy bias},''
  \href{http://dx.doi.org/10.1046/j.1365-8711.2000.03779.x}{{\em Mon. Not. Roy.
  Astron. Soc.} {\bfseries 318} (2000) 1144},
  \href{http://arxiv.org/abs/astro-ph/0005010}{{\ttfamily
  arXiv:astro-ph/0005010}}.

\bibitem{Seljak:2000gq}
U.~Seljak, ``{Analytic model for galaxy and dark matter clustering},''
  \href{http://dx.doi.org/10.1046/j.1365-8711.2000.03715.x}{{\em Mon. Not. Roy.
  Astron. Soc.} {\bfseries 318} (2000) 203},
  \href{http://arxiv.org/abs/astro-ph/0001493}{{\ttfamily
  arXiv:astro-ph/0001493}}.

\bibitem{Cooray:2002dia}
A.~Cooray and R.~K. Sheth, ``{Halo Models of Large Scale Structure},''
  \href{http://dx.doi.org/10.1016/S0370-1573(02)00276-4}{{\em Phys. Rept.}
  {\bfseries 372} (2002) 1--129},
  \href{http://arxiv.org/abs/astro-ph/0206508}{{\ttfamily
  arXiv:astro-ph/0206508}}.

\bibitem{White:2004kv}
M.~J. White, ``{Baryons and weak lensing power spectra},''
  \href{http://dx.doi.org/10.1016/j.astropartphys.2004.06.001}{{\em Astropart.
  Phys.} {\bfseries 22} (2004) 211--217},
  \href{http://arxiv.org/abs/astro-ph/0405593}{{\ttfamily
  arXiv:astro-ph/0405593}}.

\bibitem{Zhan:2004wq}
H.~Zhan and L.~Knox, ``{Effect of hot baryons on the weak-lensing shear power
  spectrum},'' \href{http://dx.doi.org/10.1086/426712}{{\em Astrophys. J.
  Lett.} {\bfseries 616} (2004) L75--L78},
  \href{http://arxiv.org/abs/astro-ph/0409198}{{\ttfamily
  arXiv:astro-ph/0409198}}.

\bibitem{Jing:2005gm}
Y.~P. Jing, P.~Zhang, W.~P. Lin, L.~Gao, and V.~Springel, ``{The influence of
  baryons on the clustering of matter and weak lensing surveys},''
  \href{http://dx.doi.org/10.1086/503547}{{\em Astrophys. J. Lett.} {\bfseries
  640} (2006) L119--L122},
  \href{http://arxiv.org/abs/astro-ph/0512426}{{\ttfamily
  arXiv:astro-ph/0512426}}.

\bibitem{Rudd:2007zx}
D.~H. Rudd, A.~R. Zentner, and A.~V. Kravtsov, ``{Effects of Baryons and
  Dissipation on the Matter Power Spectrum},''
  \href{http://dx.doi.org/10.1086/523836}{{\em Astrophys. J.} {\bfseries 672}
  (2008) 19--32}, \href{http://arxiv.org/abs/astro-ph/0703741}{{\ttfamily
  arXiv:astro-ph/0703741}}.

\bibitem{Semboloni:2011fe}
E.~Semboloni, H.~Hoekstra, J.~Schaye, M.~P. van Daalen, and I.~J. McCarthy,
  ``{Quantifying the effect of baryon physics on weak lensing tomography},''
  \href{http://dx.doi.org/10.1111/j.1365-2966.2011.19385.x}{{\em Mon. Not. Roy.
  Astron. Soc.} {\bfseries 417} (2011) 2020},
  \href{http://arxiv.org/abs/1105.1075}{{\ttfamily arXiv:1105.1075
  [astro-ph.CO]}}.

\bibitem{Natarajan:2014xba}
A.~Natarajan, A.~R. Zentner, N.~Battaglia, and H.~Trac, ``{Systematic errors in
  the measurement of neutrino masses due to baryonic feedback processes:
  Prospects for stage IV lensing surveys},''
  \href{http://dx.doi.org/10.1103/PhysRevD.90.063516}{{\em Phys. Rev. D}
  {\bfseries 90} no.~6, (2014) 063516},
  \href{http://arxiv.org/abs/1405.6205}{{\ttfamily arXiv:1405.6205
  [astro-ph.CO]}}.

\bibitem{Copeland:2019bho}
D.~Copeland, A.~Taylor, and A.~Hall, ``{Towards determining the neutrino mass
  hierarchy: weak lensing and galaxy clustering forecasts with baryons and
  intrinsic alignments},'' \href{http://dx.doi.org/10.1093/mnras/staa314}{{\em
  Mon. Not. Roy. Astron. Soc.} {\bfseries 493} no.~2, (2020) 1640--1661},
  \href{http://arxiv.org/abs/1905.08754}{{\ttfamily arXiv:1905.08754
  [astro-ph.CO]}}.

\bibitem{Schneider:2019xpf}
A.~Schneider, A.~Refregier, S.~Grandis, D.~Eckert, N.~Stoira, T.~Kacprzak,
  M.~Knabenhans, J.~Stadel, and R.~Teyssier, ``{Baryonic effects for weak
  lensing. Part II. Combination with X-ray data and extended cosmologies},''
  \href{http://dx.doi.org/10.1088/1475-7516/2020/04/020}{{\em JCAP} {\bfseries
  04} (2020) 020}, \href{http://arxiv.org/abs/1911.08494}{{\ttfamily
  arXiv:1911.08494 [astro-ph.CO]}}.

\bibitem{Chung:2019bsk}
E.~Chung, S.~Foreman, and A.~van Engelen, ``{Baryonic effects on CMB lensing
  and neutrino mass constraints},''
  \href{http://dx.doi.org/10.1103/PhysRevD.101.063534}{{\em Phys. Rev. D}
  {\bfseries 101} no.~6, (2020) 063534},
  \href{http://arxiv.org/abs/1910.09565}{{\ttfamily arXiv:1910.09565
  [astro-ph.CO]}}. [Erratum: Phys.Rev.D 102, 109903 (2020)].

\bibitem{McCarthy:2020dgq}
F.~McCarthy, S.~Foreman, and A.~van Engelen, ``{Avoiding baryonic feedback
  effects on neutrino mass measurements from CMB lensing},''
  \href{http://dx.doi.org/10.1103/PhysRevD.103.103538}{{\em Phys. Rev. D}
  {\bfseries 103} no.~10, (2021) 103538},
  \href{http://arxiv.org/abs/2011.06582}{{\ttfamily arXiv:2011.06582
  [astro-ph.CO]}}.

\bibitem{McCarthy:2021lfp}
F.~McCarthy, J.~C. Hill, and M.~S. Madhavacheril, ``{Baryonic feedback biases
  on fundamental physics from lensed CMB power spectra},''
  \href{http://arxiv.org/abs/2103.05582}{{\ttfamily arXiv:2103.05582
  [astro-ph.CO]}}.

\bibitem{Huterer:2004tr}
D.~Huterer and M.~Takada, ``{Calibrating the nonlinear matter power spectrum:
  Requirements for future weak lensing surveys},''
  \href{http://dx.doi.org/10.1016/j.astropartphys.2005.02.006}{{\em Astropart.
  Phys.} {\bfseries 23} (2005) 369--376},
  \href{http://arxiv.org/abs/astro-ph/0412142}{{\ttfamily
  arXiv:astro-ph/0412142}}.

\bibitem{LoVerde:2006cj}
M.~LoVerde, L.~Hui, and E.~Gaztanaga, ``{Magnification-Temperature Correlation:
  The Dark Side of ISW Measurements},''
  \href{http://dx.doi.org/10.1103/PhysRevD.75.043519}{{\em Phys. Rev. D}
  {\bfseries 75} (2007) 043519},
  \href{http://arxiv.org/abs/astro-ph/0611539}{{\ttfamily
  arXiv:astro-ph/0611539}}.

\bibitem{Amara:2007as}
A.~Amara and A.~Refregier, ``{Systematic Bias in Cosmic Shear: Beyond the
  Fisher Matrix},''
  \href{http://dx.doi.org/10.1111/j.1365-2966.2008.13880.x}{{\em Mon. Not. Roy.
  Astron. Soc.} {\bfseries 391} (2008) 228--236},
  \href{http://arxiv.org/abs/0710.5171}{{\ttfamily arXiv:0710.5171
  [astro-ph]}}.

\bibitem{Mead:2015yca}
A.~Mead, J.~Peacock, C.~Heymans, S.~Joudaki, and A.~Heavens, ``{An accurate
  halo model for fitting non-linear cosmological power spectra and baryonic
  feedback models},'' \href{http://dx.doi.org/10.1093/mnras/stv2036}{{\em Mon.
  Not. Roy. Astron. Soc.} {\bfseries 454} no.~2, (2015) 1958--1975},
  \href{http://arxiv.org/abs/1505.07833}{{\ttfamily arXiv:1505.07833
  [astro-ph.CO]}}.

\bibitem{Heitmann:2013bra}
K.~Heitmann, E.~Lawrence, J.~Kwan, S.~Habib, and D.~Higdon, ``{The Coyote
  Universe Extended: Precision Emulation of the Matter Power Spectrum},''
  \href{http://dx.doi.org/10.1088/0004-637X/780/1/111}{{\em Astrophys. J.}
  {\bfseries 780} (2014) 111}, \href{http://arxiv.org/abs/1304.7849}{{\ttfamily
  arXiv:1304.7849 [astro-ph.CO]}}.

\bibitem{Schaye:2009bt}
J.~Schaye, C.~Dalla~Vecchia, C.~M. Booth, R.~P.~C. Wiersma, T.~Theuns, M.~R.
  Haas, S.~Bertone, A.~R. Duffy, I.~G. McCarthy, and F.~van~de Voort, ``{The
  physics driving the cosmic star formation history},''
  \href{http://dx.doi.org/10.1111/j.1365-2966.2009.16029.x}{{\em Mon. Not. Roy.
  Astron. Soc.} {\bfseries 402} (2010) 1536},
  \href{http://arxiv.org/abs/0909.5196}{{\ttfamily arXiv:0909.5196
  [astro-ph.CO]}}.

\bibitem{vanDaalen:2011xb}
M.~P. van Daalen, J.~Schaye, C.~M. Booth, and C.~D. Vecchia, ``{The effects of
  galaxy formation on the matter power spectrum: A challenge for precision
  cosmology},'' \href{http://dx.doi.org/10.1111/j.1365-2966.2011.18981.x}{{\em
  Mon. Not. Roy. Astron. Soc.} {\bfseries 415} (2011) 3649--3665},
  \href{http://arxiv.org/abs/1104.1174}{{\ttfamily arXiv:1104.1174
  [astro-ph.CO]}}.

\bibitem{Perez:2007ipy}
F.~P\'{e}rez and B.~Granger, ``{IPython: A System for Interactive Scientific
  Computing},'' \href{http://dx.doi.org/10.1109/MCSE.2007.53}{{\em Comput. Sci.
  Eng.} {\bfseries 9} (2007) 21}.

\bibitem{Hunter:2007mat}
J.~Hunter, ``{Matplotlib: A 2D Graphics Environment},''
  \href{http://dx.doi.org/10.1109/MCSE.2007.55}{{\em Comput. Sci. Eng.}
  {\bfseries 9} (2007) 90}.

\bibitem{Harris:2020xlr}
C.~Harris {\em et~al.}, ``{Array Programming with NumPy},''
  \href{http://dx.doi.org/10.1038/s41586-020-2649-2}{{\em Nature} {\bfseries
  585} (2020) 3572}, \href{http://arxiv.org/abs/2006.10256}{{\ttfamily
  arXiv:2006.10256 [cs.MS]}}.

\bibitem{Virtanen:2019joe}
P.~Virtanen {\em et~al.}, ``{SciPy 1.0 -- Fundamental Algorithms for Scientific
  Computing in Python},''
  \href{http://dx.doi.org/10.1038/s41592-019-0686-2}{{\em Nat. Meth.}
  {\bfseries 17} (2020) 261}, \href{http://arxiv.org/abs/1907.10121}{{\ttfamily
  arXiv:1907.10121 [cs.MS]}}.

\bibitem{Hotinli:2020adc}
S.~C. Hotinli, \href{http://dx.doi.org/10.25560/85382}{{\em {New directions in
  cosmology and astrophysics}}}.
\newblock PhD thesis, Imperial Coll., London, 2020.

\end{thebibliography}\endgroup

\end{document}